\documentclass[
aps, pra, reprint,
amsfonts, amssymb, amsmath,
nofootinbib,
superscriptaddress,
floatfix]{revtex4-2}

\usepackage{graphicx}
\usepackage{physics}
\usepackage{amsthm}
\usepackage{hyperref}
\usepackage{xcolor}
\usepackage[linesnumbered,ruled,vlined]{algorithm2e}
\usepackage[nameinlink,capitalise]{cleveref}
\usepackage{booktabs}
\usepackage{tabularx}

\newtheorem{theorem}{Theorem}
\newtheorem{corollary}{Corollary}
\newtheorem{proposition}{Proposition}
\newtheorem{lemma}{Lemma}
\newtheorem{definition}{Definition}

\graphicspath{{Images/}}

\newcommand{\XC}{\mathcal{X}}
\newcommand{\GC}{\mathcal{G}}
\newcommand{\YC}{\mathcal{Y}}

\newcommand{\HC}{\mathcal{H}}

\newcommand{\ER}{R_{\mathrm{expl}}}
\newcommand{\IR}{R_{\mathrm{impl}}}
\newcommand{\Remp}{\widehat{R}}
\renewcommand{\vec}[1]{\boldsymbol{#1}}
\DeclareMathOperator{\poly}{poly}
\DeclareMathOperator*{\eb}{\mathbb{E}}
\DeclareMathOperator*{\argmin}{arg\,min}

\crefname{table}{Tab.}{Tabs.}
\crefname{section}{Sec.}{Secs.}
\crefname{appendix}{App.}{Apps.}
\crefname{algorithm}{Algo.}{Algos.}
\crefname{theorem}{Thm.}{Thms.}
\crefname{proposition}{Prop.}{Props.}
\crefname{definition}{Def.}{Defs.}
\crefname{lemma}{Lem.}{Lems.}
\crefname{corollary}{Cor.}{Cors.}

\Crefname{algorithm}{Algorithm}{Algorithms}
\Crefname{theorem}{Theorem}{Theorems}
\Crefname{lemma}{Lemma}{Lemmas}

\allowdisplaybreaks

\hypersetup{colorlinks,
citecolor=blue,
linkcolor=red,
urlcolor=purple}

\begin{document}

\title{Concept learning of parameterized quantum models from limited measurements}

\author{Beng Yee Gan}
\email{gan.bengyee@u.nus.edu}
\affiliation{Centre for Quantum Technologies, National University of Singapore, Singapore 117543}

\author{Po-Wei Huang}
\affiliation{Centre for Quantum Technologies, National University of Singapore, Singapore 117543}

\author{Elies Gil-Fuster}
\affiliation{Dahlem Center for Complex Quantum Systems, Freie Universit\"at Berlin, 14195 Berlin, Germany}
\affiliation{Fraunhofer Heinrich Hertz Institute, 10587 Berlin, Germany}

\author{Patrick Rebentrost}
\email{patrick@comp.nus.edu.sg}
\affiliation{Centre for Quantum Technologies, National University of Singapore, Singapore 117543}
\affiliation{Department of Computer Science, National University of Singapore, Singapore 117417}

\date{\today}

\begin{abstract}
Classical learning of the expectation values of observables for quantum states is a natural variant of learning quantum states or channels. 
While learning-theoretic frameworks establish the sample complexity and the number of measurement shots per sample required for learning such statistical quantities, the interplay between these two variables has not been adequately quantified before.
In this work, we take the probabilistic nature of quantum measurements into account in classical modelling and discuss these quantities under a single unified learning framework. We provide provable guarantees for learning parameterized quantum models that also quantify the asymmetrical effects and interplay of the two variables on the performance of learning algorithms.
These results show that while increasing the sample size enhances the learning performance of classical machines, even with single-shot estimates, 
the improvements from increasing measurements become asymptotically trivial beyond a constant factor.
We further apply our framework and theoretical guarantees to study the impact of measurement noise on the classical surrogation of parameterized quantum circuit models. 
Our work provides new tools to analyse the operational influence of finite measurement noise in the classical learning of quantum systems. 
\end{abstract}

\maketitle

\section{Introduction}

Understanding how classical learners can learn quantum systems sheds light on the potential and limitations of quantum information processing.
For instance, access to quantum measurement results~\cite{aaronson2020shadow, huang2020classical} in some cases allows the efficient learning of representations or models of the quantum system~\cite{huang2021power,huang2023learning,zhao2023learning,zhao2023provable,anshu2023survey,schreiber2023classical,huang2022provably,che2023exponentially,lewis2024improved}.
Examples of such \emph{quantum models} include expectation values of quantum observables given a variational quantum circuit~\cite{cerezo2020variationalreview} or ground state properties of quantum systems~\cite{huang2022provably,che2023exponentially,lewis2024improved}.
As quantum measurements on such systems are typically subjected to statistical fluctuations,
the observed outcomes may deviate from the true underlying quantum model due to the measurement or shot noise.
Thus, shot noise is an intrinsic aspect of learning quantum models.

Current works often consider shot noise as an error term that needs to be mitigated and rely on its minimization to ensure learnability~\cite{huang2022provably,che2023exponentially,schreiber2023classical}. 
In these scenarios, the effects of altering the number of training data input $N_1$ and the number of measurement shots per data input $N_s$ are typically discussed separately, with $N_s$ assumed to be sufficient enough as to not affect the analysis regarding $N_1$ and the learning performance.
In practice, however, conducting an abundant amount of measurements per data point may be too costly.
It is therefore important to understand how the interplay between the quantities $N_1$ and $N_s$ will affect the learnability performance of learning algorithms.

In statistical learning theory, such a quantum model is described as a \emph{concept}~\cite{valiant1984theory}, a function that maps inputs, say
high-dimensional
vectors, to \emph{deterministic} outputs expressed in the form of discrete class labels or continuous vectors.
However, stemming from the probabilistic nature of quantum measurements, there is no deterministic mapping of the input data to the observed outcomes that can capture the behaviour of the quantum model. 
Nonetheless, there is some structure to this uncertainty, as quantum models represent the conditional expectation of their unbiased estimators.
A natural learning-theoretic framework to capture such probabilistic models is the \emph{probabilistic concepts}~\cite{kearns1994efficient}, or $p$-concept learning framework.
The $p$-concept setting was first introduced in a quantum setup in Scott Aaronson’s seminal paper~\cite{aaronson2007learnability} and subsequently in Refs.~\cite{rocchetto2018stabiliser,rocchetto2019experimental,cheng2016learnability,caro2020pseudo}. 
These works identify shot noise as structural randomness, allowing the casting of quantum models as $p$-concepts. Fat-shattering dimension, a complexity measure that shows the expressiveness of a given set of $p$-concepts, is then used to quantify the difficulty of learning quantum states~\cite{aaronson2007learnability,rocchetto2018stabiliser,rocchetto2019experimental}, measurements~\cite{cheng2016learnability}, and quantum circuits~\cite{caro2020pseudo}.

In this work, we utilize this probabilistic framework to investigate the provable concept learning of parameterized quantum models.
Contrasting with prior works, our work utilizes kernel theory and the respective learnability results to investigate the impact of shot noise in classical learning of quantum models.
In particular, this unified learning framework enables the study of the interplay between $N_1$ and $N_s$ without isolating the discussion of sample complexity from shot noise.
That is, it allows us to tackle the following problem: \emph{given a well-defined learning setting and quantum model, can we obtain provable guarantees of learning that exemplify the relationship between $N_1$ and $N_s$?}

To answer this question, we extend the kernel method-based analysis of Ref.~\cite{goel2019learning} to showcase asymmetrical effects of $N_1$ and $N_s$ on concept learning.
Our analytical results show that increasing $N_1$ enhances the learning performance of classical models, even when observed outcomes are estimated with limited measurement shots, e.g., in the single-shot limit~\cite{recioarmengol2024singleshot} when $N_s=1$.
On the other hand, for a fixed training data size $N_1$, we find that improvements in learning performances from increasing the number of measurement shots $N_s$ become asymptotically trivial beyond a constant factor.
We also exhibit a scenario for which there exists an optimal pair of $N_1$ and $N_s$ that will maximize the performance of the classical models.

Under the lens of the bias-variance-noise decomposition of classical learning theory, we further illustrate the impact of shot noise on the bias and variance of classical models. 
Specifically, high shot noise can lead to high variance in classical models, which is consistent with observations in the literature~\cite{neal2019modern,yang2020rethinking}.
In this regard, we demonstrate that such sensitivity to shot noise can be suppressed by incorporating an $L$-Lipschitz non-decreasing function, known as a \emph{link function}~\cite{goel2019learning}, into the classical models for learning parameterized quantum models.
This suppression of variance enables us to separate and isolate the effects of the size of the training dataset and the number of measurement shots on learnability error bounds. Finally, we apply our framework to numerically study the impact of shot noise on the classical surrogation of parameterized quantum circuit models. 
The numerical results are consistent with our theoretical predictions.
Our work provides a new perspective to investigate the role of measurements in learning quantum models.

\section{Preliminaries}
\label{Sec:Preliminary}

In this section, we will first introduce two frameworks in statistical learning theory that we use to provide learning guarantees, the deterministic concept learning framework and the probabilistic concept learning framework. 
Then, 
we will provide a brief introduction to the types of quantum models considered in this work. 

\subsection{Probabilistic concept learning}
\label{Subsec:p-concept}

Let $\XC = \mathbb{R}^m$ and $\YC \subset \mathbb{R}$ be the data and label spaces, respectively.
Further, we assume that data points $\vec x$ are independently and identically distributed (i.i.d.) according to some unknown but fixed distribution $p(\vec{x})$ and the label space $\YC$ is a compact and convex subspace of $\mathbb{R}$.

In the learning-theoretic setting, there are two types of functions of interest: $\emph{concept}$ and $\emph{hypothesis}$.
A concept $c$ is a function that maps the data space to the label space, i.e., $c: \XC \rightarrow \YC$.  
A particular set of these functions with specific properties forms a \textit{concept class} $\mathcal{C}\subseteq\YC^\XC$. 
In the deterministic learning setting, a concept maps data points $\vec{x} \in \XC$ to associated labels $y \in \YC$, i.e., a data sample is given by $(\vec{x},y)$ where $\vec{x}$ is sampled from $p(\vec{x})$ and $y=c(\vec{x})$.
Similarly, a hypothesis is defined as $h: \XC \rightarrow \YC$, and a subset of these functions forms a $\emph{hypothesis class}$ $\HC\subseteq\YC^\XC$.
Then, given a collection of samples $\mathcal{S} = (\vec{x}_{i},c(\vec{x}_{i}))_{i=1}^{N_1}$where $\vec{x}_i \sim p(\vec{x})$, a learning algorithm selects a hypothesis $h \in \HC$ such that the difference between $h(\vec x)$ and the corresponding label $y=c(\vec{x})$ is low under some performance measure.

The \emph{probabilistic concept} ($p$-concept) and the $p$-concept class are defined similarly. 

\begin{definition}[$p$-concept]\label{Def:p-concept}
    Let $P_{\vec{x}}(\mathcal{Y})$ be a conditional probability distribution over the label space $\mathcal{Y}$, with probability density specified as $p(y|{\vec{x}})$ for each input $\vec{x}\in\mathcal{X}$.
    We call \emph{$p$-concept} a function $c:\mathcal{X}\to\mathcal{Y}$ defined as the conditional expectation value of $y$ given $\vec{x}$ arising from $p$:
	\begin{align}
		c:\mathcal{X} &\to\mathcal{Y} \\ \vec{x}&\mapsto c(\vec{x}) = \mathbb{E}_{y\sim p(y|\vec{x})}[y].
	\end{align}
\end{definition}

\begin{definition}[$p$-concept class]
    Let $\mathcal{P}\subseteq\{P_{\vec{x}}(\mathcal{Y})\}$ be a subset of all conditional probability distributions over the label space.
    For each distribution $P_{\vec{x}}(\mathcal{Y})\in\mathcal{P}$, which specifies the conditional distribution $p(y|\vec{x})$ of $y$ given $\vec{x}$, the corresponding $p$-concept is defined as per \cref{Def:p-concept}.
    Then, a \emph{$p$-concept class} is the class of functions $\mathcal{C}$ that corresponds to all functions arising from the set $\mathcal{P}$ of probability distributions:
	\begin{align}
        \mathcal{C} &:= \{c:\vec{x}\mapsto \mathbb{E}_{y\sim p(y|\vec{x})}[y] \,|\, P_{\vec{x}}(\mathcal{Y})\in\mathcal{P}\}.
    \end{align}
\end{definition}

Noteworthy is that defining this concept class explicitly from a set of conditional distributions does not impose any limitations on what these functions can be.
For any function $f\in\mathcal{Y}^\mathcal{X}$, one can always interpret it as a $p$-concept in an infinite number of ways.
For instance, one could consider simply the probability distribution that always returns the value of the function $p(y|\vec{x})=\delta(y-f(\vec{x}))$\footnote{In this case, the $p$-concepts reduce to the ``regular'' concepts defined in Ref.~\cite{valiant1984theory}.}.
Alternatively, one could consider any random function $\xi(\vec{x})$ with zero mean $\mathbb{E}[\xi(\vec{x})]=0$, and then one obtains a $p$-concept as the expectation value of the random function $f(\vec{x})+\xi(\vec{x})$.
Indeed, there are infinitely many different probability distributions that give rise to the same $p$-concept class\footnote{
The original definition of $p$-concepts given by \citet{kearns1994efficient} is simply a generalization of \emph{concepts} in PAC learning~\cite{valiant1984theory} in terms of the function range, while the actual probabilistic component is defined with the learnability of $p$-concept classes. 
Here we take a slightly different approach and define $p$-concepts such that the element of probability is captured within the definition of $p$-concepts itself.}.
Nonetheless, some of these distributions can be generated via physically realizable processes, which are the focus of this work.

Contrasting with the deterministic learning setting, in the $p$-concept learning setting, the samples $(\vec x,y)$ are obtained by sampling the joint distribution $\mathcal{D} = p(\vec{x})p(y|\vec{x})$ with $c(\vec{x}) = \mathbb{E}_{y\sim p(y|\vec{x})}[y]$. 
In this work, we further consider a flexible setting that allows for access to $p(y|\vec{x})$. 
That is, given a data point $\vec{x}$, we can obtain multiple i.i.d. random labels from $p(y|\vec{x})$, e.g., $y_1, \dots, y_{N_s} \sim p(y|\vec{x})$, and use these labels to estimate the empirical mean of the random labels $\bar{y}_{N_s} = \frac{1}{N_s}\sum_{i=1}^{N_s} y_i$. 
Such sampling then averaging procedure can be directly modelled as the sampling process $(\vec{x}, \bar{y}_{N_s}) \sim \bar{\mathcal{D}}_{N_s} = p({\vec{x}})p(\bar{y}_{N_s}|\vec{x})$ where $\bar{y}_{N_s}$ is distributed with variance $\sigma^2_{\bar{y}_{N_s}|\vec{x}} = \sigma^2_{y|\vec{x}}/N_s$ and $\sigma^2_{y|\vec{x}} = \operatorname{Var}_{y\sim p(y|\vec{x}_i)}[y]$. 
By construction, for all $N_s \in \mathbb{N}$, $p(\bar{y}_{N_s}|\vec{x})$ gives the same $p$-concept as $p(y|\vec{x})$, i.e.,
\begin{align}
	c(\vec{x}) = \mathbb{E}_{y\sim p(y|\vec{x})}[y] = \mathbb{E}_{\bar{y}_{N_s}\sim p(\bar{y}_{N_s}|\vec{x})}[\bar{y}_{N_s}].
\end{align}
For ease of notation, we let $c(\vec{x}) := \mathbb{E}_{\bar{y}_{N_s}}[\bar{y}_{N_s}|\vec{x}]$ and implicitly assume the dependence of $\bar{y}_{N_s}$ and $\bar{\mathcal{D}}_{N_s}$ on $N_s$, and denote them as $\bar{y}$ and $\bar{\mathcal{D}}$, respectively.

In the $p$-concept learning setting, a learning algorithm similarly selects a hypothesis $h$ from a hypothesis class $\HC$ such that the difference between $h(\vec x)$ and the corresponding $p$-concept $c(\vec{x})$ is low under some performance measure.
Here, we define two different performance measures:
explicit and implicit loss.
In particular, the explicit loss of $h$ is defined as
\begin{align}
    \ell_{\mathrm{expl}}(h) = (h(\vec{x}) - c(\vec{x}))^2
\end{align}
while the implicit loss of $h$ is defined as
\begin{align}
    \ell_{\mathrm{impl}}(h) = (h(\vec{x}) - \bar{y})^2.
\end{align}
That is, the explicit loss directly measures the performance of $h$ concerning the target $p$-concept $c(\vec{x})$, while the implicit loss indirectly quantifies the differences between $h(\vec{x})$ and $c(\vec{x})$ through the noisy labels $\bar{y}$ as $c(\vec{x})=\eb_{\bar{y}}[\bar{y}|\vec{x}]$.
Averaging both losses over all data points yields their respective risks: explicit risk
\begin{align} \label{Eq:Expected-risk}
    \ER(h) := \eb_{(\vec{x},\bar{y}) \sim \bar{\mathcal{D}}} [(h(\vec{x}) - c(\vec{x}))^2]
\end{align}
and implicit risk
\begin{align} \label{Eq:Exp-sq-err}
    \IR(h) := \eb_{(\vec{x},\bar{y}) \sim \bar{\mathcal{D}}} [(h(\vec{x}) - \bar{y})^2].
\end{align}
The decomposition of $\IR(h)$, i.e., 
\begin{align}
    \IR(h) &= \ER(h) + \eb_{(\vec{x},\bar{y}) \sim \bar{\mathcal{D}}} [(\mathbb{E}_{\bar{y}}[\bar{y}|\vec{x}] - \bar{y})^2] \\
    &= \ER(h) + \IR(\mathbb{E}_{\bar{y}}[\bar{y}|\vec{x}]),
\end{align}
shows that the implicit and explicit risks are related by a constant shift. Hence, a small $\IR(h)$ implies a small $\ER(h)$ and vice versa.

In practice, the distribution $\bar{\mathcal{D}}$ and the exact $p$-concept $c(\vec{x})$ are inaccessible as we only have access to finite samples drawn from the distribution $\mathcal{S} = (\vec{x}_i, \bar{y}_i)_{i=1}^{N_1}$ with $({\vec{x}_i, \bar{y}_i}) \sim \bar{\mathcal{D}}$ and $c(\vec{x}) = \mathbb{E}_{\bar{y}}[\bar{y}|\vec{x}]$.
Therefore, instead of minimizing the implicit or explicit risks, we will minimize the empirical (implicit) risk
\begin{align} \label{Eq:Empirical-risk}
    \Remp(h) := \frac{1}{N_1} \sum_{i=1}^{N_1} (h(\vec{x}_i) - \bar{y}_i)^2.
\end{align}
using samples $\mathcal{S}$ to obtain the optimal hypothesis that well approximates the underlying $p$-concept rather than the noisy label.
That is, we aim to achieve low $\ER$ by minimizing $\Remp$.
We have provided a glossary of error definitions in \cref{tab:Error-Glossary} for ease of reference.

There is nothing that formally distinguishes a $p$-concept from a ``regular'' concept in the probably approximately correct (PAC) framework~\cite{kearns1994efficient, valiant1984theory}.
Instead, the role of the probability distribution only comes forward when we talk about the learnability of such concepts.

\begin{table}
\begin{tabularx}{\linewidth}{>{\centering\arraybackslash}>{\hsize=.7\hsize}X
>{\centering\arraybackslash}>{\hsize=.6\hsize}X 
>{\centering\arraybackslash}X}
\toprule
Name            & Notation      & Definition \\
\midrule
Explicit loss   & $\ell_{\mathrm{expl}}(h)$        & $(h(\vec{x}) - c(\vec{x}))^2$ \\
Implicit loss   & $\ell_{\mathrm{impl}}(h)$        & $(h(\vec{x}) - \bar{y})^2$ \\
Explicit risk   & $\ER(h)$        & $\eb_{(\vec{x},\bar{y}) \sim \bar{\mathcal{D}}} [(h(\vec{x}) - c(\vec{x}))^2]$ \\
Implicit risk  & $\IR(h)$      & $\eb_{(\vec{x},\bar{y}) \sim \bar{\mathcal{D}}} [(h(\vec{x}) - \bar{y})^2]$\\
Empirical risk & $\Remp(h)$  & $\frac{1}{N_1} \sum_{i=1}^{N_1} (h(\vec{x}_i) - \bar{y}_i)^2$ \\
\bottomrule
\end{tabularx}
\caption{Glossary of error terms used in our paper. Explicit (implicit) loss and risk are associated with the $p$-concept $c(\vec{x})$ (noisy labels $\bar{y}$), while the empirical risk is the empirical version of the \emph{implicit} risk. We have omitted the subscript in the empirical risk for ease of notation.
}
\label{tab:Error-Glossary}
\end{table}

\begin{definition}[$p$-concept learning\label{Def:p-concept-learning}]
    Let $N_s \in \mathbb{N}$ be the number of random labels (per data input) and $\mathcal{P} = \{ p(\bar{y}|\vec{x}) \}$ be a set of conditional probability distributions over $\YC$ associated with a $p$-concept class
    \begin{align}
        \mathcal{C} &:= \{c:\vec{x}\mapsto \mathbb{E}_{\bar{y}\sim p(\bar{y}|\vec{x})}[\bar{y}] \,|\, p(\bar{y}|\vec{x})\in\mathcal{P}\}.
    \end{align}
    We say $\mathcal{C}$ is \emph{$p$-concept learnable} if there exists an algorithm $\mathcal{A}$ such that:
    \begin{enumerate}
        \item for any error tolerance $\varepsilon$ and success probability $\delta$,
		\item for any conditional distribution $p(\bar{y}|\vec{x}) \in \mathcal{P}$ and corresponding $p$-concept $c\in\mathcal{C}$ in the class, and
		\item for any probability distribution $p(\vec{x})$,
    \end{enumerate}
    the learning algorithm $\mathcal{A}$, when given as input a training set $\mathcal{S} = (\vec{x}_i, \bar{y}_i)_{i=1}^{N_1}$, where $(\vec{x}_i)_{i=1}^{N_1}\sim D^{N_1}$, and each $\bar{y}_i\sim p(\bar{y}|\vec{x})$, produces a hypothesis $h$ fulfilling
	\begin{align}
        \mathbb{P} \left(\ER(h) \leq \varepsilon\right) \geq 1 - \delta,
    \end{align}
    where $\ER(h)$ is the risk functional defined in \cref{Eq:Expected-risk} and the probability is over both: the sampling of training sets $(\vec{x}_i)_{i=1}^{N_1}$ of size $N_1$ and the sampling of random labels $\bar{y}$ conditional on each $\vec{x}_i$.
    
    Further, $\mathcal{C}$ is \emph{efficiently \emph{p}-concept learnable} if $\mathcal{A}$ has runtime polynomial in $1/\varepsilon$,  $1/\delta$, and $\sigma^2_{\bar{y}|\vec{x}}$, the conditional variance of $\bar{y}$ given $\vec{x}_i$ for each $i\in\{1,\ldots,N_1\}$.
    (Runtime efficiency implies sample efficiency, runtime of $\mathcal{A}$ upper-bounds $N_1$). 
\end{definition}

When there is no uncertainty in the given label, the $p$-concept learning model will reduce to the PAC learning model~\cite{valiant1984theory}.
This is captured in \cref{Def:p-concept-learning} by letting $N_s \rightarrow \infty$, and in this regime, we have $\IR(\mathbb{E}_{\bar{y}}[\bar{y}|\vec{x}]) = 0$ hence $\ER(h) = \IR(h)$.

\subsubsection{Hypothesis class for modelling probabilistic concepts}
\label{Subsec:Hypothesis-pconcepts}

To model the $p$-concepts, we consider the following hypothesis class
\begin{align} \label{Eq:Generic-hypothesis-class}
    \HC = \left \{h(\vec x) = u(\langle \vec{w}, \vec{\phi}(\vec{x})\rangle) \: ; \: \vec{w}, \vec{\phi}(\vec{x}) \in \mathbb{R}^p\right\},
\end{align}
where $u: \mathbb{R} \rightarrow \mathcal{Y}$ is an $L$-Lipschitz function that matches the label space $\mathcal{Y}$, $\vec{w}$ are weight vectors with bounded 2-norm $\|\vec{w}\|_2 \le B$ and $\vec{\phi}: \mathbb{R}^m \rightarrow \mathbb{R}^p$ is the feature map that maps $\vec{x}$ to a higher dimensional feature space with $p > m$ and $\|\phi(\vec{x})\|_2\le 1$, and $\langle\cdot\,,\cdot\rangle$ is the usual inner product. This hypothesis class consists of two components (i) the feature map $\vec{\phi}(\vec{x})$ and (ii) the link function $u$, each serving different roles. 

The feature map dictates the class of realizable functions and given two feature vectors $\vec{\phi}(\vec{x}), \vec{\phi}(\vec{x}')$, their inner product is equal to the kernel function
\begin{align}
    k(\vec{x}, \vec{x}') = \langle \vec{\phi}(\vec{x}), \vec{\phi}(\vec{x}') \rangle.
\end{align}
Interestingly, one could express the same class of functions in terms of the kernel. Computing the kernel function $k(\vec{x}, \vec{x}')$ directly without explicitly evaluating the feature vectors and their inner products is known as the kernel trick. Note that $\HC$ reduces to the typical kernel machines when $u$ is set to be the identity function. 

The function $u$, on the other hand, provides us extra flexibility to incorporate the information about the $p$-concepts. 
As discussed, one does not necessarily have access to the exact $p$-concept $c(\vec{x})$ but rather to the samples $(\vec{x},\bar{y}) \sim \bar{\mathcal{D}} = p(\vec{x})p(\bar{y}|\vec{x})$. 
Direct optimizing kernel machines with the empirical risk $\widehat{R}(\cdot)$ using the training samples $\mathcal{S} = (\vec{x}_i, \bar{y}_i)_{i=1}^{N_1}$ yields
\begin{align}
    g(\vec{x}) = \sum_{i=1}^{N_1} a_i k(\vec{x},\vec{x}_i).
\end{align}
However, this kernel-based model might be too expressive for $p$-concept modelling as it tends to overfit the noisy labels. 
Crucially, the link function $u$ can be used to restrict the size of the model class, allowing us to suppress their tendency to overfit.
For the sake of clarity, we will postpone the illustrations of the above-mentioned role of $u$ to the latter sections as the examples could be more appropriately understood in the quantum context.

Now, we are ready to express the $p$-concepts in terms of the hypothesis class. That is,
\begin{align}
    \mathbb{E}_{\bar{y}}[\bar{y}|\vec{x}] = c(\vec{x}) = u( \langle \vec{w}, \vec{\phi}(\vec{x})\rangle + \xi(\vec{x})),
\end{align}
where $\xi(\vec{x}) \in [-M, M]$ is some noise function with $\mathbb{E}_{\vec{x}}[\xi(\vec{x})^2] \le \epsilon_1$ that captures how well one can approximate $p$-concepts using hypotheses from $\HC$.
By the $L$-Lipschitz property of $u$ and $\mathbb{E}_{\vec{x}}[\xi(\vec{x})^2] \le \epsilon_1$, we have
\begin{align}
    \ER(h) &= \mathbb{E}_{\bar{\mathcal{D}}} [(h(\vec{x})  - c(\vec{x}))^2] \le L^2\epsilon_1.
\end{align}
That is, low $\epsilon_1$ implies low approximation error of $c(\vec{x})$ using $h \in \HC$. This approximation enables us to systematically reduce the learning task to the search of appropriate feature map $\vec{\phi}(\vec{x})$, the link function $u$ and the design of efficient algorithms to learn the weight vector $\vec{w}$.

\subsection{The family of parameterized quantum models}
\label{Subsec:Family-PQMs}

In this work, we are interested in learning a family of parameterized quantum models $f_{\vec\theta}(\vec{x})$
\begin{align}
    \mathcal{F} = \left\{f_{\vec\theta}(\vec{x}) = \tr(\rho_{\vec\theta}(\vec{x})O) \, |\, \vec\theta \in \Theta \right\}.
\end{align}
where $\rho_{\vec\theta}(\vec{x})$ are parameterized quantum states with parameters $\vec\theta \in \Theta = \mathbb{R}^r$ and input data $\vec{x} \in \mathcal{X} = \mathbb{R}^m$ while $O$ is an arbitrary Hermitian observable. 
The quantum states $\rho_{\vec\theta}(\vec{x}) = \mathcal{M}_{\vec\theta,\vec{x}}(\rho_0)$ could be prepared by applying parameterized quantum channels $\mathcal{M}_{\vec\theta,\vec{x}}(\cdot)$ on some initial state $\rho_0$. 
In particular, we will focus on specific quantum channels that are generated by parameterized quantum circuits (PQCs). 
We remark that our analysis can be directly extended to other quantum channels including ground state preparation channels~\cite{huang2022provably,lewis2024improved,che2023exponentially}.

\subsubsection{PQCs and their classical Fourier representations} \label{Subsubsec:Fourier-PQCs}

\begin{figure*}[tb]
\includegraphics[width=\textwidth]{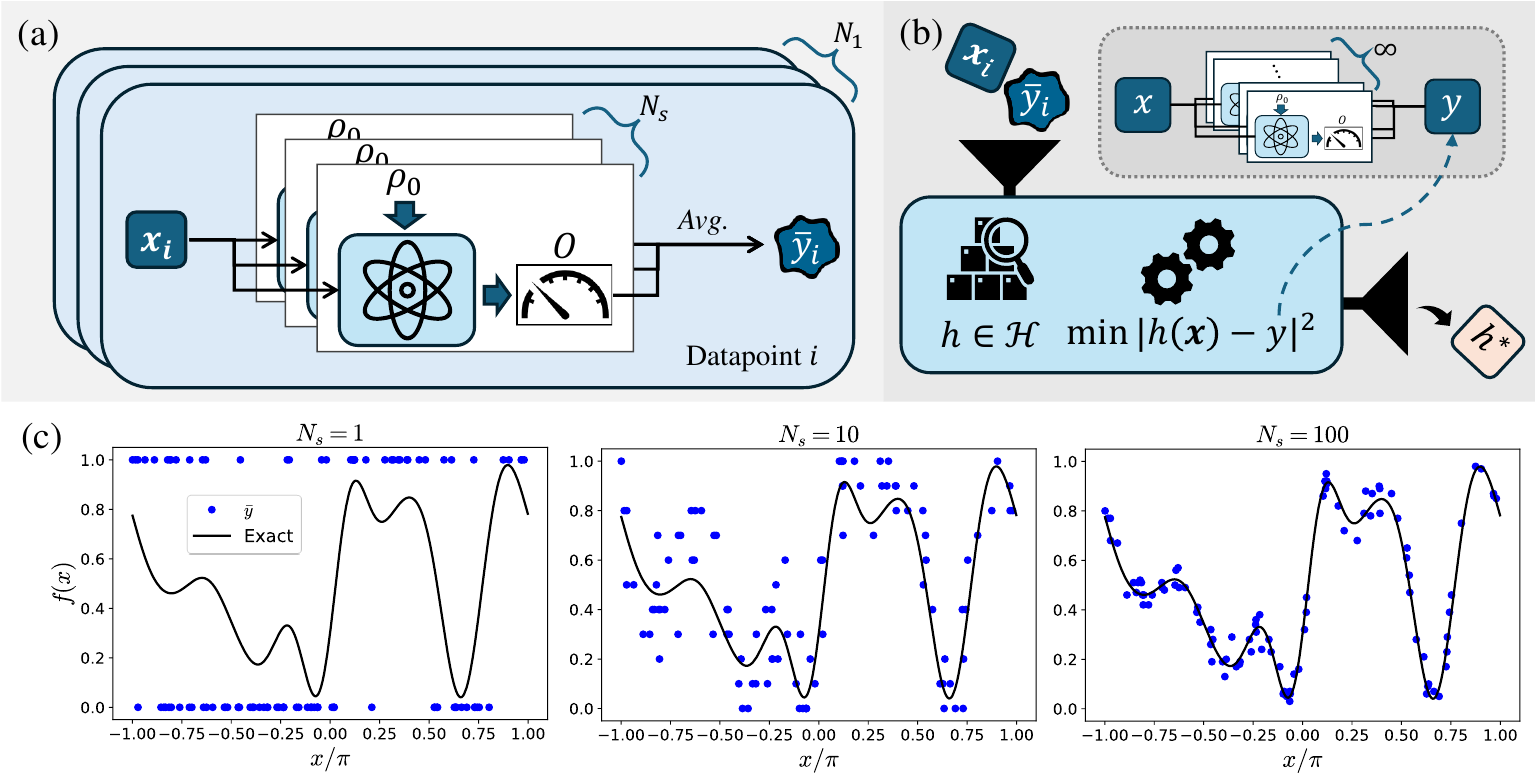}
\caption{\label{Fig:Motivation-schematics}Concept learning of parameterized quantum models. 
(a) To learn quantum models, one needs to probe the quantum model with $N$ different input data points $\vec{x}$, and construct an estimator of the quantum model $y = f(\vec{x})$ conditioned on the input. 
Such estimators $\bar{y}$ can be constructed by taking the average over $N_s$ duplicate quantum measurements.
(b) Using data pairs ($\vec{x}_i$, $\bar{y}_i$) collected from the quantum model, the task is to classically learn a representation $h^*$ of the quantum model such that the output of classical representation $h(\vec x)$ is close to the underlying expected output $y = f(\vec{x})$ of the quantum model for any arbitrary $\vec x$. 
As illustrated in (c), the number of measurement shots $N_S$ will determine the closeness between the estimator $\bar{y}$ (blue dots) and the underlying expected value $f(x)$ (black solid line). 
}
\end{figure*}

Let $\vec{x} = (x_1, \dots, x_m)\in \mathcal{X} = [0, 2\pi)^m$ be a vector of data points, $\vec{\theta} = (\theta_1, \dots, \theta_m)\in \Theta = [0, 2\pi)^r$ be a vector of parameters, $U(\vec{x}, \vec{\theta})$ be the unitary that represents the PQCs, $\ket{{\bf 0}}=\ket{0}^{\otimes n}$ and $O$ be an arbitrary Hermitian observable. 
We define the PQC model as
\begin{align}
    f_{\vec{\theta}}(\vec{x}) = \bra{{\bf 0}} U^\dagger(\vec{x}, \vec{\theta}) O U(\vec{x}, \vec{\theta}) \ket{{\bf 0}}.
\end{align}
For a given $U(\cdot,\vec{\theta})$ and $O$, we define the PQC model class $\mathcal{F}_{U,O}$ as
\begin{align} \label{Eq:PQC-concept-class}
    \mathcal{F}_{U,O} = \left \{f_{\vec{\theta}}(\cdot) = \bra{{\bf 0}} U^\dagger(\cdot, \vec{\theta}) O U(\cdot, \vec{\theta}) \ket{{\bf 0}} \, | \, \vec{\theta} \in \Theta \right \}
\end{align}
for all $\vec{x} \in \mathcal{X}$.

The parameterized unitary $U(\vec{x}, \vec{\theta})$ consists of a sequence of two different types of parameterized quantum gates. 
The first type of parameterized quantum gates is controlled by parameters $\vec{\theta}$, while the second type embeds data points $x_i$ into the PQCs via unitary evolution
\begin{align}
    V_{(j,k)}(x_j) = e^{-i H_k^{(j)} x_j},
\end{align}
generated by some Hamiltonian $H_{k}^{(j)}$. 
Given this parameterisation strategy, it is well-known that PQC models could be written as a Fourier series\footnote{The Fourier expansion in this work mainly follows the treatment in Ref.~\cite{landman2022classically,sweke2023potential} but equivalent Fourier representation of PQC models can be obtained by other Fourier expansion methods~\cite{fontana2023classical,nemkov2023fourier}}~\cite{gilvidal2020input,schuld2021effect}
\begin{align} \label{Eq:PQCMs-Fourier}
    f_{\vec{\theta}}(\vec{x}) = \sum_{\vec{\omega} \in \tilde{\Omega}} c_{\vec{\omega}}(\vec{\theta}) e^{i\langle \vec{\omega}, \vec{x} \rangle},
\end{align}
where the frequency spectrum $\tilde{\Omega}$ is determined by the ensemble of eigenvalues of embedding Hamiltonian $\{H_k^{(j)}\}_{j,k}$ and the coefficients $c_{\vec{\omega}}(\vec{\theta})$ depend on the quantum gates parameterized by $\vec{\theta}$.

\cref{Eq:PQCMs-Fourier} can be further simplified by noting that the non-zero frequencies in $\tilde{\Omega}$ come in pairs, i.e., $\vec{\omega},-\vec{\omega} \in \tilde{\Omega}$, allowing us to split $\tilde{\Omega}$ into two components. That is, $\tilde{\Omega} := \Omega \: \cup \: (-\Omega)$ with $\Omega \: \cap \: (-\Omega) =\{\vec{\omega}_0\}$, where $\vec{\omega}_0 = (0,\dots,0) \in \tilde{\Omega}$ is the vector of zero frequencies. Let $\Omega = \{\vec{\omega}_0, \vec{\omega}_1, \dots, \vec{\omega}_{|\Omega|}\}$ and for all $\vec{\omega} \in \Omega \backslash \{\vec{\omega}_0 \}$, we have
\begin{align}
    a_{\vec{\omega}}(\vec{\theta}) &:= c_{\vec{\omega}}(\vec{\theta}) + c_{-\vec{\omega}}(\vec{\theta}) \\
    b_{\vec{\omega}}(\vec{\theta}) &:= i(c_{\vec{\omega}}(\vec{\theta}) - c_{-\vec{\omega}}(\vec{\theta})).
\end{align}
Given this definition, \cref{Eq:PQCMs-Fourier} can be equivalently written as
\begin{align} \nonumber
    f_{\vec{\theta}}(\vec{x}) = c_{\vec{\omega}_0}(\vec{\theta}) + \sum_{i=1}^{|\Omega|-1} ( &a_{\vec{\omega}_i}(\vec{\theta}) \cos(\langle \vec{\omega}_i, \vec{x} \rangle) +\\
    &b_{\vec{\omega}_i}(\vec{\theta}) \sin(\langle \vec{\omega}_i, \vec{x} \rangle) )
\end{align}
Identifying the corresponding weight vectors $\vec{w}$ 
\begin{align}
    \vec{w}_F(\vec{\theta})= \sqrt{|\Omega|} \begin{pmatrix}
        c_{\vec{\omega}_0}(\vec{\theta}) \\
        a_{\vec{\omega}_1}(\vec{\theta})\\
        b_{\vec{\omega}_1}(\vec{\theta})\\
        \vdots\\
        a_{\vec{\omega}_{|\Omega|-1}}(\vec{\theta})\\
        b_{\vec{\omega}_{|\Omega|-1}}(\vec{\theta})
    \end{pmatrix}^\intercal
\end{align}
and the trigonometric polynomial feature map 
\begin{align} \label{Eq:Full-feature-map}
    \vec{\phi}_F(\vec{x}) = \frac{1}{\sqrt{|\Omega|}}
    \begin{pmatrix}
        1\\
        \cos(\langle \vec{\omega}_1, \vec{x} \rangle)\\
        \sin(\langle \vec{\omega}_1, \vec{x} \rangle)\\
        \vdots\\
        \cos(\langle \vec{\omega}_{|\Omega|-1}, \vec{x} \rangle)\\
        \sin(\langle \vec{\omega}_{|\Omega|-1}, \vec{x} \rangle)
    \end{pmatrix} 
\end{align}
enables us to express the PQC model as a linear model with respect to the feature map $\vec{\phi}_F$, i.e.,
\begin{align}
    f_{\vec{\theta}}(\vec{x}) = \langle \vec{w}_F (\vec{\theta}), \vec{\phi}_F(\vec{x}) \rangle,
\end{align}
and the associated kernel function is given by $k_F(\vec{x}, \vec{x}') = \langle \vec{\phi}_F(\vec{x}), \vec{\phi}_F(\vec{x}') \rangle$.

\subsubsection{Data extraction from parameterized quantum models}
\label{Subsubsec:Data-Extraction}
In general, one does not have direct access to $f_{\vec{\theta}}(\vec{x})$. Instead, they are estimated using finite samples from measurement procedures such as direct measurement or classical shadow methods, as described in \cref{Sec:Estimation-methods}.
We denote outputs of such estimation procedures as $\bar{y}$ and their dependency on the data point $\vec{x}$, parameter $\vec{\theta}$, and the number of measurement shots $N_s$ are implicitly assumed. In addition, they are unbiased estimators of $f_{\vec{\theta}}(\vec{x})$, i.e.,
\begin{align} \label{Eq:VQM-pconcept}
    f_{\vec{\theta}}(\vec{x}) = \mathbb{E}_{\bar{y}}[\bar{y}|\vec{x},\vec{\theta}].
\end{align}
For ease of notation, we will drop the conditional dependency of the expectation on $\vec{\theta}$ from now on.

Now, we describe the procedures for obtaining labelled data points from a given $f_{\vec{\theta}}$. 
Without loss of generality, we let $p(\vec{x})$ be a uniform distribution of input $\vec{x}$. 
As depicted in \cref{Fig:Motivation-schematics}~(a), a set of $N$ i.i.d. samples of $\vec{x}$ is first drawn from $p(\vec{x})$ and subsequently input to the quantum model to collect their associated labels $\bar{y}$ via the procedures described in \cref{Sec:Estimation-methods} using $N_s$ measurement repetitions. 
This gives the set of data $\mathcal{S} = (\vec{x}_i,\bar{y}_i)_{i=1}^{N}$.
Shown in \cref{Fig:Motivation-schematics}~(c) are the labels $\bar{y}$ estimated with $N_s = 1, 10, 100$. 

\section{Parameterized Quantum Models as Probabilistic Concepts}
\label{Sec:VQMs-p-concepts}

One can immediately deduce from \cref{Eq:VQM-pconcept} that parameterized quantum models (PQMs) are $p$-concepts. Now, we will show that the hypothesis class defined in \cref{Subsec:Hypothesis-pconcepts} is an appropriate model class for the learning of PQMs.  
Modelling PQMs using the hypothesis from $\HC$, as defined in \cref{Eq:Generic-hypothesis-class}, assumes the following: there exist a feature map $\vec{\phi}(\vec{x})$, a function $u$, a weight vector $\vec{w}$, and a noise function $\xi(\vec{x})$ such that the PQMs can be expressed as
\begin{align} \label{Eq:VQM-GLM}
    \mathbb{E}_{\bar{y}}[\bar{y}|\vec{x}] = \tr(\rho_{\vec{\theta}}(\vec{x}) O) = u( \langle \vec{w}, \vec{\phi}(\vec{x}) \rangle + \xi(\vec{x})),
\end{align}
with $\|\vec{w}\|_2 \le B$, $\xi(\vec{x}) \in [-M,M]$, and $\mathbb{E}_{\vec{x}}[\xi(\vec{x})^2] \le \epsilon_1$.
Hence, the aim here is to find an appropriate function $u$ that contains information on the PQM as well as construct the feature map $\vec{\phi}(\vec{x})$ that efficiently approximates a PQM. 

\begin{algorithm}[!t]
\caption{The learning algorithm}
\label{Algo:Learning-algorithm}
\DontPrintSemicolon
  \KwIn{Training data size $N_1$, validation data size $N_2$, number of measurement shots $N_s$, parameter setting of quantum model $\vec{\theta}$, distribution of input $p(\vec{x})$, non-decreasing $L$-Lipschitz function $u: \mathbb{R} \rightarrow \mathcal{Y}$, kernel function $k$ corresponding to feature map $\vec\phi$, learning rate $\lambda >0$, number of iterations $T$}
  
  Sample $N_1$ training data inputs $\vec{x}_1, \dots, \vec{x}_{N_1} \sim p(\vec{x})$.

  {For each $\vec{x}_i$, obtain the associated labels $\bar{y}_i$, with $\eb_{\bar{y}}[\bar{y}|\vec{x}] = \tr(\rho_{\vec{\theta}}(\vec{x})O)$, by extracting and averaging $N_s$ measurement samples of the quantum model via procedure described in \cref{Subsubsec:Data-Extraction}. This yields the training dataset $(\vec{x}_i,\bar{y}_i)_{i=1}^{N_1}$.}

  Repeat steps 1 and 2 to collect labelled validation data of size $N_2$, $(\vec{p}_j,\bar{q}_j)_{j=1}^{N_2}$.
    
  Initialize $\alpha^i := 0 \in \mathbb{R}^{N_1}$.

  \For{$t=1,\dots, T$}{
        $h^t(\vec{x}):= u\left( \sum_{i=1}^{N_1} \alpha_i^t k(\vec{x},\vec{x}_i)\right)$
        
        \For{$i = 1, 2,\dots, {N_1}$}{
            $\alpha_i^{t+1}:= \alpha_i^t + \frac{\lambda}{{N_1}}(\bar{y}_i - h^t(\vec{x}_i))$
        }
   }
   \KwOut{$h^r$ where $r = \arg \min_{t\in \{1,\dots,T\}} \frac{1}{N_2} \sum_{j=1}^{N_2} (\bar{q}_j - h^t(\vec{p}_j))^2$}
\end{algorithm}

\subsection{Algorithm for concept learning of parameterized quantum models}

By expressing the PQMs in terms of the hypothesis in $\HC$, we systematically reduce the modelling problem to the search of the appropriate feature map $\vec{\phi}(\vec{x})$, link function $u$, and optimal weight $\vec{w}^*$. As the first two attributes are highly dependent on the problem at hand, we will defer their discussion to \cref{Sec:Surrogate-PQC}. In this section, we will assume $\vec{\phi}(\vec{x})$ and $u$ are known and focus on the algorithm part of the problem. 

Consider a PQM $\tr(\rho(\vec{x})O)$ that can be approximated by a feature map $\vec{\phi}(\vec{x})$ and a known $L$-Lipschitz non-decreasing function $u : \mathbb{R} \rightarrow \mathcal{Y}$, as per \cref{Eq:VQM-GLM}. 
The task of learning PQMs could be formulated as the search of the optimal weight vector $\vec{w}^*$ such that the output hypothesis $h^*(\vec{x}) = u(\langle \vec{w}^*, \vec{\phi}(\vec{x}) \rangle)$ minimizes the explicit risk
\begin{align}
    \ER(h^*) = \mathbb{E}_{\bar{\mathcal{D}}} [(u(\langle \vec{w}^*, \vec{\phi}(\vec{x}) \rangle) - \tr(\rho(\vec{x})O))^2].
\end{align}
As discussed in \cref{Subsec:p-concept}, we will only have access to finite samples drawn from the distribution $\bar{\mathcal{D}}$. Hence, we will be minimizing the empirical risk $ \widehat{R}(h)$ in \cref{Eq:Empirical-risk} using some training samples $\mathcal{S} = (\vec{x}_i, \bar{y}_i)_{i=1}^{N_1}$ with $({\vec{x}_i,\bar{y}_i}) \sim \bar{\mathcal{D}}$ instead of $\ER(h)$. Note that the extra $u$ in the empirical risk makes the optimization non-convex.
As detailed in \cref{Fig:Motivation-schematics}~(b), while the classical machine learns from a noisy dataset consisting of shot noise from quantum measurements, our objective is to enable the classical machine to approximate the underlying $p$-concept of the PQM, i.e., the expected value of the measured outcomes of PQMs.

Shown in \cref{Algo:Learning-algorithm} is an iterative-based method that learns PQMs under some mild assumptions. 
While our algorithm is derived from the iterative method in Ref.~\cite{goel2019learning}, we extended the provable guarantee of the original algorithm to include the number of measurement shots $N_s$ used to estimate $\bar{y}$ and show its operational role in the algorithm. 
The analytical guarantee enables us to understand the contributions of errors and the intuitive explanation of the working principle of \cref{Algo:Learning-algorithm} is provided in \cref{Sec:Algo-intuition}. 

\begin{theorem}[$p$-concept learnability of PQMs]\label{Theorem:Alphatron-guarantee} 
We are given a quantum observable $O$ such that $\|O\|_\infty = \Delta$. With this observable, we have quantum model whose expected output can be expressed as a classical representation as follows: $\tr(\rho(\vec x)O) = u(\langle \vec{w}, \vec\phi(\vec{x}) \rangle + \xi(\vec{x}))$, where $u: \mathbb{R} \rightarrow [-\Delta,\Delta]$ is a known $L$-Lipschitz non-decreasing function, $\xi: \mathbb{R}^m \rightarrow [-M, M]$ such that $\mathbb{E}_{\vec{x}}[\xi(\vec{x})^2] \le \epsilon_1$, $\|\vec{w}\|_2 \le B$, and $\|\phi(\vec{x})\|_2\le 1$. Considering a training dataset of $N_1$ i.i.d. samples of $\vec{x}$ as input to the quantum model, and whose label is the sample mean of the output of the quantum model sampled over $N_s$ measurements. Let the conditional variance of an individual measurement averaged over all $\vec x$ be $\bar{\sigma}$. For $\delta \in (0,1)$, with probability $1-\delta$, setting the learning rate $\lambda = \frac{1}{L}$ and given a validation dataset size of $N_2 = \mathcal{O}(N_1\Delta^2\log(\frac{T}{\delta}))$, after $T = \mathcal{O}(\frac{BL}{\epsilon_4})$ iterations, \cref{Algo:Learning-algorithm} outputs a hypothesis $h \in \HC$ such that 
\begin{multline}
 \label{Eq.Error-bound}
    \ER(h) \le \mathcal O(L\Delta\sqrt{\epsilon_1} + L\Delta M\epsilon_2  \\+ LB\Delta\epsilon_3 + LB\epsilon_4 + \Delta^2 \epsilon_5),
\end{multline}
where $\epsilon_2 = \sqrt[\leftroot{-2}\uproot{2}4]{\frac{\log(\frac{1}{\delta})}{N_1}}$, $\epsilon_3 = \sqrt{\frac{1}{N_1}}$, $\epsilon_4 = \sqrt{\frac{\bar\sigma \log(\frac{1}{\delta})}{N_1 N_{s}}}$, $\epsilon_5 = \sqrt{\frac{\log(\frac{1}{\delta})}{N_1}}$, and $\bar{\sigma} = \mathbb{E}_{\vec{x}}[\sigma^2_{y|\vec{x}}]$.
\end{theorem}
The proof of this theorem can be found in \cref{appTheorem1Proof}. As shown in \cref{Eq.Error-bound}, four different error sources will affect the performance of the models: (i) the approximation error $\epsilon_1$, (ii) the data sampling errors $\epsilon_2$ and $\epsilon_5$, (iii) the learnability error $\epsilon_3$, and (iv) the label sampling error $\epsilon_4$. 
Firstly, the approximation error $\epsilon_1$ captures the intrinsic error that can be achieved by our hypothesis class as it tells us how far away our hypothesis $h(\vec{x})$ is from the true function $\tr(\rho(\vec{x})O)$ we wish to learn, i.e., $\mathbb{E}_{\vec{x}}[\xi(\vec{x})^2] \le \epsilon_1$. 
It is therefore impossible to obtain a small risk if the approximation error is high to begin with. 
On the other hand, the data sampling errors $\epsilon_2$ and $\epsilon_5$ capture the statistical noise arising from the finite data samples provided to the learning algorithm while the learnability error $\epsilon_3$ stems from Rademacher complexity and quantifies the hardness of learning with the given hypothesis class. 
Both of these errors can be minimized by providing more data samples. 
Lastly, the label sampling error $\epsilon_4$ is influenced by three attributes, the averaged variance $\mathbb{E}_{\vec{x}}[\sigma^2_{y|\vec{x}}]$, the number of training data points $N_1$, and the number of measurement shots $N_s$. 
Increasing either $N_1$ or $N_s$ could reduce the label sampling error. 
Additionally, a measurement scheme that results in smaller variance will require fewer training data points and measurement shots to achieve a smaller $\epsilon_4$ error.

\subsection{Asymmetrical effects of \texorpdfstring{$N_1$}{N1} and \texorpdfstring{$N_s$}{Ns}}

While the individual implications of all four types of errors are straightforward to deduce, jointly analysing the last three sources of error leads to an interesting observation regarding the asymmetrical effects of $N_1$ and $N_s$ on classical learning of quantum models. 
On the one hand, increasing $N_s$ can only decrease the label sampling error $\epsilon_4$ but not the data sampling errors $\epsilon_2$ and $\epsilon_5$ and the learnability error $\epsilon_3$. 
On the other hand, increasing $N_1$ will simultaneously decrease all three errors, and $\epsilon_4$ approaches $0$ regardless of the value of $N_s$. Consequently, one could set $N_s = 1$ when $N_1$ is sufficiently large. 
This observation aligns with intuition, as the labels are dependent on the parameters. 
By sampling across the training points, one effectively samples across various labels, thereby providing a reasonable estimation of quantum models. 
In contrast, increasing the resolution of the labels does not provide extra information on other data points. 
This observation is summarised in \cref{Corollary:Asymmetry} and numerically illustrated in \cref{Fig:Theoretical-results}~(a). For simplicity, we assume $\delta = 0.01$, and $\bar\sigma = L= B = \Delta = 1$.

\begin{corollary}[Asymmetrical effects of $N_1$ and $N_s$]\label{Corollary:Asymmetry} Let all variables defined as per \cref{Theorem:Alphatron-guarantee}. For the hypothesis class $\HC$ with link function $u$, feature map $\vec\phi(\vec{x})$ and weight vector $\vec{w}$ with $\tr(\rho(\vec{x})O) = u(\langle \vec{w}, \vec\phi(\vec{x}) \rangle) \in \HC$, i.e., $\mathbb{E}_{\vec{x}}[\xi(\vec{x})^2] = 0$, \cref{Algo:Learning-algorithm} will output a hypothesis $h \in \HC$ such that 
\begin{align}
    \ER(h) \le  c_1\sqrt{\frac{1}{N_1}} + c_2\sqrt{\frac{1}{N_1 N_{s}}} + c_3\sqrt{\frac{1}{N_1}},
\end{align}
where $c_1 = \mathcal O (LB\Delta)$, $c_2 = \mathcal O \left(LB\sqrt{\bar\sigma \log(\frac{1}{\delta})}\right)$, $c_3 =\mathcal O \left(\Delta^2 \log(\frac{1}{\delta})\right)$, and $N_1$ and $N_s$ contribute asymmetrically to $R(h)$. That is, for a constant $N_1$, $R(h) \not\to 0$ when $N_s \to\infty$, but $R(h) \to 0$ when $N_1 \to \infty$ regardless of the value of $N_s$. Note that $\epsilon_1 = 0$ implies $M = 0$ by definition.
\end{corollary}

The overall analysis shows that for a sufficiently large $N_1$, classical models can learn quantum models that have efficient classical representation even when target labels are estimated with limited measurement shots, e.g., $N_s = 1$. 
Conversely, when such efficient classical representation cannot be found, $\tr(\rho(\vec{x})O)$ is not learnable even when $N_1$ and $N_s$ are infinite. 
\cref{Corollary:Asymmetry} further shows that $N_s$ plays a less significant role than $N_1$ in the classical learning of quantum models. 
In other words, shot noise is not a fundamental concern in classical learning of quantum models as its role can be easily substituted by $N_1$.

\subsection{Trade-offs between \texorpdfstring{$N_1$}{N1} and \texorpdfstring{$N_s$}{Ns}}

In an ideal world, one would choose $N_1$ and $N_s$ as large as possible to minimize the explicit risk. 
However, external constraints like financial budgets and time limitations might significantly restrict the total number of queries to a quantum model. 
Thus, a more realistic setting is to first consider a fixed number of queries to quantum models, and $N_1$ and $N_s$ are subsequently decided.

In general, producing more samples for a fixed parameter setting in an experiment is much cheaper and faster than changing the parameter settings each time. 
Changing parameters incurs an additional cost that may stem from preprocessing subroutines, classical transpilation and optimization of the circuits or platform-dependent factors regarding the hardware we are executing the quantum circuits. 
For example, the penalty cost for superconducting quantum computers would be larger than for trapped-ion quantum computers, as it is relatively cheaper to produce more samples for a fixed parameter set than to change the parameter setting each time in the former platform~\cite{linke2017experimental}. To quantify such costs for easier discussion, we assume that these costs can be quantitatively evaluated to be some multiple $\gamma\in \mathbb{R}_+$ of the cost to run a repetition of quantum circuits that have already been configured. That is, we assume changing the parameter setting once will incur an extra cost of $\gamma$ shots.

\begin{figure}[t]
\includegraphics[width=80mm,scale=0.5]{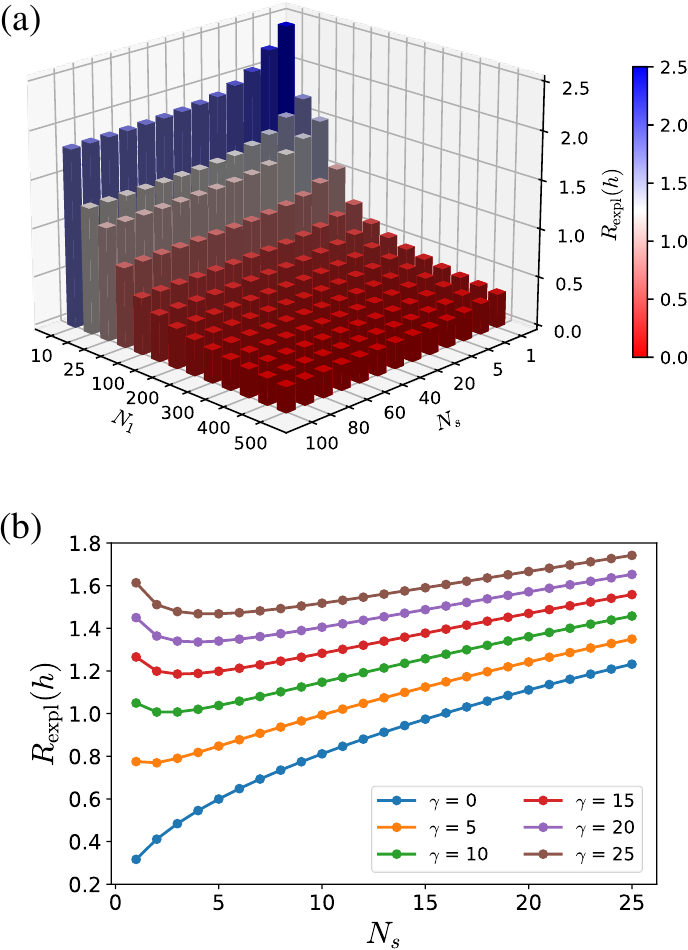}
\caption{\label{Fig:Theoretical-results}The respective numerical illustrations of \cref{Corollary:Asymmetry} and \cref{Corollary:Trade-off} with $\delta = 0.01$, and $\bar{\sigma}=L=B=\Delta=1$. (a) The plot shows the asymmetrical effect of the number of training samples $N_1$ and the number of measurement shots $N_s$ on the explicit risk $\ER(h)$.
(b) For a fixed total measurement budget $N_{tot}$, the optimal pair of $N_1$ and $N_s$ will change with $\gamma$. 
When $\gamma = 0$, the optimal shot number is $N_s = 1$ but it depends on $\gamma$ when $\gamma > 0$. 
All curves are computed with $N_{tot} = 600$ and $N_s = \{1,2,3,\dots,24,25\}$.
}
\end{figure}

Considering the total measurement budget for training, we find that $N_{\mathrm{tot}} = N_1\cdot (N_s + \gamma)$.
Fixing the total measurement budget $N_{\mathrm{tot}}$ implies a trade-off between $N_1$ and $N_s$: increasing $N_1$ will reduce $N_s$ and vice versa. 
This poses an interesting learning-theoretic question: \textit{given a fixed $N_{\mathrm{tot}}$, will classical machines learn better with training datasets consisting of more inputs/parameters with noisier labels (larger $N_1$ but smaller $N_s$) or fewer inputs with cleaner labels (smaller $N_1$ but larger $N_s$)?} As shown in \cref{Corollary:Trade-off}, when $N_1$ and $N_s$ are treated equally ($\gamma = 0$), asymptotically, it is generally better to sample more inputs, i.e., $(N_1^*, N_s^*) = (N_{tot},1)$.
When there is an extra cost for changing the parameter settings, i.e., $\gamma >0$, there exists a pair of optimal input size $N_1^*$ and the shot number $N_s^{*}$ that minimize the explicit risk.
These observations are numerically illustrated in \cref{Fig:Theoretical-results}~(b). For simplicity, we assume $\delta = 0.01$, and $\bar\sigma = L= B =\Delta = 1$.

\begin{corollary}[Trade-off between $N_1$ and $N_s$]\label{Corollary:Trade-off}Consider the setting as per \cref{Corollary:Asymmetry}. For a given fix total measurement budget for training $N_{\mathrm{tot}} \in \mathbb{N}$ and a fix penalty cost $\gamma \in \mathbb{R_+}$, $N_1$ and $N_s$ are determined by $N_{\mathrm{tot}} = N_1 \cdot (N_s + \gamma) $.
Respecting this constraint, \cref{Algo:Learning-algorithm} will output a hypothesis $h \in \HC$ such that 
\begin{align}
    \ER(h) \le c_1\sqrt{\frac{N_s + \gamma}{N_{\mathrm{tot}}}} + c_2\sqrt{\frac{N_s + \gamma}{N_{\mathrm{tot}}N_s}} + c_3\sqrt{\frac{N_s + \gamma}{N_{\mathrm{tot}}}}.
\end{align}
When $\gamma = 0$, the upper bound of the explicit risk $\ER(h)$ reduces to 
\begin{align} \label{Eq:Generalization-error-gamma}
    \ER(h) \le c_1\sqrt{\frac{N_s}{N_{\mathrm{tot}}}} + c_2\sqrt{\frac{1}{N_{\mathrm{tot}}}} + c_3\sqrt{\frac{N_s}{N_{\mathrm{tot}}}}
\end{align}
which is minimized when $N_s = 1$. When $\gamma > 0$, there exists a pair of optimal input data size and shot number $(N_1^*, N_s^*)$ where
\begin{equation}
N_1^* = \frac{N_{tot}}{N_s^* + \gamma} \: \: \text{and} \: \: N_s^* = \left(\frac{c_2 \gamma}{c_1 + c_3}\right)^{\frac{2}{3}}
\end{equation}
that minimizes our upper bound of $\ER(h)$.
\end{corollary}

Note that the optimal value of $N_s$ does not correlate with $N_{\mathrm{tot}}$, but depends on the constant penalty cost $\gamma$. 
Hence, setting $N_s=1$ regardless of the value of $\gamma$ would only increase $\ER(h)$ by a factor of $\sqrt{1+\gamma}$, retaining learnability up to a constant factor for single measurement learning.

Taking a closer examination at $N_s$, we check whether other factors apart from the device-dependent cost $\gamma$ affect the value of $N_s$. Delving into the definitions of the terms $c_1$, $c_2$, and $c_3$, we note that the terms $L$, $\Delta$, and $\delta$ are constants that either depend on the problem setting or can be set arbitrarily. $B$, on the other hand, is directly related to the expressibility of the hypothesis $h$ used to model the quantum model. \emph{Does the expressibility $B$ of the hypothesis $h$ affect the number of shots $N_s$ required?}

Plugging in $c_1 = \mathcal O (B)$, $c_2 = \mathcal O (B)$, $c_3 =\mathcal O(1)$, we note that in terms of $B$, $N_s = \mathcal O\left(\frac{B}{B+K}\right)$, where $K$ is a constant. 
While $N_s$ is indeed dependent on $B$, its dependency can be upper bounded by a constant as $N_s\to \mathcal O(1)$ as $B$ grows. 
Hence, even in cases where the expressibility of the hypothesis $h$ we use to exactly model the quantum model scales exponentially, the number of shots required to sample each data is still limited to a constant value.

\subsection{Shot-noise dependent bias-variance trade-off}
\label{Subsec:Shot-noise_Bias-Variance}

An alternative framework for analyzing the occurrences of different error terms commonly seen in machine learning analysis is the bias-variance-noise decomposition. 
Here, we provide a summary with the full introduction deferred to \cref{Sec:Bias-Variance-Noise-Decomposition}. 

Optimizing the empirical risk $\Remp(h)$ using different training datasets $\mathcal S$ would yield different trained models $h_{\mathcal S}(\vec{x})$.
The bias then measures, on average, how much $h_{\mathcal S}$ deviates from the ground truth $f$
\begin{align}
    \textbf{Bias}_{\mathcal S} := \mathbb{E}_{\mathcal S}[h_{\mathcal S}(\vec x)] - f(\vec{x}),
\end{align}
while the variance 
\begin{align}
     \textbf{Var}_{\mathcal S} & := \mathbb{E}_{\mathcal S} \left[\left(\mathbb{E}_{\mathcal S}[h_{\mathcal S}(\vec{x})] - h_{\mathcal S}(\vec{x})\right)^2 \right]
\end{align}
measures the fluctuations among the trained models. 
As the expectation is taken over all possible training datasets of the same size, therefore, the bias and variance will be dependent on the complexity of the hypothesis class, the number of training data points $N_1$ and the number of random labels $N_s$.

The average of the explicit risk over all possible training datasets will then have the following decomposition, whose derivation is also found in \cref{Sec:Bias-Variance-Noise-Decomposition}:
\begin{align} \label{Eq:Ensemble-varepsilon-bias-var}
    \mathbb{E}_{\mathcal{S}}[\ER(h_\mathcal{S})] = \mathbb{E}_{\vec{x}}\left[\textbf{Bias}^2_{\mathcal S} \right] + \mathbb{E}_{\vec{x}} \left[\textbf{Var}_{\mathcal S} \right].
\end{align}
where
\begin{align} \label{Eq:Expected-bias}
    \mathbb{E}_{\vec{x}} \left[\textbf{Bias}^2_{\mathcal S} \right] &:= \mathbb{E}_{\vec{x}}\left[\left(\mathbb{E}_{\mathcal S}[h_{\mathcal S}(\vec{x})] - f(\vec{x}) \right)^2 \right] \quad \text{and}\\ \label{Eq:Expected-var}
    \mathbb{E}_{\vec{x}} \left[\textbf{Var}_{\mathcal S}\right] & := \mathbb{E}_{\vec{x},{\mathcal S}} \left[\left(\mathbb{E}_{\mathcal S}[h_{\mathcal S}(\vec{x})] - h_{\mathcal S}(\vec{x})\right)^2 \right]
\end{align}
are the averaged bias squared and averaged variance, respectively. This decomposition shows that the shot noise has an indirect impact on the performance of classical machine learning models and this impact can be studied by analysing the statistical quantities $\textbf{Bias}^2_{\mathcal S}$ and $\textbf{Var}_{\mathcal S}$. 
Intuitively, high shot noise implies high variance in the labels, indirectly leading to high variance in the models, and further induces overfitting.
We will provide a simple example in \cref{Sec:Surrogate-PQC} to illustrate how shot noise affects the bias-variance trade-off. 

\cref{Eq.Error-bound} and \cref{Eq:Ensemble-varepsilon-bias-var} appear to be unrelated to each other. 
On the one hand, the algorithm-specific \cref{Eq.Error-bound} gives a probabilistic guarantee on the performance of each trained model. 
On the other hand, the algorithm-agnostic \cref{Eq:Ensemble-varepsilon-bias-var} provides an understanding of the average behaviour of the overall hypothesis class. 
One can however observe the similarities between the two by directly comparing \cref{Eq.Error-bound} and \cref{Eq:Ensemble-varepsilon-bias-var}. 
Specifically, the first term in \cref{Eq.Error-bound} can be understood as the bias of the models since it quantifies the asymptotic error that is achievable by the models while the second term captures the finite sampling noise of the bias; the other three terms inform on the variance of the model. 
Interestingly, the shot-noise dependent variance is captured by $\epsilon_4$ and \cref{Fig:Theoretical-results} essentially captures the variance dependence on the number of training data points and the number of measurement shots.

\section{Classical surrogates of PQC models as probabilistic concepts}
\label{Sec:Surrogate-PQC}

As a direct example, we apply our theoretical framework to create a classical surrogate of parameterized quantum circuit (PQC) models~\cite{schreiber2023classical}. 
Treating PQC models as $p$-concepts enables us to study the impact of shot noise on constructing their corresponding classical surrogates. 
In particular, we observed asymmetrical effects from both the number of training data points and the number of measurement shots, as well as the potential for using a relatively small number of measurement shots to surrogate PQC models. 
As predicted by our theoretical analysis, the bias and variance of the classical surrogates are highly dependent on the strength of the shot noise. 
Finally, we highlight the role of the link function in our surrogate models in suppressing their variance in the presence of the shot noise.

We wish to emphasize that our work aims to provide a generic framework to analyse the learnability of PQMs in the presence of shot noise. 
Therefore, in this example, we will consider the feature map proposed in the literature~\cite{gilvidal2020input,schuld2021effect}, but our framework is readily adaptable to future proposals of efficient feature maps.
In addition, our results can be easily extended to other types of PQMs by replacing the quantum channel with appropriate substitutes.

\subsection{Classical approximation of PQC models}

In this section, we will briefly discuss the existing methods for approximating PQC models classically. 
As described in \cref{Subsubsec:Fourier-PQCs}, PQC models can be written as $f_{\vec{\theta}}(\vec{x}) = \langle \vec{w}_F (\vec{\theta}), \vec{\phi}_F(\vec{x}) \rangle.$
Therefore, the immediate choice of feature map for modelling PQC models classically is the full trigonometric polynomial feature map $\vec\phi_F (\vec{x})$.
However, the associated model class could be too expressive thus it might overfit the training data points~\cite{schreiber2023classical}.
In addition, the size of the frequency spectrum could be exponential in the data dimension, which becomes intractable for classical computers.
Instead, one could hope to exploit some structure of the PQC to construct an efficient feature map $\vec\phi(\vec{x})$ to approximate the PQC models
\begin{align}
    \bra{{\bf 0}} U^\dagger(\vec{x},\vec\theta) O U(\vec{x},\vec\theta) \ket{{\bf 0}} &\approx \langle \vec w, \vec\phi (\vec{x}) \rangle.
\end{align}
According to our notation above, we would use the noise function $\xi(\vec{x})$ to refer to the approximation error $\xi(\vec{x}) = \langle \vec w_F, \vec\phi_F (\vec{x}) \rangle - \langle \vec w, \vec\phi (\vec{x}) \rangle$.
We drop the explicit $\vec\theta$ dependence of $\vec w_F$ and $\xi$ for ease of notation. 

One approach would be to construct $\vec\phi$ as a truncated version of $\vec\phi_F$.
This would take advantage of the fact that the high-frequency components of PQC models that are subjected to Pauli noise~\cite{fontana2023classical} typically make smaller contributions than lower frequency terms.
Thus, Fourier series with an appropriate level of truncation can be used to model PQC models without compromising much of the accuracy.
This approach assumes that we know which components to truncate ahead of time, though, and that might be unrealistic for practical scenarios.

As an alternative, one can utilize a popular technique from machine learning called Random Fourier Features (RFF)~\cite{rahimi2007random}, used to efficiently approximate the high-dimensional inner product $\langle \vec{w}_F (\vec{\theta}), \vec{\phi}_F(\vec{x}) \rangle$ by randomly selecting only a few of its dominant terms~\cite{landman2022classically,sweke2023potential}.
Using RFF amounts to performing a truncation of the Hilbert space, with the only difference being that the selection of components that are kept is probabilistic.
RFF has been proposed as an approach to ``dequantize" PQC-based quantum machine learning models by exploiting the efficient low-dimensional feature map $\vec{\phi}$ in Ref.~\cite{landman2022classically}.
On the other hand, Ref.~\cite{sweke2023potential} discusses the applicability of RFF in terms of which PQCs are likely to admit the efficient approximation.
In both these references, the task is not to \emph{learn} the classically efficient representation of the PQC but rather to show that a given downstream task can be learned efficiently classically, without ever running the PQC.
Even though the specific task is not the same, we observe that the main limitation in learning quantum models comes from an efficient classical representation, which deeply aligns with the use of RFF.
In \cref{a:RFF} we formally discuss how the performance guarantees of RFF bring about learnability in the sense introduced in \cref{Sec:Preliminary,Sec:VQMs-p-concepts}.
Also, we wish to emphasize that the efficient classical representation of PQCs is still under active research, but our work could be directly adapted if efficient feature maps were found.

\subsection{Modelling PQCs with and without link functions}
\label{Subsec:w-w/o-u}

In this section, we will discuss the operational role of the link function $u$ in the surrogation of PQC models. Without loss of generality, we let $O = \ket{{\bf 0}}\bra{{\bf 0}}$, hence we have
\begin{align} \label{Eq:PQC-models}
    f_{\vec{\theta}}(\vec{x}) = \lvert\bra{{\bf 0}} U(\vec{x},\vec{\theta}) \ket{{\bf 0}}\rvert^2 =  \langle \vec{w}, \vec{\phi}_{\mathrm{RFF}}(\vec{x}) \rangle + \xi(\vec{x}),
\end{align}
and $f_{\vec{\theta}}(\cdot) \in \mathcal{F}_{U,\ket{{\bf 0}}\bra{{\bf 0}}}$. 
To model the probabilistic function $f_{\vec{\theta}}(\vec{x})$, we consider the following hypothesis class
\begin{equation} \label{Eq:RFF-hypothesis-class}
 \HC_{\textrm{RFF}} = \left \{h(\vec{x}) = u(\langle \vec w, \vec{\phi}_{\mathrm{RFF}} (\vec{x})\rangle), \|\vec w\|_2 \le B\right\}
\end{equation}
where $u$ is the clipping function
\begin{equation} \label{Eq:Clipping-fn}
    u(x) = \begin{cases}
                0, & x<0\\
                x, & 0
                \le x \le 1\\
                1, & x>1
            \end{cases},
\end{equation}
a $1$-Lipschitz function that enforces the matching of co-domains of the hypothesis class $\mathcal{H}$ and $\lvert\bra{{\bf 0}} U(\vec{x},\vec\theta) \ket{{\bf 0}}\rvert^2$ while ensuring the output of the linear hypothesis $h$ within range $[0,1]$ is not distorted.
Note that the link function has no impact on $f_{\vec{\theta}}(\vec{x})$ since $f_{\vec{\theta}}(\vec{x}) \in [0,1]$.

To provide context regarding the value of $B$, we note that similar to the full Fourier representation of PQC models, the weight vectors of the RFF feature map can also be written as  
\begin{align}
    \vec{w}= \sqrt{D} \begin{pmatrix}
        a_{\vec{\tilde\omega}_1}(\vec{\theta})\\
        b_{\vec{\tilde\omega}_1}(\vec{\theta})\\
        \vdots\\
        a_{\vec{\tilde\omega}_{D}}(\vec{\theta})\\
        b_{\vec{\tilde\omega}_{D}}(\vec{\theta})\\
    \end{pmatrix}^\intercal
\end{align}
where $D$ is the dimension of the random Fourier feature map $\vec{\phi}_{\mathrm{RFF}}$ and $\vec{\tilde\omega}_i \in \Omega$ are the sampled frequencies from the original Fourier spectrum $\Omega$. 
Following general algorithm-independent results in statistical learning theory on RFFs~\cite{rahimi2008weighted, rahimi2008uniform}, we assume that the value $|a_{\vec{\tilde\omega}_i}(\vec{\theta})|$ and $|b_{\vec{\tilde\omega}_i}(\vec{\theta})|$ are bounded by some constant $K$. Note that $\|\vec w\|_2 \le KD \in \mathcal{O}(D)$.

Combining \cref{Theorem:Alphatron-guarantee} with the results from the RFF approximation yield \cref{Corollary:uPQCs-learnability}. 

\begin{corollary} 
\label{Corollary:uPQCs-learnability}
Consider the hypothesis class $\HC_{\mathrm{RFF}}$ as defined in \cref{Eq:RFF-hypothesis-class}, a target function $f(\vec{x}) \in \mathcal{F}_{U,\ket{{\bf 0}}\bra{{\bf 0}}}$, and variables as defined in \cref{Theorem:Alphatron-guarantee}. Let $\mathcal{S} = (\vec{x}_i,\bar{y}_i)_{i=1}^{N_1}$ be the training dataset with $\bar{y}_i$ estimated with $N_s$ measurement shots.
Running \cref{Algo:Learning-algorithm} with $\mathcal{S}$ will yield $h \in \HC_{\mathrm{RFF}}$ such that
\begin{align}
    \label{Eq:PQCs-gen-err}
    \ER(h) \le \mathcal O \left (\sqrt{\epsilon_1} + M\epsilon_2 + D \left[ \epsilon_3 + \epsilon_4 \right] + \epsilon_5 \right),
\end{align}
where $\epsilon_2 = \sqrt[\leftroot{-2}\uproot{2}4]{\frac{\log(\frac{1}{\delta})}{N_1}}$, $\epsilon_3 = \sqrt{\frac{1}{N_1}}$, $\epsilon_4 = \sqrt{\frac{\bar{\sigma} \log(\frac{1}{\delta})}{N_1 N_{s}}}$, $\epsilon_5 = \sqrt{\frac{ \log(\frac{1}{\delta})}{N_1}}$, and $\bar\sigma = \mathbb{E}_{\vec{x}}[\sigma^2_{\lambda|\vec{x}}]$. 
\end{corollary}

Here, we exploited the information about the co-domain of the target $p$-concepts to design an appropriate link function $u$ that restricts the size of the hypothesis class. To illustrate the impact of limiting the hypothesis class size, we relax the co-domains matching constraint, i.e., set $u$ to be the identity map, hence the hypothesis class considered becomes 
\begin{align}\label{Eq:Eq:RFF-hypothesis-class-w/o-link}
    \mathcal{G}_{\mathrm{RFF}} = \left \{g(\vec{x}) = \langle \vec w, \vec{\phi}_{\mathrm{RFF}} (\vec{x})\rangle, \|\vec w\|_2 \le B\right\}.
\end{align}

The most straightforward method to learn $f_{\vec{\theta}}(\vec{x})$ under this relaxed formulation is to directly minimize the empirical risk $\widehat{R}(h)$ given a sample $\mathcal{S}$ sampled from the distribution $\bar{\mathcal{D}}$, which we call empirical risk minimization (ERM).
We can formulate the above as a quadratically constrained quadratic program as follows:
\begin{equation}\label{Eq:ERM}
\vec w^* = \argmin_{\vec w, \|\vec{w}\|_2\le B} \frac{1}{|\mathcal S|}\sum_{(\vec{x}, \bar y) \in \mathcal{S}}|\langle \vec w, \vec{\phi}(\vec{x})\rangle - \bar{y}|^2,
\end{equation}
which can be efficiently solved by convex optimization methods such as interior point methods~\cite{boyd2004interior} or projected gradient descent. Alternatively, by including the constraint in the loss with Lagrangian multipliers, the problem can be formulated as a ridge regression task. Various prior work use this formulation to tackle learning problems involving PQCs~\cite{schreiber2023classical, landman2022classically, sweke2023potential}.

\begin{lemma} 
\label{Lemma:PQCs-learnability}Consider the hypothesis class $\GC_{\mathrm{RFF}}$ as defined in \cref{Eq:Eq:RFF-hypothesis-class-w/o-link}, a target function $f(\vec{x}) \in \mathcal{F}_{U,\ket{{\bf 0}}\bra{{\bf 0}}}$, and variables as defined in \cref{Theorem:Alphatron-guarantee}. Let $\mathcal{S} = (\vec{x}_i,\bar{y}_i)_{i=1}^{N_1}$ be the training dataset with $\bar{y}_i$ estimated with $N_s$ measurement shots.
Optimizing \cref{Eq:ERM} with $\mathcal{S}$ will yield $g_\mathcal{S}^{\mathrm{ERM}} \in \GC_{\mathrm{RFF}}$ such that
\begin{align} \label{Eq:ERM-gen-err}
\ER(g_\mathcal{S}^{\mathrm{ERM}}) \le \mathcal{O}\left( \epsilon_1 + D^2 \sqrt{\frac{\log\frac{1}{\delta}}{N_1}}\right).
\end{align}
\end{lemma}

The proof of this lemma can be found in \cref{appLemma1Proof}. Similar to \cref{Eq:PQCs-gen-err}, one could understand \cref{Eq:ERM-gen-err} from the bias and variance perspective, i.e., the first term informs the bias of the model while the second term tells us about the model's variance. Firstly, the inclusion of the link function $u$ in the hypothesis class $\HC$ results in a class of model that has higher bias as compared to hypothesis class $\mathcal{G}$ hence leads to a quadratic increase in error $\epsilon_1$. Consequently, $\HC$ that is higher in bias will have lower variance, and we can observe the separate and asymmetrical effects of the data sampling and shot noises on the explicit risk. Removing the link function leads to higher variance in $\mathcal{G}$, and the sensitivity to shot and data sampling noises becomes indistinguishable. The relationship between errors in these two generalization bounds essentially manifests the bias-variance trade-off. Further, such results showcase the fact the ERM-based hypothesis selection can still generalize with a constant number of measurements provided that we have abundant data points.

We note that without the application of the link function $u$ in our modelling, classical models are much more susceptible to shot noise. Our theoretical results of \cref{Corollary:Trade-off} and \cref{Lemma:PQCs-learnability} imply that to learn labels obtained from constant number measurements, without the link function $u$, classical algorithms may require up to data points $N_1$ that are square of what is needed for models with the link function $u$. In the following section, we numerically showcase this property.

\begin{figure*}[t]
\includegraphics[width=\textwidth]{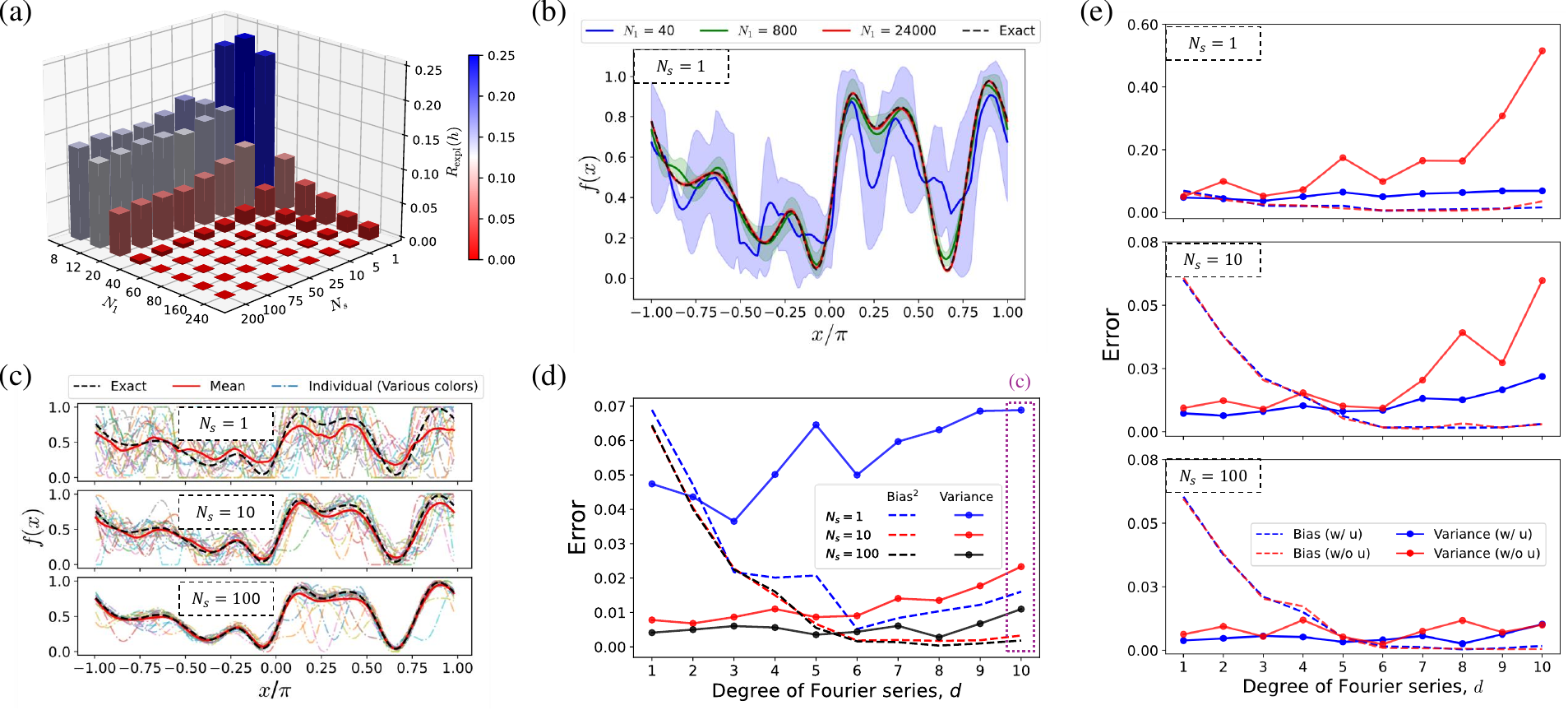}
\caption{\label{Fig:Numerical-results}(a) The averaged explicit risk for different numbers of training data points $N_1$ and number of measurement shots $N_s$. The overall trends agreed with the theoretical prediction in \cref{Fig:Theoretical-results}: for a fixed $N_1$, the explicit risk saturated after some threshold value of $N_s$, but the explicit risk can be reduced by increasing $N_1$ regardless of the value of $N_s$. 
(b) When the model in $\HC_{10}$ are presented with a sufficiently large dataset, i.e., $N_1 = 24000$, the exact function (black dashed line) can be learned even if the labels are estimated with one measurement shot. 
(c) Twenty different trained models (dotted dashed line of various colours) from $\HC_{10}$ and their mean predictors (solid red line) for $N_1 = 1, 10, 100$. 
Increasing $N_s$ reduces the shot noise, hence reducing the spread of the trained models. 
(d) The bias-variance trade-off curve. The bias and variance of the trained models in (c) are calculated and plotted in the purple dotted box. 
The rest of the values are computed using similar procedures as per (c) for $\HC_d$ with $d = \{ 1,2,3,4,5,6,7,8,9 \}$. 
Both the bias and variance decrease when $N_s$ increases, illustrating the shot-noise dependent bias-variance trade-off.
(e) Bias and variance for models with and without the link function $u$. 
The models without the link function are more expressive, hence they are more susceptible to the shot noise, i.e., they have a higher tendency to overfit the shot noise. 
Increasing $N_s$ will reduce the shot noise, hence suppressing the shot-noise-induced variance. 
Note that the same target function is considered in all these numerical experiments. 
}
\end{figure*}

\section{Numerical validation on the role of shot noise}

In this section, we will provide numerical verifications of our theoretical results, validating the operational roles of shot noise in learning quantum models. 

\subsection{Numerical settings}
We consider the data re-uploading model~\cite{perezsalinas2020data} for one-dimensional data points $x$ in our numerical demonstration
\begin{align}
    f_{\vec{\theta}}(x) = \left|\bra{0} \mathrm{Rot} \left(\vec{\theta}^{L_r+1} \right) \: \Pi_{l=1}^{L_r} \left[R_X(x) \mathrm{Rot} \left(\vec{\theta}^l \right) \right] \ket{0} \right|^2,
\end{align}
where $\vec{\theta} = (\vec{\theta}^1, \dots, \vec{\theta}^{L_r+1})$ is the set of rotational angles, $\mathrm{Rot}(\vec{\theta}^l) = R_Z (\theta_3^l) R_Y (\theta_2^l) R_Z (\theta_1^l)$ is the universal single qubit unitary gate, $L_r$ is the number of layer repetitions, and $R_{P}(\cdot)$ are the Pauli rotation unitary gates with $P \in \{ X, Y, Z \}$. While we considered only the single qubit data re-uploading model with one-dimensional data points, it is straightforward to generalize our results to the multi-qubit model or with multi-dimensional data points. As described, the data re-uploading model can be expressed as a truncated Fourier series
\begin{align}
    f_{\vec{\theta}}(x) = c_{0} (\vec{\theta}) + \sum_{\omega=1}^{L_r} a_{\omega}(\vec{\theta}) \cos(\omega x) + b_{\omega}(\vec{\theta}) \sin(\omega x) , 
\end{align}
with Fourier spectrum $\Omega^+_{L_r} = \{1, \dots, L_r\}$ while the Fourier coefficients are dependent solely on $\vec{\theta}$ and their associated unitaries. 

Now, we will describe our numerical setting. Firstly, we considered fixed randomly generated angles $\vec{\theta}$ in our numerical demonstrations, and this set of angles is used for all numerical experiments. Therefore, we will drop the dependency on $\vec{\theta}$ from now on. In addition, we set $L_r=10$ for the data re-uploading model to generate a degree $10$ Fourier series. Such a target function is sufficiently complex for us to observe various impacts of shot noise in learning $f(x)$.  Finally, this numerical example does not require the utilization of random Fourier features, as the model under consideration is rather straightforward; therefore, the truncation method suffices. 
Specifically, we consider the following truncated Fourier series as our hypothesis class
\begin{align}
    \HC_d = \left \{ h_d(x) = u \left(\nu_{0}  + \sum_{\omega=1}^{d} \alpha_{\omega} \cos(\omega x) + \beta_{\omega} \sin(\omega x) \right) \right \}
\end{align}
where $\nu_0, \alpha_\omega, \beta_\omega \in \mathbb{R}$, $u(\cdot)$ is the clipping function as defined in \cref{Eq:Clipping-fn} and $d \in \mathbb{N}$ controls the degree of the truncated Fourier series, hence the complexity of $\HC_d$. 
For all numerical experiments, the number of training steps $T$ is fixed as 50, and 500 testing data points are used to evaluate the performance of trained models. To distinguish the hypothesis class with and without the link function, we denote $\HC_d$ with the identity link function as $\mathcal{G}_d$. Note that all datasets are extracted using the procedures described in \cref{Subsubsec:Data-Extraction}.

\subsection{Asymmetrical effects of \texorpdfstring{$N_1$}{N1} and \texorpdfstring{$N_s$}{Ns}}
To begin with, we investigate the asymmetry dependent on the explicit risk of the number of training data points $N_1$ and the number of measurement shots $N_s$. In particular, we set $d = 10$ such that the approximation error $\epsilon_1 = 0$, i.e., when $\nu_0 = c_0$, $\alpha_\omega = a_\omega$, and $\beta_\omega = b_\omega$ for all $\omega \in \Omega^+_{10}$, enabling us to isolate the impact of these two attributes. In this example, the ratio of the number of training data points $N_1$ to the number of validation data points $N_2$ is $N_1$:$N_2$ $=$ 8$:$2 with total data points $N = \{10, 15, 25, 50, 75, 100, 200, 300\}$. That is, we trained the model in $\HC_{10}$ with different pairwise combinations of $N_1 = \{8, 12, 20, 40, 60, 80, 160, 240\}$ training data points and $N_s = \{1, 5, 10, 25, 50, 75, 100, 200\}$ measurement shots using \cref{Algo:Learning-algorithm} under $T = 50$ training iterations, and the optimal model is chosen using $N_2 = \{2, 3, 5, 10, 15, 20, 40, 60\}$ validation data points for the respective value of $N_1$. Finally, the explicit risk is estimated with $500$ testing data points and we averaged the explicit risk over 5 random instances of training and validation datasets. 

The results shown in \cref{Fig:Numerical-results}~(a) agreed with our theoretical prediction in \cref{Fig:Theoretical-results}, validating the asymmetrical effects of $N_1$ and $N_s$ as described in \cref{Corollary:Asymmetry}. In particular, it shows the decreasing trend of explicit risk with the increase of $N_1$ while keeping $N_s = 1$. This observation is further validated by \cref{Fig:Numerical-results}~(b), where the exact function can be learned when the model is presented with sufficiently large training data points with labels estimated using one measurement repetition, i.e., $N_1 = 2.4 \times 10^4$ and $N_s = 1$. The three solid curves in \cref{Fig:Numerical-results}~(b) are the mean predictors obtained using training datasets of size $N_1 = \{ 40, 800, 24000 \}$ and validation datasets of size $N_2 = \{10, 200, 600\}$ respectively. Each of these mean predictors is averaged over 5 different training instances and the shaded regions are the standard deviations of the predictions. As expected, increasing $N_1$ reduces the standard deviations of the predictors and improves the mean predictions. 

\subsection{Shot-noise dependent bias-variance trade-off}

As discussed in \cref{Subsec:Shot-noise_Bias-Variance}, training models with different finite-size training datasets will yield different trained models.
This phenomenon is observed in \cref{Fig:Numerical-results}~(c) where 20 distinct trained models from $\HC_{10}$, i.e., dashed-dotted lines of different colours, each trained with different training datasets of size 40 are different across $N_s = 1, 10, 100$. 
In addition, the reducing fluctuations of the trained models with increasing $N_s$ demonstrated the $N_s$-dependent relationship between these trained models. 
As $N_1$ is sufficiently large, the prediction accuracy can be improved by increasing $N_s$ and this is reflected in \cref{Fig:Numerical-results}~(c) where the mean predictor is approaching the exact function as $N_s$ increases. 
These two observations can otherwise be captured by computing two statistical quantities, the squared bias
\begin{align}
    \mathbb{E}_{\vec{x}} \left[\textbf{Bias}^2_{\mathcal S} \right] &:= \mathbb{E}_{\vec{x}}\left[\left(\mathbb{E}_{\mathcal S}[h_{\mathcal S}(\vec{x})] - f(\vec{x}) \right)^2 \right]
\end{align}
and the variance
\begin{align}
    \mathbb{E}_{\vec{x}} \left[\textbf{Var}_{\mathcal S}\right] & := \mathbb{E}_{\vec{x},{\mathcal S}} \left[\left(\mathbb{E}_{\mathcal S}[h_{\mathcal S}(\vec{x})] - h_{\mathcal S}(\vec{x})\right)^2 \right].
\end{align}
In particular, we compute their empirical versions using the trained models as per \cref{Fig:Numerical-results}~(c) using 500 testing data points. The computed values are plotted in \cref{Fig:Numerical-results}~(d) at $d = 10$, i.e., the points in the purple dotted box. As expected the bias and variance reduce when $N_s$ increases.

These exact settings and procedures as per \cref{Fig:Numerical-results}~(c) are repeated to obtain trained models from $\HC_d$ for $d = \{1, 2, 3, 4, 5, 6, 7, 8, 9\}$, and these models are then used to compute their respective bias and variance. 
Plotting their bias and variance yields the bias-variance trade-off curve, as shown in \cref{Fig:Numerical-results}~(d). 
Across $N_s = 1, 10, 100$, the bias (variance) consistently decreases (increases) with increasing $d$. 
This observation is consistent with the bias-variance trade-off concept, where less complex models (in our case, $\HC_d$ with lower $d$) will have higher bias but with lower variance. In contrast, the highly complicated models will have lower bias but with higher variance. 
The former type of model tends to underfit the training data while the latter is more likely to overfit the training data. 
In addition, \cref{Fig:Numerical-results}~(d) illustrates the shot-noise dependent bias-variance trade-off as described in \cref{Subsec:Shot-noise_Bias-Variance}: Increase $N_s$ will reduce the bias and variance of the models. 

Using the same settings as per \cref{Fig:Numerical-results}~(d) but a different training procedure, we extract the bias and variance of the hypothesis without the link function $u$, i.e., $\mathcal{G}_d$. 
Specifically, there is an exact analytical solution if we solve \cref{Eq:ERM} using kernel ridge regression and we use the validation dataset to choose the optimal regularization strength out of $C = \{0.006, 0.015, 0.03, 0.0625, 0.125, 0.25, 0.5, 1.0, 2.0, 5.0, 8.0,$ $16.0, 32.0, 64.0, 128.0, 256, 512, 1024\}$. 
Then, their bias and variance are compared against the hypothesis equipped with the link function in \cref{Fig:Numerical-results}~(e) and these numerical results are in agreement with the theoretical analysis in \cref{Subsec:w-w/o-u}. 
That is, the variance of $\mathcal{G}_d$ is significantly higher than their counterpart when $N_s$ is low. High shot noise implies that the estimated labels would be very different from their exact values and a more expressive model class like $\mathcal{G}_d$ will have a higher tendency to overfit the shot noise, lending to a higher variance. Increasing $N_s$ reduces the shot noise but the finite data sampling noise remains. This explains the reducing but non-vanishing variance for both $\HC_d$ and $\mathcal{G}_d$ when $N_s$ increases as well as the success of current learning protocol using $\mathcal{G}_d$~\cite{che2023exponentially,lewis2024improved,schreiber2023classical}. 
Interestingly, the model's bias with the link function matches well with the one without. In summary, the link function helps suppress the shot noise-induced variance by restricting the expressivity of the hypothesis class.

\begin{figure}[t]
 \includegraphics[width=80mm,scale=0.5]{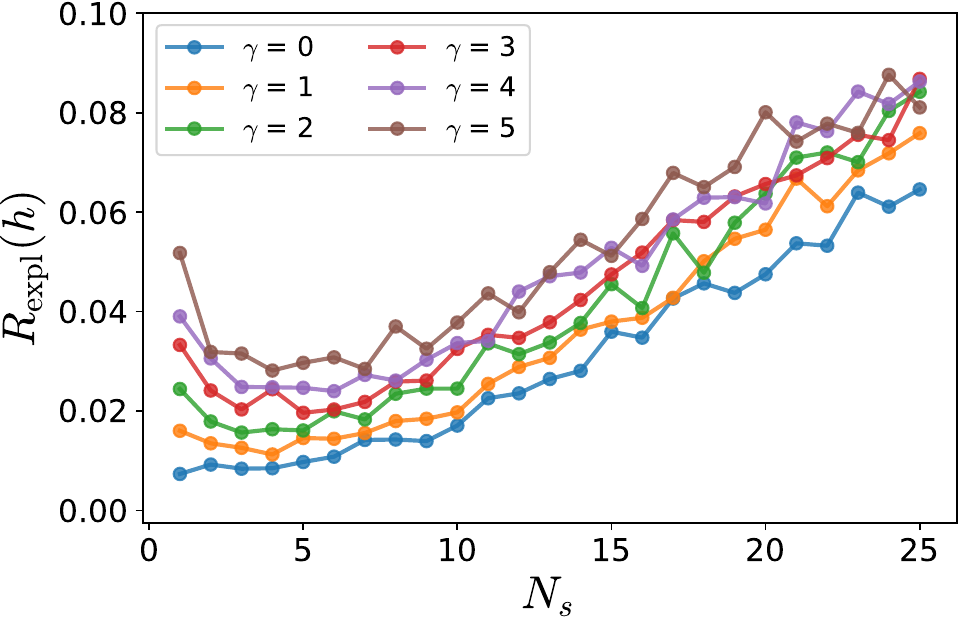}
\caption{\label{Fig:Trade-off}
The trade-off between $N_1$ and $N_s$ is considered under a fixed total measurement budget of $N_{\mathrm{tot}} = 600$ for $\gamma = \{0,1,2,3,4,5\}$ and $N_s = \{1,2,3,\dots,24,25\}$.
When $N_1$ and $N_s$ are treated equally, i.e., $\gamma = 0$, the optimal pair of $N_1$ and $N_s$ is given by $(N_1^*,N_s^*) = (600,1)$. As $\gamma$ increases, more measurement shots are required, hence smaller $N_1$, to achieve better model performance. However, there will be a threshold beyond which the performance of models worsens.
}
\end{figure}

\subsection{Trade-off between \texorpdfstring{$N_1$}{N1} and \texorpdfstring{$N_s$}{Ns}}

Finally, we numerically investigate the trade-off between $N_1$ and $N_s$ under a fixed total measurement budget of $N_{\mathrm{tot}}$. 
Recall that the relationship between $N_{\mathrm{tot}}$, $N_1$, and $N_s$ is given by $N_{\mathrm{tot}} = (N_1 + N_2) (N_s + \gamma)$, where $\gamma$ is the penalty cost and $N_2$ is the size of validation dataset. 
The inclusion of $N_2$ captures the resource constraint for choosing the optimal time step $T^*$. 
Here, we let $N_{\mathrm{tot}} = 600$, $\gamma = \{0,1,2,3,4,5\}$, and the ratio of $N_1$ to $N_2$ be 8:2. 
Furthermore, we set $N_s = \{1,2,3,\dots,24,25\}$ giving different combinations of $N_1$ and $N_2$. 
Repeating the similar procedures as per \cref{Fig:Numerical-results}~(a) over 60 random instances of training and validation datasets for the above-mentioned settings yields \cref{Fig:Trade-off}. 
As predicted by \cref{Corollary:Trade-off}, for a fixed $N_{\mathrm{tot}}$ and when $\gamma = 0$, the performance of classical machines can be enhanced by reducing $N_s$, and the optimality is achieved when $N_s = 1$. On the other hand, the optimal pair of $N_1$ and $N_s$ depends on the penalty cost when $\gamma > 0$; the larger the $\gamma$, the higher the $N_s$ required to achieve optimal model performance. 

\section{Discussion}
\label{Sec:Discussion}

Finite measurement or shot noise is an intrinsic quantum phenomenon. 
Such noise is always present in the estimation of quantum models; hence, classical machines will unavoidably encounter shot noise when learning quantum models. 
Therefore, it is crucial to understand whether shot noise could increase the difficulty for classical machines to learn quantum models, or if it is just a statistical feature that can be well-handled by classical models. 

By formulating parameterized quantum models as probabilistic concepts, we show that classical machines can learn quantum models with efficient classical representation in the presence of shot noise. 
Said otherwise, the fundamental hardness of learning quantum models depends on the existence of efficient classical representation, while the impact of shot noise is only prominent when there is an insufficient number of training data points. 
When sufficient training data points are provided, classical learning of quantum models is possible even when the labels are estimated with limited measurement shots. 
This asymmetrical effect of the number of training data and measurement repetitions arises from the differences in information gained when sampling each component.
That is, one effectively samples across various labels when sampling across the training points but increasing the resolution of the labels does not provide extra information on other data points.

Each quantum measurement, be it on a fixed or different parameter setting, counts as a query to quantum models.
While unlimited queries to quantum models are desired, our limited time and monetary resources force us to wisely distribute our budget to maximize the information extracted from quantum models.
That is, one has to choose to train the classical machines with datasets consisting of either more inputs with noisier labels or fewer inputs with cleaner labels. 
If sampling across parameter settings does not incur extra cost compared to sampling quantum models with the same parameter setting, then the classical machine would learn better with datasets consisting of more inputs but with the noisiest labels. 
Otherwise, the optimal budget partition would depend on the cost differences between measurements with fixed and different parameter settings.

While the hardness of learning quantum models classically is not dictated by the shot noise, it has an impact on the actual training of classical machines. 
For a given set of training data points $\{\vec{x}_i\}_{i=1}^{N_1}$, different label sampling instances will yield different training datasets, i.e., $\mathcal{S} = (\vec{x}_i, \bar{y}_{i})_{i=1}^{N_1}$ or $\mathcal{S}' = (\vec{x}_i, \bar{y}'_{i})_{i=1}^{N_1}$. Each dataset will produce an associated trained model. 
We capture this model's sensitivity to variation of labels through the bias-variance-noise decomposition and show that the link function can suppress this undesired sensitivity by restricting the size of the hypothesis class.
We further use our framework for the classical surrogation of parameterized quantum circuit models, and our theoretical analysis correctly predicts the behaviours of classical surrogates in the presence of shot noise.

Viewed from other angles, our work provides a framework to study the impact of classical approximation and shot noise on learning quantum models classically. 
Future works could focus on searching for good classical approximations, and our framework could be directly adapted to handle shot noise. 
An interesting direction is to combine our framework with the analysis in Ref.~\cite{cerezo2023does} to investigate the classical learnability of the parameterized quantum circuit models that are free of barren plateaus. 
This will provide an alternative perspective on the relationship between classical simulability and learnability of parameterized quantum models~\cite{hinsche2023one}. 
Shallow parameterized quantum circuits usually admit efficient classical representation, yet they might experience exponential concentration if observables are not chosen carefully ~\cite{cerezo2021cost}. 
This setting is suitable to push the limits of our framework to check if classical machines can still learn such models under the influence of exponential concentration.

Finally, the core of our framework is the assumption that the parameterized quantum models represent the conditional unbiased expectation of their unbiased estimators. 
However, estimators might not be unbiased after some post-processing operations. 
An example of such post-processing operations is quantum error mitigation. 
It will be interesting to investigate the role of shot noise in the biased regime. 

\begin{acknowledgments}
The authors thank Yuxuan Du, Naixu Guo, Hela Mhiri, and Manuel Rudolph for discussions. This work is supported by the National Research Foundation, Singapore, and A*STAR under its CQT Bridging Grant and its Quantum Engineering Programme under grant NRF2021-QEP2-02-P05. EGF is supported by a Google PhD Fellowship, the Einstein Foundation (Einstein Research Unit on Quantum Devices), BMBF (Hybrid), and BMWK (EniQmA).
\end{acknowledgments}

\bibliography{main}

\begin{thebibliography}{50}%
\makeatletter
\providecommand \@ifxundefined [1]{%
 \@ifx{#1\undefined}
}%
\providecommand \@ifnum [1]{%
 \ifnum #1\expandafter \@firstoftwo
 \else \expandafter \@secondoftwo
 \fi
}%
\providecommand \@ifx [1]{%
 \ifx #1\expandafter \@firstoftwo
 \else \expandafter \@secondoftwo
 \fi
}%
\providecommand \natexlab [1]{#1}%
\providecommand \enquote  [1]{``#1''}%
\providecommand \bibnamefont  [1]{#1}%
\providecommand \bibfnamefont [1]{#1}%
\providecommand \citenamefont [1]{#1}%
\providecommand \href@noop [0]{\@secondoftwo}%
\providecommand \href [0]{\begingroup \@sanitize@url \@href}%
\providecommand \@href[1]{\@@startlink{#1}\@@href}%
\providecommand \@@href[1]{\endgroup#1\@@endlink}%
\providecommand \@sanitize@url [0]{\catcode `\\12\catcode `\$12\catcode `\&12\catcode `\#12\catcode `\^12\catcode `\_12\catcode `\%12\relax}%
\providecommand \@@startlink[1]{}%
\providecommand \@@endlink[0]{}%
\providecommand \url  [0]{\begingroup\@sanitize@url \@url }%
\providecommand \@url [1]{\endgroup\@href {#1}{\urlprefix }}%
\providecommand \urlprefix  [0]{URL }%
\providecommand \Eprint [0]{\href }%
\providecommand \doibase [0]{https://doi.org/}%
\providecommand \selectlanguage [0]{\@gobble}%
\providecommand \bibinfo  [0]{\@secondoftwo}%
\providecommand \bibfield  [0]{\@secondoftwo}%
\providecommand \translation [1]{[#1]}%
\providecommand \BibitemOpen [0]{}%
\providecommand \bibitemStop [0]{}%
\providecommand \bibitemNoStop [0]{.\EOS\space}%
\providecommand \EOS [0]{\spacefactor3000\relax}%
\providecommand \BibitemShut  [1]{\csname bibitem#1\endcsname}%
\let\auto@bib@innerbib\@empty
\bibitem [{\citenamefont {Aaronson}(2020)}]{aaronson2020shadow}%
  \BibitemOpen
  \bibfield  {author} {\bibinfo {author} {\bibfnamefont {S.}~\bibnamefont {Aaronson}},\ }\bibfield  {title} {\bibinfo {title} {Shadow tomography of quantum states},\ }\href {https://doi.org/10.1137/18M120275X} {\bibfield  {journal} {\bibinfo  {journal} {SIAM J. Comput.}\ }\textbf {\bibinfo {volume} {49}},\ \bibinfo {pages} {STOC18} (\bibinfo {year} {2020})}\BibitemShut {NoStop}%
\bibitem [{\citenamefont {Huang}\ \emph {et~al.}(2020{\natexlab{a}})\citenamefont {Huang}, \citenamefont {Zhang}, \citenamefont {Newman}, \citenamefont {Cai}, \citenamefont {Gao}, \citenamefont {Tian}, \citenamefont {Wu}, \citenamefont {Xu}, \citenamefont {Yu}, \citenamefont {Yuan}, \citenamefont {Szegedy}, \citenamefont {Shi},\ and\ \citenamefont {Chen}}]{huang2020classical}%
  \BibitemOpen
  \bibfield  {author} {\bibinfo {author} {\bibfnamefont {C.}~\bibnamefont {Huang}}, \bibinfo {author} {\bibfnamefont {F.}~\bibnamefont {Zhang}}, \bibinfo {author} {\bibfnamefont {M.}~\bibnamefont {Newman}}, \bibinfo {author} {\bibfnamefont {J.}~\bibnamefont {Cai}}, \bibinfo {author} {\bibfnamefont {X.}~\bibnamefont {Gao}}, \bibinfo {author} {\bibfnamefont {Z.}~\bibnamefont {Tian}}, \bibinfo {author} {\bibfnamefont {J.}~\bibnamefont {Wu}}, \bibinfo {author} {\bibfnamefont {H.}~\bibnamefont {Xu}}, \bibinfo {author} {\bibfnamefont {H.}~\bibnamefont {Yu}}, \bibinfo {author} {\bibfnamefont {B.}~\bibnamefont {Yuan}}, \bibinfo {author} {\bibfnamefont {M.}~\bibnamefont {Szegedy}}, \bibinfo {author} {\bibfnamefont {Y.}~\bibnamefont {Shi}},\ and\ \bibinfo {author} {\bibfnamefont {J.}~\bibnamefont {Chen}},\ }\href@noop {} {\bibinfo {title} {Classical simulation of quantum supremacy circuits}} (\bibinfo {year} {2020}{\natexlab{a}}),\ \Eprint {https://arxiv.org/abs/2005.06787} {arXiv:2005.06787 [quant-ph]} \BibitemShut
  {NoStop}%
\bibitem [{\citenamefont {Huang}\ \emph {et~al.}(2021)\citenamefont {Huang}, \citenamefont {Broughton}, \citenamefont {Mohseni}, \citenamefont {Babbush}, \citenamefont {Boixo}, \citenamefont {Neven},\ and\ \citenamefont {McClean}}]{huang2021power}%
  \BibitemOpen
  \bibfield  {author} {\bibinfo {author} {\bibfnamefont {H.-Y.}\ \bibnamefont {Huang}}, \bibinfo {author} {\bibfnamefont {M.}~\bibnamefont {Broughton}}, \bibinfo {author} {\bibfnamefont {M.}~\bibnamefont {Mohseni}}, \bibinfo {author} {\bibfnamefont {R.}~\bibnamefont {Babbush}}, \bibinfo {author} {\bibfnamefont {S.}~\bibnamefont {Boixo}}, \bibinfo {author} {\bibfnamefont {H.}~\bibnamefont {Neven}},\ and\ \bibinfo {author} {\bibfnamefont {J.~R.}\ \bibnamefont {McClean}},\ }\bibfield  {title} {\bibinfo {title} {Power of data in quantum machine learning},\ }\href {https://doi.org/10.1038/s41467-021-22539-9} {\bibfield  {journal} {\bibinfo  {journal} {Nat. Commun.}\ }\textbf {\bibinfo {volume} {12}},\ \bibinfo {pages} {2631} (\bibinfo {year} {2021})}\BibitemShut {NoStop}%
\bibitem [{\citenamefont {Huang}\ \emph {et~al.}(2023)\citenamefont {Huang}, \citenamefont {Chen},\ and\ \citenamefont {Preskill}}]{huang2023learning}%
  \BibitemOpen
  \bibfield  {author} {\bibinfo {author} {\bibfnamefont {H.-Y.}\ \bibnamefont {Huang}}, \bibinfo {author} {\bibfnamefont {S.}~\bibnamefont {Chen}},\ and\ \bibinfo {author} {\bibfnamefont {J.}~\bibnamefont {Preskill}},\ }\bibfield  {title} {\bibinfo {title} {Learning to predict arbitrary quantum processes},\ }\href {https://doi.org/10.1103/PRXQuantum.4.040337} {\bibfield  {journal} {\bibinfo  {journal} {PRX Quantum}\ }\textbf {\bibinfo {volume} {4}},\ \bibinfo {pages} {040337} (\bibinfo {year} {2023})}\BibitemShut {NoStop}%
\bibitem [{\citenamefont {Zhao}\ \emph {et~al.}(2023{\natexlab{a}})\citenamefont {Zhao}, \citenamefont {Lewis}, \citenamefont {Kannan}, \citenamefont {Quek}, \citenamefont {Huang},\ and\ \citenamefont {Caro}}]{zhao2023learning}%
  \BibitemOpen
  \bibfield  {author} {\bibinfo {author} {\bibfnamefont {H.}~\bibnamefont {Zhao}}, \bibinfo {author} {\bibfnamefont {L.}~\bibnamefont {Lewis}}, \bibinfo {author} {\bibfnamefont {I.}~\bibnamefont {Kannan}}, \bibinfo {author} {\bibfnamefont {Y.}~\bibnamefont {Quek}}, \bibinfo {author} {\bibfnamefont {H.-Y.}\ \bibnamefont {Huang}},\ and\ \bibinfo {author} {\bibfnamefont {M.~C.}\ \bibnamefont {Caro}},\ }\href@noop {} {\bibinfo {title} {Learning quantum states and unitaries of bounded gate complexity}} (\bibinfo {year} {2023}{\natexlab{a}}),\ \Eprint {https://arxiv.org/abs/2310.19882} {arXiv:2310.19882 [quant-ph]} \BibitemShut {NoStop}%
\bibitem [{\citenamefont {Zhao}\ \emph {et~al.}(2023{\natexlab{b}})\citenamefont {Zhao}, \citenamefont {Guo}, \citenamefont {Luo},\ and\ \citenamefont {Rebentrost}}]{zhao2023provable}%
  \BibitemOpen
  \bibfield  {author} {\bibinfo {author} {\bibfnamefont {L.}~\bibnamefont {Zhao}}, \bibinfo {author} {\bibfnamefont {N.}~\bibnamefont {Guo}}, \bibinfo {author} {\bibfnamefont {M.-X.}\ \bibnamefont {Luo}},\ and\ \bibinfo {author} {\bibfnamefont {P.}~\bibnamefont {Rebentrost}},\ }\href@noop {} {\bibinfo {title} {Provable learning of quantum states with graphical models}} (\bibinfo {year} {2023}{\natexlab{b}}),\ \Eprint {https://arxiv.org/abs/2309.09235} {arXiv:2309.09235 [quant-ph]} \BibitemShut {NoStop}%
\bibitem [{\citenamefont {Anshu}\ and\ \citenamefont {Arunachalam}(2023)}]{anshu2023survey}%
  \BibitemOpen
  \bibfield  {author} {\bibinfo {author} {\bibfnamefont {A.}~\bibnamefont {Anshu}}\ and\ \bibinfo {author} {\bibfnamefont {S.}~\bibnamefont {Arunachalam}},\ }\bibfield  {title} {\bibinfo {title} {A survey on the complexity of learning quantum states},\ }\href {http://doi.org/10.1038/s42254-023-00662-4} {\bibfield  {journal} {\bibinfo  {journal} {Nat. Rev. Phys.}\ }\textbf {\bibinfo {volume} {6}},\ \bibinfo {pages} {59–69} (\bibinfo {year} {2023})}\BibitemShut {NoStop}%
\bibitem [{\citenamefont {Schreiber}\ \emph {et~al.}(2023)\citenamefont {Schreiber}, \citenamefont {Eisert},\ and\ \citenamefont {Meyer}}]{schreiber2023classical}%
  \BibitemOpen
  \bibfield  {author} {\bibinfo {author} {\bibfnamefont {F.~J.}\ \bibnamefont {Schreiber}}, \bibinfo {author} {\bibfnamefont {J.}~\bibnamefont {Eisert}},\ and\ \bibinfo {author} {\bibfnamefont {J.~J.}\ \bibnamefont {Meyer}},\ }\bibfield  {title} {\bibinfo {title} {Classical surrogates for quantum learning models},\ }\href {https://doi.org/10.1103/PhysRevLett.131.100803} {\bibfield  {journal} {\bibinfo  {journal} {Phys. Rev. Lett.}\ }\textbf {\bibinfo {volume} {131}},\ \bibinfo {pages} {100803} (\bibinfo {year} {2023})}\BibitemShut {NoStop}%
\bibitem [{\citenamefont {Huang}\ \emph {et~al.}(2022)\citenamefont {Huang}, \citenamefont {Kueng}, \citenamefont {Torlai}, \citenamefont {Albert},\ and\ \citenamefont {Preskill}}]{huang2022provably}%
  \BibitemOpen
  \bibfield  {author} {\bibinfo {author} {\bibfnamefont {H.-Y.}\ \bibnamefont {Huang}}, \bibinfo {author} {\bibfnamefont {R.}~\bibnamefont {Kueng}}, \bibinfo {author} {\bibfnamefont {G.}~\bibnamefont {Torlai}}, \bibinfo {author} {\bibfnamefont {V.~V.}\ \bibnamefont {Albert}},\ and\ \bibinfo {author} {\bibfnamefont {J.}~\bibnamefont {Preskill}},\ }\bibfield  {title} {\bibinfo {title} {Provably efficient machine learning for quantum many-body problems},\ }\href {http://doi.org/10.1126/science.abk3333} {\bibfield  {journal} {\bibinfo  {journal} {Science}\ }\textbf {\bibinfo {volume} {377}},\ \bibinfo {pages} {eabk3333} (\bibinfo {year} {2022})}\BibitemShut {NoStop}%
\bibitem [{\citenamefont {Che}\ \emph {et~al.}(2023)\citenamefont {Che}, \citenamefont {Gneiting},\ and\ \citenamefont {Nori}}]{che2023exponentially}%
  \BibitemOpen
  \bibfield  {author} {\bibinfo {author} {\bibfnamefont {Y.}~\bibnamefont {Che}}, \bibinfo {author} {\bibfnamefont {C.}~\bibnamefont {Gneiting}},\ and\ \bibinfo {author} {\bibfnamefont {F.}~\bibnamefont {Nori}},\ }\href@noop {} {\bibinfo {title} {Exponentially improved efficient machine learning for quantum many-body states with provable guarantees}} (\bibinfo {year} {2023}),\ \Eprint {https://arxiv.org/abs/2304.04353} {arXiv:2304.04353 [quant-ph]} \BibitemShut {NoStop}%
\bibitem [{\citenamefont {Lewis}\ \emph {et~al.}(2024)\citenamefont {Lewis}, \citenamefont {Huang}, \citenamefont {Tran}, \citenamefont {Lehner}, \citenamefont {Kueng},\ and\ \citenamefont {Preskill}}]{lewis2024improved}%
  \BibitemOpen
  \bibfield  {author} {\bibinfo {author} {\bibfnamefont {L.}~\bibnamefont {Lewis}}, \bibinfo {author} {\bibfnamefont {H.-Y.}\ \bibnamefont {Huang}}, \bibinfo {author} {\bibfnamefont {V.~T.}\ \bibnamefont {Tran}}, \bibinfo {author} {\bibfnamefont {S.}~\bibnamefont {Lehner}}, \bibinfo {author} {\bibfnamefont {R.}~\bibnamefont {Kueng}},\ and\ \bibinfo {author} {\bibfnamefont {J.}~\bibnamefont {Preskill}},\ }\bibfield  {title} {\bibinfo {title} {Improved machine learning algorithm for predicting ground state properties},\ }\href {http://doi.org/10.1038/s41467-024-45014-7} {\bibfield  {journal} {\bibinfo  {journal} {Nat. Commun.}\ }\textbf {\bibinfo {volume} {15}},\ \bibinfo {pages} {895} (\bibinfo {year} {2024})}\BibitemShut {NoStop}%
\bibitem [{\citenamefont {Cerezo}\ \emph {et~al.}(2021{\natexlab{a}})\citenamefont {Cerezo}, \citenamefont {Arrasmith}, \citenamefont {Babbush}, \citenamefont {Benjamin}, \citenamefont {Endo}, \citenamefont {Fujii}, \citenamefont {McClean}, \citenamefont {Mitarai}, \citenamefont {Yuan}, \citenamefont {Cincio},\ and\ \citenamefont {Coles}}]{cerezo2020variationalreview}%
  \BibitemOpen
  \bibfield  {author} {\bibinfo {author} {\bibfnamefont {M.}~\bibnamefont {Cerezo}}, \bibinfo {author} {\bibfnamefont {A.}~\bibnamefont {Arrasmith}}, \bibinfo {author} {\bibfnamefont {R.}~\bibnamefont {Babbush}}, \bibinfo {author} {\bibfnamefont {S.~C.}\ \bibnamefont {Benjamin}}, \bibinfo {author} {\bibfnamefont {S.}~\bibnamefont {Endo}}, \bibinfo {author} {\bibfnamefont {K.}~\bibnamefont {Fujii}}, \bibinfo {author} {\bibfnamefont {J.~R.}\ \bibnamefont {McClean}}, \bibinfo {author} {\bibfnamefont {K.}~\bibnamefont {Mitarai}}, \bibinfo {author} {\bibfnamefont {X.}~\bibnamefont {Yuan}}, \bibinfo {author} {\bibfnamefont {L.}~\bibnamefont {Cincio}},\ and\ \bibinfo {author} {\bibfnamefont {P.~J.}\ \bibnamefont {Coles}},\ }\bibfield  {title} {\bibinfo {title} {Variational quantum algorithms},\ }\href {https://doi.org/10.1038/s42254-021-00348-9} {\bibfield  {journal} {\bibinfo  {journal} {Nat. Rev. Phys.}\ }\textbf {\bibinfo {volume} {3}},\ \bibinfo {pages} {625–644} (\bibinfo {year}
  {2021}{\natexlab{a}})}\BibitemShut {NoStop}%
\bibitem [{\citenamefont {Valiant}(1984)}]{valiant1984theory}%
  \BibitemOpen
  \bibfield  {author} {\bibinfo {author} {\bibfnamefont {L.~G.}\ \bibnamefont {Valiant}},\ }\bibfield  {title} {\bibinfo {title} {A theory of the learnable},\ }\href {http://doi.org/10.1145/1968.1972} {\bibfield  {journal} {\bibinfo  {journal} {Commun. ACM}\ }\textbf {\bibinfo {volume} {27}},\ \bibinfo {pages} {1134–1142} (\bibinfo {year} {1984})}\BibitemShut {NoStop}%
\bibitem [{\citenamefont {Kearns}\ and\ \citenamefont {Schapire}(1994)}]{kearns1994efficient}%
  \BibitemOpen
  \bibfield  {author} {\bibinfo {author} {\bibfnamefont {M.~J.}\ \bibnamefont {Kearns}}\ and\ \bibinfo {author} {\bibfnamefont {R.~E.}\ \bibnamefont {Schapire}},\ }\bibfield  {title} {\bibinfo {title} {Efficient distribution-free learning of probabilistic concepts},\ }\href {http://doi.org/10.1016/S0022-0000(05)80062-5} {\bibfield  {journal} {\bibinfo  {journal} {J. Comput. Syst. Sci.}\ }\textbf {\bibinfo {volume} {48}},\ \bibinfo {pages} {464–497} (\bibinfo {year} {1994})}\BibitemShut {NoStop}%
\bibitem [{\citenamefont {Aaronson}(2007)}]{aaronson2007learnability}%
  \BibitemOpen
  \bibfield  {author} {\bibinfo {author} {\bibfnamefont {S.}~\bibnamefont {Aaronson}},\ }\bibfield  {title} {\bibinfo {title} {The learnability of quantum states},\ }\href {https://doi.org/10.1098/rspa.2007.0113} {\bibfield  {journal} {\bibinfo  {journal} {Proc. R. Soc. A: Math. Phys. Eng. Sci.}\ }\textbf {\bibinfo {volume} {463}},\ \bibinfo {pages} {3089} (\bibinfo {year} {2007})}\BibitemShut {NoStop}%
\bibitem [{\citenamefont {Rocchetto}(2018)}]{rocchetto2018stabiliser}%
  \BibitemOpen
  \bibfield  {author} {\bibinfo {author} {\bibfnamefont {A.}~\bibnamefont {Rocchetto}},\ }\bibfield  {title} {\bibinfo {title} {Stabiliser states are efficiently {PAC}-learnable},\ }\href {https://doi.org/10.26421/QIC18.7-8-1} {\bibfield  {journal} {\bibinfo  {journal} {Quantum Inf. Comput.}\ }\textbf {\bibinfo {volume} {18}},\ \bibinfo {pages} {541} (\bibinfo {year} {2018})}\BibitemShut {NoStop}%
\bibitem [{\citenamefont {Rocchetto}\ \emph {et~al.}(2019)\citenamefont {Rocchetto}, \citenamefont {Aaronson}, \citenamefont {Severini}, \citenamefont {Carvacho}, \citenamefont {Poderini}, \citenamefont {Agresti}, \citenamefont {Bentivegna},\ and\ \citenamefont {Sciarrino}}]{rocchetto2019experimental}%
  \BibitemOpen
  \bibfield  {author} {\bibinfo {author} {\bibfnamefont {A.}~\bibnamefont {Rocchetto}}, \bibinfo {author} {\bibfnamefont {S.}~\bibnamefont {Aaronson}}, \bibinfo {author} {\bibfnamefont {S.}~\bibnamefont {Severini}}, \bibinfo {author} {\bibfnamefont {G.}~\bibnamefont {Carvacho}}, \bibinfo {author} {\bibfnamefont {D.}~\bibnamefont {Poderini}}, \bibinfo {author} {\bibfnamefont {I.}~\bibnamefont {Agresti}}, \bibinfo {author} {\bibfnamefont {M.}~\bibnamefont {Bentivegna}},\ and\ \bibinfo {author} {\bibfnamefont {F.}~\bibnamefont {Sciarrino}},\ }\bibfield  {title} {\bibinfo {title} {Experimental learning of quantum states},\ }\href {https://doi.org/10.1126/sciadv.aau1946} {\bibfield  {journal} {\bibinfo  {journal} {Sci. Adv.}\ }\textbf {\bibinfo {volume} {5}},\ \bibinfo {pages} {eaau1946} (\bibinfo {year} {2019})}\BibitemShut {NoStop}%
\bibitem [{\citenamefont {Cheng}\ \emph {et~al.}(2016)\citenamefont {Cheng}, \citenamefont {Hsieh},\ and\ \citenamefont {Yeh}}]{cheng2016learnability}%
  \BibitemOpen
  \bibfield  {author} {\bibinfo {author} {\bibfnamefont {H.-C.}\ \bibnamefont {Cheng}}, \bibinfo {author} {\bibfnamefont {M.-H.}\ \bibnamefont {Hsieh}},\ and\ \bibinfo {author} {\bibfnamefont {P.-C.}\ \bibnamefont {Yeh}},\ }\bibfield  {title} {\bibinfo {title} {The learnability of unknown quantum measurements},\ }\href {https://doi.org/10.26421/QIC16.7-8-4} {\bibfield  {journal} {\bibinfo  {journal} {Quantum Inf. Comput.}\ }\textbf {\bibinfo {volume} {16}},\ \bibinfo {pages} {615–656} (\bibinfo {year} {2016})}\BibitemShut {NoStop}%
\bibitem [{\citenamefont {Caro}\ and\ \citenamefont {Datta}(2020)}]{caro2020pseudo}%
  \BibitemOpen
  \bibfield  {author} {\bibinfo {author} {\bibfnamefont {M.~C.}\ \bibnamefont {Caro}}\ and\ \bibinfo {author} {\bibfnamefont {I.}~\bibnamefont {Datta}},\ }\bibfield  {title} {\bibinfo {title} {Pseudo-dimension of quantum circuits},\ }\href {https://doi.org/10.1007/s42484-020-00027-5} {\bibfield  {journal} {\bibinfo  {journal} {Quantum Mach. Intell.}\ }\textbf {\bibinfo {volume} {2}},\ \bibinfo {pages} {14} (\bibinfo {year} {2020})}\BibitemShut {NoStop}%
\bibitem [{\citenamefont {Goel}\ and\ \citenamefont {Klivans}(2019)}]{goel2019learning}%
  \BibitemOpen
  \bibfield  {author} {\bibinfo {author} {\bibfnamefont {S.}~\bibnamefont {Goel}}\ and\ \bibinfo {author} {\bibfnamefont {A.~R.}\ \bibnamefont {Klivans}},\ }\bibfield  {title} {\bibinfo {title} {Learning neural networks with two nonlinear layers in polynomial time},\ }in\ \href {https://proceedings.mlr.press/v99/goel19b.html} {\emph {\bibinfo {booktitle} {Proceedings of the Thirty-Second Conference on Learning Theory}}},\ \bibinfo {series} {Proceedings of Machine Learning Research}, Vol.~\bibinfo {volume} {99},\ \bibinfo {editor} {edited by\ \bibinfo {editor} {\bibfnamefont {A.}~\bibnamefont {Beygelzimer}}\ and\ \bibinfo {editor} {\bibfnamefont {D.}~\bibnamefont {Hsu}}}\ (\bibinfo  {publisher} {PMLR},\ \bibinfo {year} {2019})\ pp.\ \bibinfo {pages} {1470--1499}\BibitemShut {NoStop}%
\bibitem [{\citenamefont {Recio-Armengol}\ \emph {et~al.}(2024)\citenamefont {Recio-Armengol}, \citenamefont {Eisert},\ and\ \citenamefont {Meyer}}]{recioarmengol2024singleshot}%
  \BibitemOpen
  \bibfield  {author} {\bibinfo {author} {\bibfnamefont {E.}~\bibnamefont {Recio-Armengol}}, \bibinfo {author} {\bibfnamefont {J.}~\bibnamefont {Eisert}},\ and\ \bibinfo {author} {\bibfnamefont {J.~J.}\ \bibnamefont {Meyer}},\ }\href@noop {} {\bibinfo {title} {Single-shot quantum machine learning}} (\bibinfo {year} {2024}),\ \Eprint {https://arxiv.org/abs/2406.13812} {arXiv:2406.13812 [quant-ph]} \BibitemShut {NoStop}%
\bibitem [{\citenamefont {Neal}\ \emph {et~al.}(2019)\citenamefont {Neal}, \citenamefont {Mittal}, \citenamefont {Baratin}, \citenamefont {Tantia}, \citenamefont {Scicluna}, \citenamefont {Lacoste-Julien},\ and\ \citenamefont {Mitliagkas}}]{neal2019modern}%
  \BibitemOpen
  \bibfield  {author} {\bibinfo {author} {\bibfnamefont {B.}~\bibnamefont {Neal}}, \bibinfo {author} {\bibfnamefont {S.}~\bibnamefont {Mittal}}, \bibinfo {author} {\bibfnamefont {A.}~\bibnamefont {Baratin}}, \bibinfo {author} {\bibfnamefont {V.}~\bibnamefont {Tantia}}, \bibinfo {author} {\bibfnamefont {M.}~\bibnamefont {Scicluna}}, \bibinfo {author} {\bibfnamefont {S.}~\bibnamefont {Lacoste-Julien}},\ and\ \bibinfo {author} {\bibfnamefont {I.}~\bibnamefont {Mitliagkas}},\ }\href@noop {} {\bibinfo {title} {A modern take on the bias-variance tradeoff in neural networks}} (\bibinfo {year} {2019}),\ \Eprint {https://arxiv.org/abs/1810.08591} {arXiv:1810.08591 [cs.LG]} \BibitemShut {NoStop}%
\bibitem [{\citenamefont {Yang}\ \emph {et~al.}(2020)\citenamefont {Yang}, \citenamefont {Yu}, \citenamefont {You}, \citenamefont {Steinhardt},\ and\ \citenamefont {Ma}}]{yang2020rethinking}%
  \BibitemOpen
  \bibfield  {author} {\bibinfo {author} {\bibfnamefont {Z.}~\bibnamefont {Yang}}, \bibinfo {author} {\bibfnamefont {Y.}~\bibnamefont {Yu}}, \bibinfo {author} {\bibfnamefont {C.}~\bibnamefont {You}}, \bibinfo {author} {\bibfnamefont {J.}~\bibnamefont {Steinhardt}},\ and\ \bibinfo {author} {\bibfnamefont {Y.}~\bibnamefont {Ma}},\ }\bibfield  {title} {\bibinfo {title} {Rethinking bias-variance trade-off for generalization of neural networks},\ }in\ \href {https://proceedings.mlr.press/v119/yang20j.html} {\emph {\bibinfo {booktitle} {Proceedings of the 37th International Conference on Machine Learning}}},\ \bibinfo {series} {Proceedings of Machine Learning Research}, Vol.\ \bibinfo {volume} {119},\ \bibinfo {editor} {edited by\ \bibinfo {editor} {\bibfnamefont {H.~D.}\ \bibnamefont {III}}\ and\ \bibinfo {editor} {\bibfnamefont {A.}~\bibnamefont {Singh}}}\ (\bibinfo  {publisher} {PMLR},\ \bibinfo {year} {2020})\ pp.\ \bibinfo {pages} {10767--10777}\BibitemShut {NoStop}%
\bibitem [{\citenamefont {Landman}\ \emph {et~al.}(2022)\citenamefont {Landman}, \citenamefont {Thabet}, \citenamefont {Dalyac}, \citenamefont {Mhiri},\ and\ \citenamefont {Kashefi}}]{landman2022classically}%
  \BibitemOpen
  \bibfield  {author} {\bibinfo {author} {\bibfnamefont {J.}~\bibnamefont {Landman}}, \bibinfo {author} {\bibfnamefont {S.}~\bibnamefont {Thabet}}, \bibinfo {author} {\bibfnamefont {C.}~\bibnamefont {Dalyac}}, \bibinfo {author} {\bibfnamefont {H.}~\bibnamefont {Mhiri}},\ and\ \bibinfo {author} {\bibfnamefont {E.}~\bibnamefont {Kashefi}},\ }\href@noop {} {\bibinfo {title} {Classically approximating variational quantum machine learning with random {Fourier} features}} (\bibinfo {year} {2022}),\ \Eprint {https://arxiv.org/abs/2210.13200} {arXiv:2210.13200 [quant-ph]} \BibitemShut {NoStop}%
\bibitem [{\citenamefont {Sweke}\ \emph {et~al.}(2023)\citenamefont {Sweke}, \citenamefont {Recio}, \citenamefont {Jerbi}, \citenamefont {Gil-Fuster}, \citenamefont {Fuller}, \citenamefont {Eisert},\ and\ \citenamefont {Meyer}}]{sweke2023potential}%
  \BibitemOpen
  \bibfield  {author} {\bibinfo {author} {\bibfnamefont {R.}~\bibnamefont {Sweke}}, \bibinfo {author} {\bibfnamefont {E.}~\bibnamefont {Recio}}, \bibinfo {author} {\bibfnamefont {S.}~\bibnamefont {Jerbi}}, \bibinfo {author} {\bibfnamefont {E.}~\bibnamefont {Gil-Fuster}}, \bibinfo {author} {\bibfnamefont {B.}~\bibnamefont {Fuller}}, \bibinfo {author} {\bibfnamefont {J.}~\bibnamefont {Eisert}},\ and\ \bibinfo {author} {\bibfnamefont {J.~J.}\ \bibnamefont {Meyer}},\ }\href@noop {} {\bibinfo {title} {Potential and limitations of random {Fourier} features for dequantizing quantum machine learning}} (\bibinfo {year} {2023}),\ \Eprint {https://arxiv.org/abs/2309.11647} {arXiv:2309.11647 [quant-ph]} \BibitemShut {NoStop}%
\bibitem [{\citenamefont {Fontana}\ \emph {et~al.}(2023)\citenamefont {Fontana}, \citenamefont {Rudolph}, \citenamefont {Duncan}, \citenamefont {Rungger},\ and\ \citenamefont {C{\^\i}rstoiu}}]{fontana2023classical}%
  \BibitemOpen
  \bibfield  {author} {\bibinfo {author} {\bibfnamefont {E.}~\bibnamefont {Fontana}}, \bibinfo {author} {\bibfnamefont {M.~S.}\ \bibnamefont {Rudolph}}, \bibinfo {author} {\bibfnamefont {R.}~\bibnamefont {Duncan}}, \bibinfo {author} {\bibfnamefont {I.}~\bibnamefont {Rungger}},\ and\ \bibinfo {author} {\bibfnamefont {C.}~\bibnamefont {C{\^\i}rstoiu}},\ }\href@noop {} {\bibinfo {title} {Classical simulations of noisy variational quantum circuits}} (\bibinfo {year} {2023}),\ \Eprint {https://arxiv.org/abs/2306.05400} {arXiv:2306.05400 [quant-ph]} \BibitemShut {NoStop}%
\bibitem [{\citenamefont {Nemkov}\ \emph {et~al.}(2023)\citenamefont {Nemkov}, \citenamefont {Kiktenko},\ and\ \citenamefont {Fedorov}}]{nemkov2023fourier}%
  \BibitemOpen
  \bibfield  {author} {\bibinfo {author} {\bibfnamefont {N.~A.}\ \bibnamefont {Nemkov}}, \bibinfo {author} {\bibfnamefont {E.~O.}\ \bibnamefont {Kiktenko}},\ and\ \bibinfo {author} {\bibfnamefont {A.~K.}\ \bibnamefont {Fedorov}},\ }\bibfield  {title} {\bibinfo {title} {{Fourier} expansion in variational quantum algorithms},\ }\href {http://doi.org/10.1103/PhysRevA.108.032406} {\bibfield  {journal} {\bibinfo  {journal} {Phys. Rev. A}\ }\textbf {\bibinfo {volume} {108}},\ \bibinfo {pages} {032406} (\bibinfo {year} {2023})}\BibitemShut {NoStop}%
\bibitem [{\citenamefont {Gil~Vidal}\ and\ \citenamefont {Theis}(2020)}]{gilvidal2020input}%
  \BibitemOpen
  \bibfield  {author} {\bibinfo {author} {\bibfnamefont {F.~J.}\ \bibnamefont {Gil~Vidal}}\ and\ \bibinfo {author} {\bibfnamefont {D.~O.}\ \bibnamefont {Theis}},\ }\bibfield  {title} {\bibinfo {title} {Input redundancy for parameterized quantum circuits},\ }\href {https://doi.org/10.3389/fphy.2020.00297} {\bibfield  {journal} {\bibinfo  {journal} {Front. Phys.}\ }\textbf {\bibinfo {volume} {8}},\ \bibinfo {pages} {297} (\bibinfo {year} {2020})}\BibitemShut {NoStop}%
\bibitem [{\citenamefont {Schuld}\ \emph {et~al.}(2021)\citenamefont {Schuld}, \citenamefont {Sweke},\ and\ \citenamefont {Meyer}}]{schuld2021effect}%
  \BibitemOpen
  \bibfield  {author} {\bibinfo {author} {\bibfnamefont {M.}~\bibnamefont {Schuld}}, \bibinfo {author} {\bibfnamefont {R.}~\bibnamefont {Sweke}},\ and\ \bibinfo {author} {\bibfnamefont {J.~J.}\ \bibnamefont {Meyer}},\ }\bibfield  {title} {\bibinfo {title} {Effect of data encoding on the expressive power of variational quantum-machine-learning models},\ }\href {https://doi.org/10.1103/PhysRevA.103.032430} {\bibfield  {journal} {\bibinfo  {journal} {Phys. Rev. A}\ }\textbf {\bibinfo {volume} {103}},\ \bibinfo {pages} {032430} (\bibinfo {year} {2021})}\BibitemShut {NoStop}%
\bibitem [{\citenamefont {Linke}\ \emph {et~al.}(2017)\citenamefont {Linke}, \citenamefont {Maslov}, \citenamefont {Roetteler}, \citenamefont {Debnath}, \citenamefont {Figgatt}, \citenamefont {Landsman}, \citenamefont {Wright},\ and\ \citenamefont {Monroe}}]{linke2017experimental}%
  \BibitemOpen
  \bibfield  {author} {\bibinfo {author} {\bibfnamefont {N.~M.}\ \bibnamefont {Linke}}, \bibinfo {author} {\bibfnamefont {D.}~\bibnamefont {Maslov}}, \bibinfo {author} {\bibfnamefont {M.}~\bibnamefont {Roetteler}}, \bibinfo {author} {\bibfnamefont {S.}~\bibnamefont {Debnath}}, \bibinfo {author} {\bibfnamefont {C.}~\bibnamefont {Figgatt}}, \bibinfo {author} {\bibfnamefont {K.~A.}\ \bibnamefont {Landsman}}, \bibinfo {author} {\bibfnamefont {K.}~\bibnamefont {Wright}},\ and\ \bibinfo {author} {\bibfnamefont {C.}~\bibnamefont {Monroe}},\ }\bibfield  {title} {\bibinfo {title} {Experimental comparison of two quantum computing architectures},\ }\href {https://doi.org/10.1073/pnas.1618020114} {\bibfield  {journal} {\bibinfo  {journal} {Proc. Natl. Acad. Sci.}\ }\textbf {\bibinfo {volume} {114}},\ \bibinfo {pages} {3305} (\bibinfo {year} {2017})}\BibitemShut {NoStop}%
\bibitem [{\citenamefont {Rahimi}\ and\ \citenamefont {Recht}(2007)}]{rahimi2007random}%
  \BibitemOpen
  \bibfield  {author} {\bibinfo {author} {\bibfnamefont {A.}~\bibnamefont {Rahimi}}\ and\ \bibinfo {author} {\bibfnamefont {B.}~\bibnamefont {Recht}},\ }\bibfield  {title} {\bibinfo {title} {Random features for large-scale kernel machines},\ }in\ \href {https://proceedings.neurips.cc/paper_files/paper/2007/hash/013a006f03dbc5392effeb8f18fda755-Abstract.html} {\emph {\bibinfo {booktitle} {Advances in Neural Information Processing Systems}}},\ Vol.~\bibinfo {volume} {20},\ \bibinfo {editor} {edited by\ \bibinfo {editor} {\bibfnamefont {J.}~\bibnamefont {Platt}}, \bibinfo {editor} {\bibfnamefont {D.}~\bibnamefont {Koller}}, \bibinfo {editor} {\bibfnamefont {Y.}~\bibnamefont {Singer}},\ and\ \bibinfo {editor} {\bibfnamefont {S.}~\bibnamefont {Roweis}}}\ (\bibinfo  {publisher} {Curran Associates, Inc.},\ \bibinfo {year} {2007})\BibitemShut {NoStop}%
\bibitem [{\citenamefont {Rahimi}\ and\ \citenamefont {Recht}(2008{\natexlab{a}})}]{rahimi2008weighted}%
  \BibitemOpen
  \bibfield  {author} {\bibinfo {author} {\bibfnamefont {A.}~\bibnamefont {Rahimi}}\ and\ \bibinfo {author} {\bibfnamefont {B.}~\bibnamefont {Recht}},\ }\bibfield  {title} {\bibinfo {title} {Weighted sums of random kitchen sinks: Replacing minimization with randomization in learning},\ }in\ \href {https://proceedings.neurips.cc/paper_files/paper/2008/hash/0efe32849d230d7f53049ddc4a4b0c60-Abstract.html} {\emph {\bibinfo {booktitle} {Advances in Neural Information Processing Systems}}},\ Vol.~\bibinfo {volume} {21},\ \bibinfo {editor} {edited by\ \bibinfo {editor} {\bibfnamefont {D.}~\bibnamefont {Koller}}, \bibinfo {editor} {\bibfnamefont {D.}~\bibnamefont {Schuurmans}}, \bibinfo {editor} {\bibfnamefont {Y.}~\bibnamefont {Bengio}},\ and\ \bibinfo {editor} {\bibfnamefont {L.}~\bibnamefont {Bottou}}}\ (\bibinfo  {publisher} {Curran Associates, Inc.},\ \bibinfo {year} {2008})\BibitemShut {NoStop}%
\bibitem [{\citenamefont {Rahimi}\ and\ \citenamefont {Recht}(2008{\natexlab{b}})}]{rahimi2008uniform}%
  \BibitemOpen
  \bibfield  {author} {\bibinfo {author} {\bibfnamefont {A.}~\bibnamefont {Rahimi}}\ and\ \bibinfo {author} {\bibfnamefont {B.}~\bibnamefont {Recht}},\ }\bibfield  {title} {\bibinfo {title} {Uniform approximation of functions with random bases},\ }in\ \href {https://doi.org/10.1109/ALLERTON.2008.4797607} {\emph {\bibinfo {booktitle} {2008 46th Annual Allerton Conference on Communication, Control, and Computing}}}\ (\bibinfo {year} {2008})\ pp.\ \bibinfo {pages} {555--561}\BibitemShut {NoStop}%
\bibitem [{\citenamefont {Boyd}\ and\ \citenamefont {Vandenberghe}(2004)}]{boyd2004interior}%
  \BibitemOpen
  \bibfield  {author} {\bibinfo {author} {\bibfnamefont {S.}~\bibnamefont {Boyd}}\ and\ \bibinfo {author} {\bibfnamefont {L.}~\bibnamefont {Vandenberghe}},\ }\bibinfo {title} {Interior-point methods},\ in\ \href {https://doi.org/10.1017/CBO9780511804441.012} {\emph {\bibinfo {booktitle} {Convex Optimization}}}\ (\bibinfo  {publisher} {Cambridge University Press},\ \bibinfo {year} {2004})\ p.\ \bibinfo {pages} {561–630}\BibitemShut {NoStop}%
\bibitem [{\citenamefont {P{\'{e}}rez-Salinas}\ \emph {et~al.}(2020)\citenamefont {P{\'{e}}rez-Salinas}, \citenamefont {Cervera-Lierta}, \citenamefont {Gil-Fuster},\ and\ \citenamefont {Latorre}}]{perezsalinas2020data}%
  \BibitemOpen
  \bibfield  {author} {\bibinfo {author} {\bibfnamefont {A.}~\bibnamefont {P{\'{e}}rez-Salinas}}, \bibinfo {author} {\bibfnamefont {A.}~\bibnamefont {Cervera-Lierta}}, \bibinfo {author} {\bibfnamefont {E.}~\bibnamefont {Gil-Fuster}},\ and\ \bibinfo {author} {\bibfnamefont {J.~I.}\ \bibnamefont {Latorre}},\ }\bibfield  {title} {\bibinfo {title} {Data re-uploading for a universal quantum classifier},\ }\href {https://doi.org/10.22331/q-2020-02-06-226} {\bibfield  {journal} {\bibinfo  {journal} {Quantum}\ }\textbf {\bibinfo {volume} {4}},\ \bibinfo {pages} {226} (\bibinfo {year} {2020})}\BibitemShut {NoStop}%
\bibitem [{\citenamefont {Cerezo}\ \emph {et~al.}(2023)\citenamefont {Cerezo}, \citenamefont {Larocca}, \citenamefont {Garc{\'\i}a-Mart{\'\i}n}, \citenamefont {Diaz}, \citenamefont {Braccia}, \citenamefont {Fontana}, \citenamefont {Rudolph}, \citenamefont {Bermejo}, \citenamefont {Ijaz}, \citenamefont {Thanasilp}, \citenamefont {Anschuetz},\ and\ \citenamefont {Holmes}}]{cerezo2023does}%
  \BibitemOpen
  \bibfield  {author} {\bibinfo {author} {\bibfnamefont {M.}~\bibnamefont {Cerezo}}, \bibinfo {author} {\bibfnamefont {M.}~\bibnamefont {Larocca}}, \bibinfo {author} {\bibfnamefont {D.}~\bibnamefont {Garc{\'\i}a-Mart{\'\i}n}}, \bibinfo {author} {\bibfnamefont {N.}~\bibnamefont {Diaz}}, \bibinfo {author} {\bibfnamefont {P.}~\bibnamefont {Braccia}}, \bibinfo {author} {\bibfnamefont {E.}~\bibnamefont {Fontana}}, \bibinfo {author} {\bibfnamefont {M.~S.}\ \bibnamefont {Rudolph}}, \bibinfo {author} {\bibfnamefont {P.}~\bibnamefont {Bermejo}}, \bibinfo {author} {\bibfnamefont {A.}~\bibnamefont {Ijaz}}, \bibinfo {author} {\bibfnamefont {S.}~\bibnamefont {Thanasilp}}, \bibinfo {author} {\bibfnamefont {E.~R.}\ \bibnamefont {Anschuetz}},\ and\ \bibinfo {author} {\bibfnamefont {Z.}~\bibnamefont {Holmes}},\ }\href@noop {} {\bibinfo {title} {Does provable absence of barren plateaus imply classical simulability? {Or}, why we need to rethink variational quantum computing}} (\bibinfo {year} {2023}),\ \Eprint
  {https://arxiv.org/abs/2312.09121} {arXiv:2312.09121 [quant-ph]} \BibitemShut {NoStop}%
\bibitem [{\citenamefont {Hinsche}\ \emph {et~al.}(2023)\citenamefont {Hinsche}, \citenamefont {Ioannou}, \citenamefont {Nietner}, \citenamefont {Haferkamp}, \citenamefont {Quek}, \citenamefont {Hangleiter}, \citenamefont {Seifert}, \citenamefont {Eisert},\ and\ \citenamefont {Sweke}}]{hinsche2023one}%
  \BibitemOpen
  \bibfield  {author} {\bibinfo {author} {\bibfnamefont {M.}~\bibnamefont {Hinsche}}, \bibinfo {author} {\bibfnamefont {M.}~\bibnamefont {Ioannou}}, \bibinfo {author} {\bibfnamefont {A.}~\bibnamefont {Nietner}}, \bibinfo {author} {\bibfnamefont {J.}~\bibnamefont {Haferkamp}}, \bibinfo {author} {\bibfnamefont {Y.}~\bibnamefont {Quek}}, \bibinfo {author} {\bibfnamefont {D.}~\bibnamefont {Hangleiter}}, \bibinfo {author} {\bibfnamefont {J.-P.}\ \bibnamefont {Seifert}}, \bibinfo {author} {\bibfnamefont {J.}~\bibnamefont {Eisert}},\ and\ \bibinfo {author} {\bibfnamefont {R.}~\bibnamefont {Sweke}},\ }\bibfield  {title} {\bibinfo {title} {One {$T$} gate makes distribution learning hard},\ }\href {https://doi.org/10.1103/PhysRevLett.130.240602} {\bibfield  {journal} {\bibinfo  {journal} {Phys. Rev. Lett.}\ }\textbf {\bibinfo {volume} {130}},\ \bibinfo {pages} {240602} (\bibinfo {year} {2023})}\BibitemShut {NoStop}%
\bibitem [{\citenamefont {Cerezo}\ \emph {et~al.}(2021{\natexlab{b}})\citenamefont {Cerezo}, \citenamefont {Sone}, \citenamefont {Volkoff}, \citenamefont {Cincio},\ and\ \citenamefont {Coles}}]{cerezo2021cost}%
  \BibitemOpen
  \bibfield  {author} {\bibinfo {author} {\bibfnamefont {M.}~\bibnamefont {Cerezo}}, \bibinfo {author} {\bibfnamefont {A.}~\bibnamefont {Sone}}, \bibinfo {author} {\bibfnamefont {T.}~\bibnamefont {Volkoff}}, \bibinfo {author} {\bibfnamefont {L.}~\bibnamefont {Cincio}},\ and\ \bibinfo {author} {\bibfnamefont {P.~J.}\ \bibnamefont {Coles}},\ }\bibfield  {title} {\bibinfo {title} {Cost function dependent barren plateaus in shallow parametrized quantum circuits},\ }\href {https://doi.org/10.1038/s41467-021-21728-w} {\bibfield  {journal} {\bibinfo  {journal} {Nat. Commun.}\ }\textbf {\bibinfo {volume} {12}} (\bibinfo {year} {2021}{\natexlab{b}})}\BibitemShut {NoStop}%
\bibitem [{\citenamefont {Huang}\ \emph {et~al.}(2020{\natexlab{b}})\citenamefont {Huang}, \citenamefont {Kueng},\ and\ \citenamefont {Preskill}}]{huang2020predicting}%
  \BibitemOpen
  \bibfield  {author} {\bibinfo {author} {\bibfnamefont {H.-Y.}\ \bibnamefont {Huang}}, \bibinfo {author} {\bibfnamefont {R.}~\bibnamefont {Kueng}},\ and\ \bibinfo {author} {\bibfnamefont {J.}~\bibnamefont {Preskill}},\ }\bibfield  {title} {\bibinfo {title} {Predicting many properties of a quantum system from very few measurements},\ }\href {http://doi.org/10.1038/s41567-020-0932-7} {\bibfield  {journal} {\bibinfo  {journal} {Nat. Phys.}\ }\textbf {\bibinfo {volume} {16}},\ \bibinfo {pages} {1050} (\bibinfo {year} {2020}{\natexlab{b}})}\BibitemShut {NoStop}%
\bibitem [{\citenamefont {Guo}\ \emph {et~al.}(2023)\citenamefont {Guo}, \citenamefont {Pan},\ and\ \citenamefont {Rebentrost}}]{guo2023estimating}%
  \BibitemOpen
  \bibfield  {author} {\bibinfo {author} {\bibfnamefont {N.}~\bibnamefont {Guo}}, \bibinfo {author} {\bibfnamefont {F.}~\bibnamefont {Pan}},\ and\ \bibinfo {author} {\bibfnamefont {P.}~\bibnamefont {Rebentrost}},\ }\href@noop {} {\bibinfo {title} {Estimating properties of a quantum state by importance-sampled operator shadows}} (\bibinfo {year} {2023}),\ \Eprint {https://arxiv.org/abs/2305.09374} {arXiv:2305.09374 [quant-ph]} \BibitemShut {NoStop}%
\bibitem [{\citenamefont {Kohler}\ and\ \citenamefont {Lucchi}(2017)}]{kohler2017sub}%
  \BibitemOpen
  \bibfield  {author} {\bibinfo {author} {\bibfnamefont {J.~M.}\ \bibnamefont {Kohler}}\ and\ \bibinfo {author} {\bibfnamefont {A.}~\bibnamefont {Lucchi}},\ }\bibfield  {title} {\bibinfo {title} {Sub-sampled cubic regularization for non-convex optimization},\ }in\ \href {https://proceedings.mlr.press/v70/kohler17a.html} {\emph {\bibinfo {booktitle} {Proceedings of the 34th International Conference on Machine Learning}}},\ \bibinfo {series} {Proceedings of Machine Learning Research}, Vol.~\bibinfo {volume} {70},\ \bibinfo {editor} {edited by\ \bibinfo {editor} {\bibfnamefont {D.}~\bibnamefont {Precup}}\ and\ \bibinfo {editor} {\bibfnamefont {Y.~W.}\ \bibnamefont {Teh}}}\ (\bibinfo  {publisher} {PMLR},\ \bibinfo {year} {2017})\ pp.\ \bibinfo {pages} {1895--1904}\BibitemShut {NoStop}%
\bibitem [{\citenamefont {Bartlett}\ and\ \citenamefont {Mendelson}(2002)}]{bartlett2002rademacher}%
  \BibitemOpen
  \bibfield  {author} {\bibinfo {author} {\bibfnamefont {P.~L.}\ \bibnamefont {Bartlett}}\ and\ \bibinfo {author} {\bibfnamefont {S.}~\bibnamefont {Mendelson}},\ }\bibfield  {title} {\bibinfo {title} {Rademacher and {Gaussian} complexities: risk bounds and structural results},\ }\href {https://www.jmlr.org/papers/v3/bartlett02a.html} {\bibfield  {journal} {\bibinfo  {journal} {J. Mach. Learn. Res.}\ }\textbf {\bibinfo {volume} {3}},\ \bibinfo {pages} {463–482} (\bibinfo {year} {2002})}\BibitemShut {NoStop}%
\bibitem [{\citenamefont {Kakade}\ \emph {et~al.}(2008)\citenamefont {Kakade}, \citenamefont {Sridharan},\ and\ \citenamefont {Tewari}}]{kakade2008complexity}%
  \BibitemOpen
  \bibfield  {author} {\bibinfo {author} {\bibfnamefont {S.~M.}\ \bibnamefont {Kakade}}, \bibinfo {author} {\bibfnamefont {K.}~\bibnamefont {Sridharan}},\ and\ \bibinfo {author} {\bibfnamefont {A.}~\bibnamefont {Tewari}},\ }\bibfield  {title} {\bibinfo {title} {On the complexity of linear prediction: Risk bounds, margin bounds, and regularization},\ }in\ \href {https://proceedings.neurips.cc/paper_files/paper/2008/hash/5b69b9cb83065d403869739ae7f0995e-Abstract.html} {\emph {\bibinfo {booktitle} {Advances in Neural Information Processing Systems}}},\ Vol.~\bibinfo {volume} {21},\ \bibinfo {editor} {edited by\ \bibinfo {editor} {\bibfnamefont {D.}~\bibnamefont {Koller}}, \bibinfo {editor} {\bibfnamefont {D.}~\bibnamefont {Schuurmans}}, \bibinfo {editor} {\bibfnamefont {Y.}~\bibnamefont {Bengio}},\ and\ \bibinfo {editor} {\bibfnamefont {L.}~\bibnamefont {Bottou}}}\ (\bibinfo  {publisher} {Curran Associates, Inc.},\ \bibinfo {year} {2008})\BibitemShut {NoStop}%
\bibitem [{\citenamefont {Ledoux}\ and\ \citenamefont {Talagrand}(1991)}]{ledoux1991probability}%
  \BibitemOpen
  \bibfield  {author} {\bibinfo {author} {\bibfnamefont {M.}~\bibnamefont {Ledoux}}\ and\ \bibinfo {author} {\bibfnamefont {M.}~\bibnamefont {Talagrand}},\ }\href {http://doi.org/10.1007/978-3-642-20212-4} {\emph {\bibinfo {title} {Probability in Banach Spaces}}}\ (\bibinfo  {publisher} {Springer Berlin Heidelberg},\ \bibinfo {year} {1991})\BibitemShut {NoStop}%
\bibitem [{\citenamefont {Mohri}\ \emph {et~al.}(2018)\citenamefont {Mohri}, \citenamefont {Rostamizadeh},\ and\ \citenamefont {Talwalkar}}]{mohri2018foundations}%
  \BibitemOpen
  \bibfield  {author} {\bibinfo {author} {\bibfnamefont {M.}~\bibnamefont {Mohri}}, \bibinfo {author} {\bibfnamefont {A.}~\bibnamefont {Rostamizadeh}},\ and\ \bibinfo {author} {\bibfnamefont {A.}~\bibnamefont {Talwalkar}},\ }\href {https://mitpress.mit.edu/9780262039406/foundations-of-machine-learning} {\emph {\bibinfo {title} {Foundations of Machine Learning}}},\ \bibinfo {edition} {2nd}\ ed.\ (\bibinfo  {publisher} {The MIT Press},\ \bibinfo {year} {2018})\BibitemShut {NoStop}%
\bibitem [{\citenamefont {Childs}\ and\ \citenamefont {Wiebe}(2012)}]{childs2012hamiltonian}%
  \BibitemOpen
  \bibfield  {author} {\bibinfo {author} {\bibfnamefont {A.~M.}\ \bibnamefont {Childs}}\ and\ \bibinfo {author} {\bibfnamefont {N.}~\bibnamefont {Wiebe}},\ }\bibfield  {title} {\bibinfo {title} {Hamiltonian simulation using linear combinations of unitary operations},\ }\href {http://doi.org/10.26421/QIC12.11-12} {\bibfield  {journal} {\bibinfo  {journal} {Quantum Inf. Comput.}\ }\textbf {\bibinfo {volume} {12}} (\bibinfo {year} {2012})}\BibitemShut {NoStop}%
\bibitem [{\citenamefont {Gil-Fuster}\ \emph {et~al.}(2024{\natexlab{a}})\citenamefont {Gil-Fuster}, \citenamefont {Eisert},\ and\ \citenamefont {Dunjko}}]{GilFuster2024expressivity}%
  \BibitemOpen
  \bibfield  {author} {\bibinfo {author} {\bibfnamefont {E.}~\bibnamefont {Gil-Fuster}}, \bibinfo {author} {\bibfnamefont {J.}~\bibnamefont {Eisert}},\ and\ \bibinfo {author} {\bibfnamefont {V.}~\bibnamefont {Dunjko}},\ }\bibfield  {title} {\bibinfo {title} {On the expressivity of embedding quantum kernels},\ }\href {http://doi.org/10.1088/2632-2153/ad2f51} {\bibfield  {journal} {\bibinfo  {journal} {Mach. Learn. Sci. Technol.}\ }\textbf {\bibinfo {volume} {5}},\ \bibinfo {pages} {025003} (\bibinfo {year} {2024}{\natexlab{a}})}\BibitemShut {NoStop}%
\bibitem [{\citenamefont {Gil-Fuster}\ \emph {et~al.}(2024{\natexlab{b}})\citenamefont {Gil-Fuster}, \citenamefont {Eisert},\ and\ \citenamefont {Bravo-Prieto}}]{gilfuster2024understanding}%
  \BibitemOpen
  \bibfield  {author} {\bibinfo {author} {\bibfnamefont {E.}~\bibnamefont {Gil-Fuster}}, \bibinfo {author} {\bibfnamefont {J.}~\bibnamefont {Eisert}},\ and\ \bibinfo {author} {\bibfnamefont {C.}~\bibnamefont {Bravo-Prieto}},\ }\bibfield  {title} {\bibinfo {title} {Understanding quantum machine learning also requires rethinking generalization},\ }\href {https://doi.org/10.1038/s41467-024-45882-z} {\bibfield  {journal} {\bibinfo  {journal} {Nat. Commun.}\ }\textbf {\bibinfo {volume} {15}} (\bibinfo {year} {2024}{\natexlab{b}})}\BibitemShut {NoStop}%
\bibitem [{\citenamefont {Caro}\ \emph {et~al.}(2021)\citenamefont {Caro}, \citenamefont {Gil-Fuster}, \citenamefont {Meyer}, \citenamefont {Eisert},\ and\ \citenamefont {Sweke}}]{caro2021encoding}%
  \BibitemOpen
  \bibfield  {author} {\bibinfo {author} {\bibfnamefont {M.~C.}\ \bibnamefont {Caro}}, \bibinfo {author} {\bibfnamefont {E.}~\bibnamefont {Gil-Fuster}}, \bibinfo {author} {\bibfnamefont {J.~J.}\ \bibnamefont {Meyer}}, \bibinfo {author} {\bibfnamefont {J.}~\bibnamefont {Eisert}},\ and\ \bibinfo {author} {\bibfnamefont {R.}~\bibnamefont {Sweke}},\ }\bibfield  {title} {\bibinfo {title} {Encoding-dependent generalization bounds for parametrized quantum circuits},\ }\href {https://doi.org/10.22331/q-2021-11-17-582} {\bibfield  {journal} {\bibinfo  {journal} {Quantum}\ }\textbf {\bibinfo {volume} {5}},\ \bibinfo {pages} {582} (\bibinfo {year} {2021})}\BibitemShut {NoStop}%
\bibitem [{\citenamefont {Jerbi}\ \emph {et~al.}(2023)\citenamefont {Jerbi}, \citenamefont {Fiderer}, \citenamefont {Poulsen~Nautrup}, \citenamefont {K\"{u}bler}, \citenamefont {Briegel},\ and\ \citenamefont {Dunjko}}]{jerbi2023quantum}%
  \BibitemOpen
  \bibfield  {author} {\bibinfo {author} {\bibfnamefont {S.}~\bibnamefont {Jerbi}}, \bibinfo {author} {\bibfnamefont {L.~J.}\ \bibnamefont {Fiderer}}, \bibinfo {author} {\bibfnamefont {H.}~\bibnamefont {Poulsen~Nautrup}}, \bibinfo {author} {\bibfnamefont {J.~M.}\ \bibnamefont {K\"{u}bler}}, \bibinfo {author} {\bibfnamefont {H.~J.}\ \bibnamefont {Briegel}},\ and\ \bibinfo {author} {\bibfnamefont {V.}~\bibnamefont {Dunjko}},\ }\bibfield  {title} {\bibinfo {title} {Quantum machine learning beyond kernel methods},\ }\href {http://doi.org/10.1038/s41467-023-36159-y} {\bibfield  {journal} {\bibinfo  {journal} {Nat. Commun.}\ }\textbf {\bibinfo {volume} {14}},\ \bibinfo {pages} {517} (\bibinfo {year} {2023})}\BibitemShut {NoStop}%
\end{thebibliography}%

\onecolumngrid

\appendix
\counterwithin{theorem}{section}
\counterwithin{proposition}{section}
\counterwithin{lemma}{section}

\newpage

\section{Sampling and estimation methods on PQMs}

\label{Sec:Estimation-methods}

\subsection{Direct sampling-based estimation}
The most straightforward method is to generate estimations by directly conducting measurements on the target observable. Plugging in the eigendecomposition of the observable $O$, i.e., $O = \sum_{k=1}^K \lambda^{(k)} \ket{\lambda^{(k)}} \bra{\lambda^{(k)}}$ in $f(\vec{x}) = \tr(\rho(\vec{x})O)$ yields
\begin{align}
    f(\vec{x}) = \sum_{k=1}^K \lambda^{(k)} \bra{\lambda^{(k)}}\rho(\vec{x})\ket{\lambda^{(k)}},
\end{align}
where $0 \le \bra{\lambda^{(k)}}\rho(\vec{x})\ket{\lambda^{(k)}} \le 1$ $\forall k$, $\lambda^{(k)} \in \mathbb{R}$, and $\sum_{k=1}^K \bra{\lambda^{(k)}}\rho(\vec{x})\ket{\lambda^{(k)}} = 1$. The random measurement processes of $\rho(\vec{x})$ in the eigenbasis $\ket{\lambda^{(k)}}$ can be modelled as sampling of eigenvalue $\lambda^{(k)}$ from the associated probability distribution, i.e., $\lambda \sim p_{\vec{x}}(\lambda)$ with $\lambda \in \Lambda  = \{ \lambda^{(k)}\}_{k=1}^K$ and $p_{\vec{x}}(\lambda) = \bra{\lambda}\rho(\vec{x})\ket{\lambda}$.

Given $N_s$ i.i.d. measurement outcomes $\{\lambda_i\}_{i=1}^{N_s}$, one could estimate $f(\vec{x})$
by the empirical mean
\begin{align}
    \bar{y}_D = \frac{1}{N_s} \sum_{i=1}^{N_s} \lambda_i.
\end{align}
where we have implictly assumed the dependence of $\bar{y}_D$ on $\vec{x}$, i.e., $\bar{y}_D \equiv \bar{y}_D(\vec{x})$. The finite-measurement-outcome mean $\bar{y}_D$ is an unbiased estimator of $f(\vec{x})$
\begin{align} \label{Eq:Avg-estimator}
    f(\vec{x}) = \mathbb{E}_{\bar{y}_D} \left[\bar{y}_D | \vec{x} \right]
\end{align}
with variance $\sigma^2_{\bar{y}_D|\vec{x}} = \sigma^2_{\lambda|\vec{x}}/N_s$. 

It is however generally hard to measure $\rho(\vec{x})$ in the eigenbasis of the observable. In typical scenarios, one would normally consider a linear combination of $M$ Pauli observables $P_i$, i.e., $O = \sum_{i=1}^{M} a_i P_i$ with $a_i \in \mathbb{R}$, and the eigenbasis of such an observable is non-trivial to find. To estimate the original observable expectation value, one will typically measure the expectation value of $\rho(\vec{x})$ against each Pauli observable and then linearly combine them with the associated weights $a_i$
\begin{align}
    \tr(\rho(\vec{x})O) = \sum_{i=1}^{M} a_i \tr(\rho(\vec{x}) P_i).
\end{align}
While each Pauli estimator is unbiased for the associated Pauli observable, the joint estimators of $\tr(\rho(\vec{x})O)$ constructed by summing these Pauli estimators need not be unbiased.

\subsection{Shadow-based estimation}

Alternatively, estimating measurement results using classical shadows~\cite{huang2020predicting} also introduces structural noise to our framework, allowing training based on random measurements as opposed to direct measurements, which may be much more costly in practice.

Recall that in the classical shadows protocol, to estimate properties of a quantum state $\rho$, one first evolves the quantum state using a unitary $U$ sampled from a tomographically complete unitary ensemble $\mathcal U$. It performs measurements on the computational basis $|\hat b\rangle \in \{0,1\}^n$ and the bit-string $b$ is observed with probability $\Pr[\hat b = b] = \langle b|U\rho U^\dagger|b\rangle$. From the measurement outcome, one can construct a classical snapshot $\hat{\rho} = \mathcal M_{\mathcal{U}}^{-1}(U^\dagger|\hat b\rangle\langle\hat b|U)$ where we apply an inverted quantum channel $\mathcal M_{\mathcal{U}}^{-1}$ determined by the unitary ensemble $\mathcal U$. This classical snapshot is an unbiased estimator for the density matrix, i.e., $\rho = \mathbb{E}_{U,\ket{\hat{b}}}[\hat{\rho}]$. 

The classical snapshot is a good estimator for the parameterized quantum state when applied to a suitable set of Hermitian observables $\{H_1, H_2, \cdots, H_{M}\}$. 
For example, the classical snapshots can be used to estimate the function $f(\vec{x}) = \tr(\rho(\vec{x}) O)$ with $O = \sum_{i=1}^{M} a_i P_i$, i.e., $\bar{y}_{CS} = \tr(\bar{\rho} (\vec{x}) O)$ where $\bar{\rho} = \frac{1}{N_s}\sum_{i=1}^{N_s} \hat{\rho}_i$ is the averaged sum of $N_s$ classical snapshots. It is straightforward to show that such an estimator is an unbiased estimator of $f(\vec{x})$
\begin{align}
    f(\vec{x}) = \mathbb{E}_{\bar{y}_{CS}}\left[\bar{y}_{CS}| \vec{x} \right].
\end{align}
Other unbiased shadow estimation techniques based on random sampling such as operator shadows~\cite{guo2023estimating} can also be used to produce unbiased estimators for the quantum models.

\section{Details on the learning algorithm}

\subsection{Intuitive understanding on the working principle of \texorpdfstring{\Cref{Algo:Learning-algorithm}}{Algorithm 1}}
\label{Sec:Algo-intuition}

\cref{Algo:Learning-algorithm} works similarly to gradient descent algorithm.  
Take the following empirical risk
\begin{equation}
\Remp  = \frac{1}{2N_1}\sum_{j=1}^{N_1}\left|y_j - u\left(\langle \vec{w}, \vec{\phi} (\vec{x}_j)\rangle\right)\right|^2 
\end{equation}
We can then upper bound the gradient as follows: 
\begin{equation}
\frac{\partial \Remp}{\partial \vec w} \le \frac{L}{N_1} \sum_{j=1}^{N_1}(u(\langle \vec w, \vec{\phi}(\vec{x}_j)\rangle) - y_j)\vec{\phi}(\vec{x}_j)
\end{equation}
By introducing kernelization to linear models, one can set $\vec w = \sum_{i=1}^{N_1}\alpha_i \vec{\phi}(\vec{x}_i)$. Setting the upper bound as the gradient step with a learning rate of $\frac{1}{L^2}$, we see that in each step, the value of $\alpha_j$ has an update of 
\begin{equation}
    \Delta \alpha_j = \frac{u(\langle \vec w, \vec{\phi}(\vec{x}_j)\rangle) - y_j}{LN_1} = \frac{u(\langle \vec w, \vec{\phi}(\vec{x}_j)\rangle) - y_j}{LN_1}= \frac{u\left(\sum_{i=1}^{N_1} \alpha_i k(\vec{x_i}, \vec{x_j})\right) - y_j}{LN_1},
\end{equation}
giving us the update in \cref{Algo:Learning-algorithm}.

Due to the initialization of parameters to zero, which is akin to interior point methods, the algorithm provides implicit norm regularization of the parameters. This property, in addition to the limitation of gradient steps taken, provides the theoretical guarantees as shown in \cref{Theorem:Alphatron-guarantee}, which we show in the following section.

\subsection{Proof of \texorpdfstring{\Cref{Theorem:Alphatron-guarantee}}{Theorem 1}}~\label{appTheorem1Proof}

We are given the output range $\mathcal{Y} = [-\Delta,\Delta]$.
Let $\bar{\Gamma} := \frac{1}{N_1} \sum_{i=1}^{N_1} (\bar{y}_i - u(\langle \vec{w}, \vec{\phi}(\vec{x}_i) \rangle + \xi(\vec{x}_i)))\vec{\phi}(\vec{x}_i)$, $\bar{\Gamma}^t := \frac{1}{N_1} \sum_{i=1}^{N_1} (\bar{y}_i - u(\langle \vec{w}^t, \vec{\phi}(\vec{x}_i) \rangle ))\vec{\phi}(\vec{x}_i)$ and $\chi := \frac{1}{N_1} \sum_{i=1}^{N_1} \xi(\vec{x}_i)^2$. We apply the Lem. 11 from Ref.~\cite{goel2019learning} to the empirical mean $\bar{y}$. 

\begin{lemma}\label{Lemma:Bound-diff} At iterative $t$ in \cref{Algo:Learning-algorithm}, suppose $\| \vec{w}^t - \vec{w} \| \le B$ for $B > 1$, then if $\|\bar{\Gamma}\| \le \epsilon_4 < 1$, then
\begin{align}
    \| \vec{w}^{t} - \vec{w} \|^2 - \| \vec{w}^{t+1} - \vec{w} \|^2 \ge \lambda \left( \left(\frac{2}{L} - \lambda \right) \hat{R}(h^t) -2\Delta\sqrt{\chi} - 2B\epsilon_4 - \lambda \epsilon_4^2 -2\Delta\lambda \epsilon_4 \right),
\end{align}
where $\lambda$ is the regularization parameter.
\end{lemma}
Using \cref{Lemma:Bound-diff} with $\lambda = 1/L$, we have
\begin{align} \label{Eq:Iterative-bound}
    \| \vec{w}^{t} - \vec{w} \|^2 - \| \vec{w}^{t+1} - \vec{w} \|^2 \ge \frac{1}{L} \left( \frac{\hat{R}(h^t)}{L} -2\Delta\sqrt{\chi} - 2B\epsilon_4 - \frac{\epsilon_4^2}{L} - \frac{2\Delta\epsilon_4}{L} \right).
\end{align}
For each iteration $t$ of \cref{Algo:Learning-algorithm}, one of the following two cases needs to be satisfied
\begin{align}
    \text{Case 1:} & \:\| \vec{w}^{t} - \vec{w} \|^2 - \| \vec{w}^{t+1} - \vec{w} \|^2 > \frac{B \epsilon_4}{L}\\ 
    \text{Case 2:} & \:  \:\| \vec{w}^{t} - \vec{w} \|^2 - \| \vec{w}^{t+1} - \vec{w} \|^2 \le \frac{B \epsilon_4}{L}.
\end{align}
Let $t^*$ be the first iteration where Case 2 holds. We show that such an iteration exists. Assume the contradictory, that is, Case 2 fails for each iteration. Since $\|\vec{w}^0 - \vec{w}\|_2^2 = \| \vec{0} - \vec{w}\|_2^2 \le B^2$ by assumption, however,
\begin{align}
    B^2 & \ge \|\vec{w}^0 - \vec{w}\|_2^2 \ge \|\vec{w}^0 - \vec{w}\|_2^2 - \|\vec{w}^k - \vec{w}\|_2^2 \\
    &= \sum_{t=0}^{k-1} \left(\|\vec{w}^t - \vec{w}\|_2^2 - \|\vec{w}^{t+1} - \vec{w}\|_2^2 \right) \ge \frac{kB \epsilon_4}{L},
\end{align}
for $k$ iterations. Hence, in at most $T \ge \frac{BL}{\epsilon_4}$ iterations Case 1 will be violated and Case 2 will have to be true. Combining \cref{Eq:Iterative-bound} and Case 2 yields
\begin{align}
    \hat{R}(h^t) \le 2L\Delta\sqrt{\chi} + 3BL \epsilon_4 + \epsilon_4^2 + 2\Delta\epsilon_4
\end{align}
What remains to be done is to bound $\chi$ and to obtain $N_s$ in terms of $\epsilon_4$. Similar to Ref.~\cite{goel2019learning}, we could bound $\chi$ using Hoeffding's inequality
\begin{align}
    \sqrt{\chi} \le \sqrt{\epsilon_1} + \mathcal O\left( M \sqrt[\leftroot{-2}\uproot{2}4]{\frac{\log(1/\delta)}{N_1}} \right),
\end{align}
and therefore by observing that $B \propto \Delta$, we have
\begin{align}
    \hat{R}(h^t) \le \mathcal O \left(L\Delta\sqrt{\epsilon_1} + LM \sqrt[\leftroot{-2}\uproot{2}4]{\frac{\log(1/\delta)}{N_1}} + BL \epsilon_4\right).
\end{align}
By definition, we have that $(\bar{y}_i - u(\langle \vec{w}, \vec{\phi}(\vec{x}_i) \rangle + \xi(\vec{x}_i)))\vec{\phi}(\vec{x}_i)$ are zero mean i.i.d. random variables with bounded norm, so we can use the following vector Bernstein inequality to bound the norm of $\bar{\Gamma}$.

\begin{lemma}[Vector Bernstein inequality; Lem. 18,~\cite{kohler2017sub}] Let $\vec{x}_1,\dots,\vec{x}_N$ be independent zero-mean vector-valued random variables with common dimension $d$ and they are uniformly bounded and also the variance is bounded above $\mathbb{E}[\|\vec{x}_i\|^2] \le \sigma^2$. Let
\begin{align}
    \vec{z} = \frac{1}{N} \left| \left|\sum_{i=1}^N \vec{x}_i \right| \right|_2.
\end{align}
Then we have for $0 \le \epsilon \le \sigma^2/\mu$
\begin{align}
    P[\| \vec{z}\| \ge \epsilon] \le \exp\left(-\frac{N\epsilon^2}{8\sigma^2} + \frac{1}{4} \right).
\end{align}
\end{lemma}

Before using the vector Bernstein inequality, we need to compute the variance of $\|(\bar{y} - \mathbb{E}_{\bar{y}}[\bar{y}|\vec{x}])\vec{\phi}(\vec{x})\|$ where $\mathbb{E}_{\bar{y}}[\bar{y}|\vec{x}] = u(\langle \vec{w}, \vec{\phi}(\vec{x}) \rangle + \xi(\vec{x}))$, i.e., $\mathbb{E}_{\bar{\mathcal{D}}}[\|(\bar{y} - \mathbb{E}_{\bar{y}}[\bar{y}|\vec{x}])\vec{\phi}(\vec{x})\|^2]$ as $\mathbb{E}_{\bar{\mathcal{D}}}[\|(\bar{y} - \mathbb{E}_{\bar{y}}[\bar{y}|\vec{x}])\vec{\phi}(\vec{x})\|] = 0$ by definition. Therefore, 
\begin{align}
    \mathbb{E}_{\bar{\mathcal{D}}} \left[\|(\bar{y} - \mathbb{E}_{\bar{y}}[\bar{y}|\vec{x}])\vec{\phi}(\vec{x})\|^2 \right] 
    &= \mathbb{E}_{\bar{\mathcal{D}}}\left[(\bar{y} - \mathbb{E}_{\bar{y}}[\bar{y}|\vec{x}])^2\|\vec{\phi}(\vec{x})\|^2 \right]\\
    &\le \mathbb{E}_{\bar{\mathcal{D}}}\left[(\bar{y} - \mathbb{E}_{\bar{y}}[\bar{y}|\vec{x}])^2 \right]\\
    &= \mathbb{E}_{\bar{\mathcal{D}}}\left[\bar{y}^2 -2\bar{y}\mathbb{E}_{\bar{y}}[\bar{y}|\vec{x}] + \mathbb{E}_{\bar{y}}[\bar{y}|\vec{x}]^2 \right]\\
    &= \mathbb{E}_{\vec{x}} \mathbb{E}_{\bar{y}|\vec{x}}\left[\bar{y}^2 -2\bar{y}\mathbb{E}_{\bar{y}}[\bar{y}|\vec{x}] + \mathbb{E}_{\bar{y}}[\bar{y}|\vec{x}]^2 \right]\\
    &= \mathbb{E}_{\vec{x}} \left[ \mathbb{E}_{\bar{y}}[\bar{y}^2 |\vec{x}] - \mathbb{E}_{\bar{y}}[\bar{y}|\vec{x}]^2 \right]\\
    &= \mathbb{E}_{\vec{x}} \left[ \sigma^2_{\bar{y}|\vec{x}} \right]\\
    &= \frac{\mathbb{E}_{\vec{x}} \left[ \sigma^2_{y|\vec{x}} \right]}{N_s}
\end{align}
We can therefore bound $\|\bar{\Gamma}\|$ by letting $\sigma = \frac{\mathbb{E}_{\vec{x}} \left[ \sigma^2_{y|\vec{x}} \right]}{N_s}$
\begin{align}
    P(\|\bar{\Gamma}\| \le \epsilon_4) \ge 1 - \exp\left(-\frac{N_1 N_s\epsilon_4^2}{8\mathbb{E}_{\vec{x}} \left[ \sigma^2_{y|\vec{x}} \right]} + \frac{1}{4} \right).
\end{align}
For a probability of $1-\delta$, number of training samples $N_1$, and number of measurement repetitions $N_s$, we can achieve $\|\bar{\Gamma}\| \le \epsilon_4$ with
\begin{align}
    \epsilon_4 = \sqrt{\frac{8 \mathbb{E}_{\vec{x}} \left[ \sigma^2_{y|\vec{x}} \right]}{N_1 N_s} \left(\log(\frac{1}{\delta}) + \frac{1}{4} \right)}.
\end{align}

To bound $R(h^{t^*})$ with $\hat{R}(h^{t^*})$, we require the following results:
\begin{theorem}[Thm. 8,~\cite{bartlett2002rademacher}; Formulation of Thm. 21,~\cite{goel2019learning}]
Let $\mathcal{L}: \mathcal{Y}' \times \mathcal{Y} \to \mathbb{R}_+$ be a loss function upper bounded by $b>0$ and such that for any fixed $y$, $y' \to \mathcal{L} (y', y)$ is $L$-Lipschitz for some $L >0$. Given function class $\mathcal{F} \subset (\mathcal{Y}')^{\mathcal{X}}$, for any $f: \mathcal{X} \to \mathcal{Y}' \in \mathcal{F}$, and for any sample $\mathcal{S}$ from distribution $\mathcal{D}$ of size $N$,
\begin{equation}
\left|\mathbb{E}_{\mathcal{D}}\left[\mathcal{L}(f(x), y)\right] - \frac{1}{N} \sum_{(x, y) \in \mathcal{S}} \mathcal{L}(f(x), y)\right| \le 4L \mathfrak{R}_{N} (\mathcal{F}) + 2b\sqrt{\frac{\log 2/ \delta}{2N}},
\end{equation}
where $\mathfrak{R}_{N} (\mathcal{H})$ is the expected value of the empirical Rademacher complexity of the function class $\mathcal{F}$ over all samples of size $N$.
\end{theorem}
Plugging the generalization, we have the following result for all hypotheses $h \in \mathcal{H}$.
\begin{equation}
\ER(h) \le \hat{R}(h) + 8\Delta \mathfrak{R}_{N_1} (\mathcal{H}) + 2\Delta^2\sqrt{\frac{\log (\frac{2}{\delta})}{2N_1}}.
\end{equation}

Next, to compute the empirical Rademacher complexity of $\mathcal{H}$, we use the following results:
\begin{theorem}[Lem. 22,~\cite{bartlett2002rademacher} or Thm. 1,~\cite{kakade2008complexity}; Formulation of Thm. 22,~\cite{goel2019learning}]
Let $\mathcal{X}$ be a subset of an inner product space such that for all $\vec x \in \mathcal{X}$, $\|\vec{x}\|_2 \le X$, and let $\mathcal{W} = \{\vec x \to \langle \vec w, \vec x\rangle,  \|\vec{w}\|_2 \le W\}$. Then it holds that
\begin{equation}
    \mathfrak{R}_N(\mathcal{W}) \le \frac{XW}{\sqrt{N}},
\end{equation}
\end{theorem}

\begin{lemma}[Talagrand's lemma, Cor. 3.17,~\cite{ledoux1991probability}; Formulation of Lem. 5.7,~\cite{mohri2018foundations}]
Let $\Phi:\mathbb{R} \to \mathbb{R}$ be a $L$-Lipschitz function. Then for any hypothesis set $\mathcal{F}$ of real-valued functions, the following holds:
\begin{equation}
    \mathfrak{R}_N(\Psi\circ\mathcal{F}) \le L\mathfrak{R}_N(\mathcal{F}),
\end{equation}
\end{lemma}
Noting that our hypothesis class $\HC$ in question is a linear class with a $L$-Lipschitz function applied to it, with the constraints $\|\vec w\|_2 \le B$ and $\|\vec{\phi(x)}\|_2 \le 1$ combining the above two results, we get
\begin{equation}
\mathfrak{R}_{N_1} (\mathcal{H}) \le \frac{BL}{\sqrt{N_1}},
\end{equation}

Finally, we can make use of Rademacher complexity to bound $R(h^t)$
\begin{align} 
    \ER(h^{t^*}) &\le \hat{R}(h^{t^*}) + \mathcal O\left(BL\Delta\sqrt{\frac{1}{N_1}} + \Delta^2\sqrt{\frac{\log(1/\delta)}{N_1}} \right) \\ \label{Eq:Learning-guarantee}
    &= \mathcal O \left(L\Delta\sqrt{\epsilon_1} + L\Delta M \sqrt[\leftroot{-2}\uproot{2}4]{\frac{\log(1/\delta)}{N_1}} + BL\Delta\sqrt{\frac{1}{N_1}} + BL \sqrt{\frac{ \bar\sigma}{N_1 N_s} \left(\log(\frac{1}{\delta})\right)} + \Delta^2\sqrt{\frac{\log(1/\delta)}{N_1}}\right)
\end{align}
where $\bar\sigma = \mathbb{E}_{\vec{x}} \left[ \sigma^2_{y|\vec{x}} \right].$ The last step is to show that we can indeed find a hypothesis satisfying the above guarantee. Using the Hoeffding inequality and union bound, one could show that $N_2 \ge O\left(N_1 \Delta^2 \log(\frac{T}{\delta^2}) \right)$ validation data points suffice to choose the optimal hypothesis $h^{t^*}$ at time step $t^*$ that satisfies \cref{Eq:Learning-guarantee}.

\section{Bias-variance-noise decomposition}
\label{Sec:Bias-Variance-Noise-Decomposition}

For a given training dataset $\mathcal S = (\vec{x}_i, \bar{y}_i)_{i=1}^{N_1}$, one would obtain an associated trained model $h_{\mathcal S}(\vec{x})$ by optimizing the empirical risk $\Remp(h)$
using some optimization methods such as the gradient descent algorithm. 
Further, different training datasets $\mathcal S'$ would yield different trained models $h_{\mathcal S'}(\vec{x})$, each associated with explicit risk $\ER(h_{\mathcal{S}})$ and $\ER(h_{\mathcal{S}'})$, respectively. 
This observation begs the question of how these trained models are related to each other and the target concept $c(\vec{x}) = \mathbb{E}_{\bar{y}}[\bar{y}|\vec{x}]$. 

To address the above-mentioned questions, we introduce a machine learning concept known as the bias-variance trade-off. The bias of machine learning models informs their consistent errors and it is defined as
\begin{align}
    \textbf{Bias}_{\mathcal S} := \mathbb{E}_{\mathcal S}[h_{\mathcal S}(\vec x)] - f(\vec{x}).
\end{align}
Since $f(\vec{x})$ is independent of $\mathcal S$, one could express the bias as $\textbf{Bias}_{\mathcal S} = \mathbb{E}_{\mathcal S}[h_{\mathcal S}(\vec x) - f(\vec{x})]$. 
Low bias suggests that on average the trained models $h_{\mathcal S}(\vec x)$ are close to the target function, and typically machine learning models with larger model class sizes will have lower bias.
Yet, models with low bias need not be optimal as they tend to be more sensitive to variations in training data; such models are said to have high variance where the variance of the model is defined as
\begin{align}
     \textbf{Var}_{\mathcal S} & := \mathbb{E}_{\mathcal S} \left[\left(\mathbb{E}_{\mathcal S}[h_{\mathcal S}(\vec{x})] - h_{\mathcal S}(\vec{x})\right)^2 \right].
\end{align}
It should be clear from the definition that the bias and variance are dependent on the complexity of the hypothesis class, the number of training data points $N_1$ and the number of random labels $N_s$.

The above-mentioned bias-variance trade-off can be studied by analysing the average behaviour of the trained models under the constraints of finite training data points and labels, which we capture by taking the expectation value over all possible training data sets $\mathcal S$ with the same $N_1$ and $N_s$, which we denote $\mathbb{E}_{\mathcal S| N_1, N_s}[\IR(h_{\mathcal S})]$
where $\IR(h_S) = \mathbb{E}_{\mathcal{\bar{D}}}[(h_S(\vec{x})- \bar{y})^2]$ is the implicit risk of $h_S(\cdot)$ as defined in \cref{Eq:Exp-sq-err}. 

This averaged risk quantifies the overall performance of the hypothesis class $\HC$ under all realization of training datasets of size $N_1$, with empirical means estimated using $N_s$ random labels. 
Furthermore, $\IR(h) = \ER(h) + \bar{\sigma}_{N_S}$ implies
\begin{align} \label{Eq:Ensemble-err_Ensemble-varepsilon}
    \mathbb{E}_{\mathcal S| N_1, N_s}[\IR(h_{\mathcal S})] = \mathbb{E}_{\mathcal S| N_1, N_s}[\ER(h_\mathcal{S})] + \bar{\sigma}_{N_S},
\end{align}
where $\mathbb{E}_{\mathcal S| N_1, N_s}[\ER(h_{\mathcal S})]$ is the averaged explicit risk and $\IR(\mathbb{E}_{\bar{y}}[\bar{y}|\vec{x}]) := \bar{\sigma}_{N_S} = \mathbb{E}_{\vec{x}}\left[ \sigma^2_{\bar{y}|\vec{x}} \right] = \frac{1}{N_s} \mathbb{E}_{\vec{x}}\left[ \sigma^2_{y|\vec{x}} \right]$ is the irreducible error that lower bounds the implicit risk on unseen sample. 
Note that in the asymptotic regime ($N_1 \rightarrow \infty$), the variance goes to 0 as the finite data sampling noise diminishes, and one would consistently obtain the optimal model in $\HC$ that achieves the optimal explicit risk $\ER(h)$. 
In addition, the strength of the irreducible error is controllable by the number of random samples $N_s$. 
In particular, $\bar{\sigma}_{N_S} \rightarrow 0$ as $N_s \rightarrow \infty$ and therefore $\mathbb{E}_{\mathcal S| N_1, N_s\rightarrow \infty}[\IR(h_{\mathcal S})] \rightarrow \mathbb{E}_{\mathcal S| N_1, N_s \rightarrow \infty}[\ER(h_{\mathcal S})]$.

We can then further decompose the explicit risk averaged over all datasets.
\begin{align}
    \mathbb{E}_{\mathcal S}[\ER(h_{\mathcal{S}})]
    &= \mathbb{E}_{\vec x, \mathcal S}\left[\left(h_{\mathcal S}(\vec x) - c(\vec x)\right)^2\right] \\
    &= \mathbb{E}_{\vec x, \mathcal S}\left[\left( h_{\mathcal S}(\vec x)- \mathbb{E}_{\mathcal S}[h_{\mathcal S}(\vec x)] + \mathbb{E}_{\mathcal S}[h_{\mathcal S}(\vec x)] - c(\vec x)\right)^2\right]\\
    &=\mathbb{E}_{\vec x}\left[\left(\mathbb{E}_{\mathcal S}[h_{\mathcal S}(\vec x)] - c(\vec x)\right)^2\right] + \mathbb{E}_{\vec x, \mathcal S}\left[\left( \mathbb{E}_{\mathcal S}[h_{\mathcal S}(\vec x)] - h_{\mathcal S}(\vec x) ]\right)^2\right]\\
    &= \mathbb{E}_{\vec{x}}\left[\textbf{Bias}^2_{\mathcal S} \right] + \mathbb{E}_{\vec{x}} \left[\textbf{Var}_{\mathcal S} \right].
\end{align}
In summary, we have
\begin{align}
    \mathbb{E}_{\mathcal S| N_1, N_s}[\ER(h_{\mathcal{S}})] &= \mathbb{E}_{\vec{x}}\left[\textbf{Bias}^2_{\mathcal S} \right] + \mathbb{E}_{\vec{x}} \left[\textbf{Var}_{\mathcal S} \right], \quad \text{and} \\
    \mathbb{E}_{\mathcal S| N_1, N_s}[\IR(h_{\mathcal{S}})] &= \mathbb{E}_{\vec{x}}\left[\textbf{Bias}^2_{\mathcal S} \right] + \mathbb{E}_{\vec{x}} \left[\textbf{Var}_{\mathcal S} \right] + \bar{\sigma}_{N_s}.
\end{align}
\section{Random Fourier feature models}\label{a:RFF}

\subsection{Classical approximation of PQC functions}

Let $\mathcal{F}_{U, O}$ and $\vec{\phi}_F(\vec{x})$ be the PQC concept class and feature map as defined in \cref{Eq:PQC-concept-class} and \cref{Eq:Full-feature-map}, respectively. In addition, let $k_F(\vec{x}, \vec{x}') = \langle \vec{\phi}_F(\vec{x}), \vec{\phi}_F(\vec{x}') \rangle$ be the kernel of $\vec{\phi}_F(\vec{x})$. 
Our goal of learning PQCs corresponds to taking $\mathcal{F}_{U, O}$ as our concept class. 
It is a still-unresolved question which PQCs provably give rise to function families that can or cannot be well approximated by kernel-based function families, but it is known that PQCs exist which cannot.
Nonetheless, we offer a generic PQC construction whose functions are guaranteed to be well-approximated by a kernel-based hypothesis family.
The recipe we propose does not exactly match the typical PQCs used by practitioners, but it is generic enough that it may become useful in the future.
The construction relies on the Linear Combination of Unitaries (LCU) framework~\cite{childs2012hamiltonian} and resembles constructions proposed in e.g. Refs.~\cite{GilFuster2024expressivity,gilfuster2024understanding}.

Let $k_F$ be a kernel that can be well-approximated as an Embedding Quantum Kernel (EQK)~\cite{GilFuster2024expressivity} on $n$ qubits, meaning there exists a data-dependent unitary gate $U(\vec{x})$ such that
\begin{align}
    \sup_{\vec x,\vec x'\in\mathcal{X}} \left\lvert k_F(\vec x,\vec x')-\lvert\langle\vec{0}\rvert U^\dagger(\vec x)U(\vec x')\lvert\vec{0}\rangle\rvert^2\right\rvert^2 &\leq \epsilon
\end{align}
for almost every $\vec x,\vec x'\in\mathcal{X}$.
Then, given $N\in\mathbb{N}$, a vector of real numbers $\vec\alpha=(\alpha_i)_{i=1}^N$, and a set of inputs $\vec x_1,\ldots, \vec x_N$, consider a PQC over $n+\lceil\log(N)\rceil$ qubits.
The circuit starts on the all-$0$ state, and the unitary $U(\vec x)$ is applied on the first $n$-qubits.
In parallel, we perform amplitude encoding of $\vec\alpha$ on the other $\lceil\log(N)\rceil$ qubits of the auxiliary register.
Next, we define a controlled operation $CU_i$ which, conditional on the auxiliary register being in state $\lvert i\rangle$ for $i\in\{1,\ldots, N\}$ applies $U^\dagger(\vec x_i)$ on the main register.
It follows that these controlled operations commute $[CU_i, CU_j]=0$.
We need only apply all such controlled gates in sequence, then: $\prod_{i=1}^N CU_i$, and measure the probability of the first $n$-qubits being in the all-$0$ state at the end (together with a diagonal observable on the auxiliary register that takes care of the negative signs in $\vec\alpha$).
For notational ease, we do not explicitly write the extra observable on the auxiliary register, and we write only the projector on the all-$0$ state.
This means that the functions can take negative values even though they are defined as the absolute square of a complex number.
This way, given $\vec\alpha$ and $\vec x, \vec x_1,\ldots, \vec x_N$ we have defined a PQC in the form of a unitary $W(\vec\alpha, \vec x, (\vec x_i)_i)$, and produces as output a function in the kernel-based hypothesis family:
\begin{align}
    \lvert\langle\vec{0}\rvert W(\vec\alpha, \vec x, (\vec x_i)_i)\lvert\vec{0}\rangle\rvert^2 &= \sum_{i=1}^N \alpha_i \lvert\langle\vec{0}\rvert U^\dagger(\vec x)U(\vec x')\lvert\vec{0}\rangle\rvert^2.
\end{align}
From the $\epsilon$-approximation of the initial kernel $k_F$ via the EQK defined by $U$, it follows that each function of the form $\sum_{i=1}^N \alpha_i k(\vec x,\vec x_i)$ can be approximated by a function in $\mathcal{F}_{W,\lvert\vec0\rangle\!\langle\vec0\rvert}$, by taking the same $\vec \alpha$ vector and the same set of inputs $(\vec x_i)_i$.
Without loss of generality, we assume the parameter vector $\vec\alpha$ has bounded norm $\lVert\vec\alpha\rVert_2^2\leq B$:
\begin{align}
    & \sup_{\vec x\in\mathcal{X}} \left\lvert \left(\sum_{i=1}^N \alpha_i k_F(\vec x,\vec x_i)\right) - \lvert\langle\vec{0}\rvert W(\vec\alpha, \vec x, (\vec x_i)_i)\lvert\vec{0}\rangle\rvert^2\right\rvert^2 \\
    = & \sup_{\vec x\in\mathcal{X}}\left\lvert \sum_{i=1}^N \alpha_i \left(k_F(\vec x,\vec x_i) - \lvert\langle\vec{0}\rvert U^\dagger(\vec x)U(\vec x_i)\lvert\vec{0}\rangle\rvert^2\right) \right\rvert^2 \\
    \leq & \norm{\vec\alpha}^2 \sum_{i=1}^N \sup_{x\in\mathcal{X}}\left\lvert k_F(\vec x,\vec x_i) - \lvert\langle\vec{0}\rvert U^\dagger(\vec x)U(\vec x_i)\lvert\vec{0}\rangle\rvert^2\right\rvert^2 \\
    \leq & B^2\epsilon
\end{align}

Altogether, this recipe allows us to construct a PQC whose associated function family is the same as a given kernel-based function family.
For \cref{Algo:Learning-algorithm} to succeed as a classical learner of this function family, then, we need only be able to evaluate the kernel $k_F$ efficiently classically.
It is known that the complexity of evaluating the trigonometric kernels that result from quantum embeddings is upper-bounded by the cardinality of the frequency spectrum $\Tilde\Omega$ arising from the encoding strategy.

\subsection{Approximating PQCs with random Fourier features}
\label{Subsubsec:PQCs-RFF}

If the PQC $U(\vec x,\vec\theta)$ is such that its corresponding feature map is of polynomial dimension $\lvert\Tilde\Omega\rvert\in\mathcal{O}(\poly(m))$, then we know we can classically express the corresponding function exactly: $\bra{{\bf 0}} U^\dagger(\vec{x}, \vec\theta) O U(\vec{x}, \vec\theta) \ket{{\bf 0}} = \langle \vec w_F, \vec\phi_F (\vec{x}) \rangle$, where the real-valued vector $\vec w_F$ is efficiently storable in classical memory.
Refs.~\cite{caro2021encoding,schreiber2023classical} offer a discussion on what encoding strategies connected to families of PQCs will result in Fourier spectra of polynomial size.
It is nevertheless known that many natural encoding strategies result in an exponentially large Fourier spectrum, where we cannot rely on an exact realization of the PQC function as a classical linear map.
Some of these cases have been recently analyzed in Refs.~\cite{fontana2023classical,sweke2023potential} under the lens of Random Fourier Features (RFF)~\cite{rahimi2007random}.
The main idea in RFF is to efficiently approximate the high-dimensional inner product $\langle\vec w_F,\vec\phi_F(\vec x)\rangle$ by sampling a few of its dominant terms.

For instance, consider an encoding strategy which gives rise to an exponentially large Fourier spectrum $\lvert\Omega\rvert\propto\exp(m)$.
Then, the inner product
\begin{align}
    \langle\vec{w},\vec{\phi}_F(\vec {x})\rangle &= \sum_{j=1}^{2\lvert\Omega\rvert-1} w_j\phi_{F,j}(\vec{x})
\end{align}
cannot be classically evaluated in general due to its containing many terms.
Now consider a specific PQC $U(\vec{x}, \vec\theta)$ with this encoding strategy, but which is structured enough that we know that some entries of the weight vector are more dominant than others, in that they contribute more to the sum.
One way to capture this would be by considering a probability distribution over the Fourier spectrum $P(\Omega)$, where the probability associated with a specific frequency is proportional to the magnitude squared of its coefficients $p(\vec\omega)=a_{\vec\omega}^2+b_{\vec\omega}^2$. 
Without loss of generality, we assume the coefficients are already properly normalized.
Then, what the RFF algorithm prescribes we do is sample a number $D$ of frequencies from such a distribution $\tilde{\vec{\omega}}\sim P^D$, and then consider the classical efficient feature map $\vec{\phi}_{\mathrm{RFF}}(\vec{x})$ consisting of only those frequencies:
\begin{align}
    \vec{\phi}_{\mathrm{RFF}}(\vec{x}) = \frac{1}{\sqrt{D}}
    \begin{pmatrix}
        \cos\langle \tilde{\vec\omega}_1, \vec{x} \rangle\\
        \sin\langle \tilde{\vec\omega}_1, \vec{x} \rangle\\
        \vdots\\
        \cos\langle \tilde{\vec\omega}_D, \vec{x} \rangle\\
        \sin\langle \tilde{\vec\omega}_D, \vec{x} \rangle
    \end{pmatrix}.
\end{align}
The sampled frequencies $\tilde{\vec\omega}$ are all in the original Fourier spectrum $\tilde{\vec\omega}_i\in\Omega$, so $\vec\phi_{\mathrm{RFF}}(\vec x)$ is just an appropriately renormalized subvector of the full $\vec\phi_F(\vec x)$.
This smaller feature map gives rise to the hypothesis family:
\begin{align}
    \mathcal H_{\mathrm{RFF}} &= \{ u(\langle \vec{w},\vec{\phi}_{\mathrm{RFF}}(\vec{x})\rangle) \,|\, \vec{w}\in\mathbb R^D\}.
\end{align}
Then, if the PQC function is such that it can in principle be approximated as a linear map of rank $D$~\cite{jerbi2023quantum}, it follows from Refs.~\cite{rahimi2007random,landman2022classically,sweke2023potential} that the RFF hypothesis family should contain a good approximation to the function.
In the context of this work, this means that \cref{Algo:Learning-algorithm} should be able to learn the initial PQC function by using $\mathcal H_{\mathrm{RFF}}$ as a hypothesis class.

The remaining question is, again, how to specify the right $\mathcal H_{\mathrm{RFF}}$ for a given PQC of interest, modelled by the function family $\mathcal F_{U,O}$.
The ultimate general-case answer is not fully resolved~\cite{sweke2023potential}, but we provide a recipe to, given an RFF-approximable EQK $k$, construct a corresponding PQC.

Let $k$ be a kernel that can be approximated as the inner product of a feature map $\vec\phi$ of polynomial size (in particular, this could be the randomized feature map produced by the RFF algorithm):
\begin{align}
    \sup_{\vec x,\vec x'\in\mathcal{X}} \left\lvert k(\vec{x},\vec{x}') - \langle\vec\phi(\vec x),\vec\phi(\vec x')\rangle\right\rvert^2 &\leq \epsilon,
\end{align}
for almost every $\vec x, \vec x'\in\mathcal{X}$.
Let further $U(\vec x)$ be a quantum embedding that implements the feature map $\vec\phi(\vec x)$ (w.l.o.g. we take $\vec\phi$ to be properly normalized):
\begin{align}
    \sup_{\vec x,\vec x'\in\mathcal{X}} \left\lvert \langle \vec{0} \lvert U^\dagger(\vec x)U(\vec x')\lvert \vec{0}\rangle\rvert^2 - \langle\vec\phi(\vec x),\vec\phi(\vec x')\rangle\right\rvert^2 &\leq \epsilon'.
\end{align}
Then the same LCU construction we used before also results in a PQC whose function family approximates the function family of the RFF-based kernel, which in turn approximates the function family of the original kernel.
With triangular inequality, it follows that this PQC construction approximates the original kernel-based function family.
Since \cref{Algo:Learning-algorithm} can provably learn the kernel-based function family, it follows it can also learn this PQC family.

\section{Proof of \texorpdfstring{\Cref{Lemma:PQCs-learnability}}{Lemma 1}}~\label{appLemma1Proof}

To discuss the explicit risk of the ERM given the hypothesis class $\mathcal{H}$ in \cref{Eq:RFF-hypothesis-class}, we use the following result in statistical learning theory:
\begin{proposition}[Prop. 4.1,~\cite{mohri2018foundations}]\label{Prop:erm}
Let $\mathcal{D}$ be a distribution over $\mathcal{X}\times \mathcal{Y}$. Let $\mathcal{L}: \mathcal{Y}' \times \mathcal{Y} \to \mathbb{R}_+$ be a loss function. Considering a hypothesis class $\mathcal{F}$ that maps $\mathcal{X}$ to $\mathcal{Y}'$, For any sample $\mathcal{S}$ from distribution $\mathcal{D}$, for hypotheses $f \in \mathcal{F}$, the following holds:
\begin{equation}
\mathbb{E}_{\mathcal{D}}[\mathcal{L}(f_\mathcal{S}^{\mathrm{ERM}}(\vec{x}), y)] - \inf_{f\in\mathcal{F}} \mathbb{E}_{\mathcal{D}}[\mathcal{L}(f(\vec{x}), y)] \le 2\sup_{f\in\mathcal{F}}\left|\mathbb{E}_{\mathcal{D}} [\mathcal{L}(f(\vec{x}), y)] - \frac{1}{|\mathcal{S}|}\sum_{(\vec{x}, y)\in \mathcal{S}}\mathcal{L}(f(\vec{x}), y)\right|.
\end{equation}
\end{proposition}
Considering the implicit loss function $\ell_\mathrm{impl}$ for the loss function $\mathcal{L}$ in the above proposition, we get
\begin{equation}
\IR(h_\mathcal{S}^{\mathrm{ERM}}) - \inf_{h\in\mathcal{H}} \IR(h) \le 2\sup_{h\in\mathcal{H}}|\IR(h) - \Remp(h)|.
\end{equation}
Noting that $\IR(h) = \ER(h) + \IR(\mathbb{E}_{\bar{y}}[\bar{y}|\vec{x}])$, we can see that 
\begin{equation}
\ER(h_\mathcal{S}^{\mathrm{ERM}}) - \inf_{h\in\mathcal{H}} \ER(h) \le 2\sup_{h\in\mathcal{H}}|\IR(h) - \Remp(h)|.
\end{equation}
By definition, we know that 
\begin{equation}
\inf_{h\in\mathcal{H}} \ER(h) = \inf_{h\in\mathcal{H}} \mathbb{E}_{\bar{\mathcal{D}}}[(h(\vec x) - c(\vec x))^2] \le \mathbb{E}_{\bar{\mathcal{D}}}[(\langle \vec{w}, \vec{\phi(x)}\rangle - \langle \vec{w}, \vec{\phi(x)}\rangle + \xi(\vec{x}))^2]  = \mathbb{E}_{\bar{\mathcal{D}}}[\xi(\vec{x})^2] \le \epsilon_1.
\end{equation}
Hence we see that 
\begin{equation}\label{eq:erm_main}
\ER(h_\mathcal{S}^{\mathrm{ERM}}) \le \epsilon_1 + 2\sup_{h\in\mathcal{H}}|\IR(h) - \Remp(h)|.
\end{equation}

We now find error bounds on the right-hand side of the previous proposition. To do so, we require the following results obtainable by combining Thm. 8 and Lem. 22 from Ref.~\cite{bartlett2002rademacher}:
\begin{theorem}[Thm. 11.11,~\cite{mohri2018foundations}]
Given distribution $\mathcal D$ over $\mathcal X \times \mathcal Y$, let $k: \mathcal X \times \mathcal X \to \mathbb{R}$ be a positive semidefinite kernel, $\Phi:\mathcal{X} \to \mathbb H$ be the feature map associated with kernel $k$, and hypothesis class $\mathcal{H} = \{\vec x \to \langle \vec w, \vec{\phi}(\vec x)\rangle, \|\vec w\|_\mathbb H \le \Lambda\}$. Assume there exists $r, M >0$ such that $k(\vec x, \vec x) \le r^2$ and for all hypotheses $h: \mathcal{X} \to \mathcal{Y}' \in\mathcal{H}$ and all $(\vec x,y) \in \mathcal{D}$, $|h(\vec x) - y| \le M$. Then for any sample $\mathcal{S}$ from $\mathcal{D}$ of size $N$, the generalization bound is as follows:
\begin{align}
\left|\eb_{(\vec x, y)\sim\mathcal{D}}\left[|h(\vec x)- y|^2\right] - \frac{1}{N} \sum_{(\vec x, y) \in \mathcal{S}} |h(\vec x)- y|^2\right| 
\in\mathcal{O} \left(\frac{M\Lambda r}{\sqrt{N}}+ M^2 \sqrt{\frac{\log\frac{1}{\delta}}{N}}\right).
\end{align}
\end{theorem}

Plugging in our error losses and range of $\mathcal{H}$, we note that $\Lambda, M \in \mathcal{O}(D)$ and $r = 1$. We then obtain the following generalization bound for $\mathcal{H}$.
\begin{equation}
|\IR(h) - \Remp(h)| \in \mathcal{O}\left(D^2\sqrt{\frac{\log\frac{1}{\delta}}{N_1}}\right),
\end{equation}
Note that this result yields a better result than the Rademacher-based generalization bound proposed by \citet{caro2021encoding} by a logarithmic factor if we use the entire Fourier spectrum instead of RFF. We can then write the explicit risk for the ERM as follows:
\begin{equation}
\ER(h_\mathcal{S}^{\mathrm{ERM}}) \in\mathcal{O}\left( \epsilon_1 + D^2\sqrt{\frac{\log\frac{1}{\delta}}{N_1}}\right).
\end{equation}

\end{document}